\documentclass[preprint,sort&compress,times,12pt]{elsarticle}
\usepackage{amsmath,colortbl,graphicx,url}
\newcommand{\bX}{\mathbf{X}}
\newcommand{\bx}{\mathbf{x}}

\begin{document}
\title{Will the US Economy Recover in 2010?\\ A Minimal Spanning Tree Study}
\author[mas]{Yiting Zhang\corref{cor1}}
\author[mas]{Gladys Hui Ting Lee\corref{cor1}}
\author[mas]{Jian Cheng Wong}
\author[pap]{Jun Liang Kok}
\author[pap]{Manamohan Prusty}
\author[pap]{Siew Ann Cheong\corref{cor2}}
\ead{cheongsa@ntu.edu.sg}

\cortext[cor1]{These authors contributed equally to this work.}
\cortext[cor2]{Corresponding author}

\address[mas]{Division of Mathematical Sciences,
School of Physical and Mathematical Sciences,
Nanyang Technological University,
21 Nanyang Link, Singapore 637371,
Republic of Singapore}

\address[pap]{Division of Physics and Applied Physics,
School of Physical and Mathematical Sciences,
Nanyang Technological University,
21 Nanyang Link, Singapore 637371,
Republic of Singapore}

\begin{abstract}
We calculated the cross correlations between the half-hourly times series of the
ten Dow Jones US economic sectors over the period February 2000 to August 2008,
the two-year intervals 2002--2003, 2004--2005, 2008--2009, and also over 11
segments within the present financial crisis, to construct minimal spanning
trees (MSTs) of the US economy at the sector level. In all MSTs, a core-fringe
structure is found, with consumer goods, consumer services, and the industrials
consistently making up the core, and basic materials, oil \& gas, healthcare,
telecommunications, and utilities residing predominantly on the fringe.  More
importantly, we find that the MSTs can be classified into two distinct,
statistically robust, topologies: (i) star-like, with the industrials at the
center, associated with low-volatility economic growth; and (ii) chain-like,
associated with high-volatility economic crisis.  Finally, we present
statistical evidence, based on the emergence of a star-like MST in Sep 2009, and
the MST staying robustly star-like throughout the Greek Debt Crisis, that the US
economy is on track to a recovery. 

\end{abstract}


\maketitle

\section{Introduction}

In the CBS `60 Minutes' interview televised on 15 March 2009, Ben Bernanke
predicted that the recession triggered by the global financial crisis will end
in 2009, and the US economy will recover in 2010 \cite{Reuters16Mar2009}.  While
we will never know whether Bernanke made the prediction based on his gut
feelings, or on simulation results from some sophisticated macroeconomic model,
what we do know is that the prediction sparked intense public debate on whether
the Chairman of the US Federal Reserve was overly optimistic.  Given that the
financial industry was still reeling from the massive October 2008 slide,
reactions to Bernanke's statement must be especially strong.  We also know that
the US Federal Reserve does not appear to be behind its Chairman: up till
September 2010, the interest rate has not been raised \cite{FRB}, even though
there has been calls from within the Federal Reserve system to tighten the money
supply \cite{Reuters3Jun2010}.  This has led to mounting concerns from
economists that the oversupply of government money, in the form of an interest
rate that is nearly zero, will cause an inflation when the economy recovers
\cite{Rose2009, Hanke2009}.  In fact, a commentator argued that US stimulus
money is fueling property bubbles all over Asia, and warned that the global
economy will crash once again in 2012 when the Feds rein in their easy money
\cite{Today19Aug2010}.

In January 2004, there was a similar call by economists to raise interest rates
\cite{Armentano2004}, when the US economy was showing signs of coming out of the
technology bubble crisis.  The Federal Reserve responded hesitantly only in June
2004 \cite{FRB}.  We can understand the concern of the US government then, and
possibly also now: how do we know that the early signs will lead on to a
recovery that will strengthen and stay the course?  From these historical and
contemporary lessons, we know that a more sensitive and more robust indicator of
economic recovery is needed.  While much work has been done on developing and
validating reliable precursor signatures (also called \emph{leading indicators})
for the onset of financial crises (see for example,
Refs.~\cite{Chauvet1998JEconSocMeasure25p235, Camacho2004JForecast23p173,
Marcellino2006HandbookEconomicForecasting1p879, Kannan2009IMFWP252,
Chalamandaris2010ApplFinEcon20p73, Simpson2010ApplFinEcon20p45,
Tamirisa2010IMFResBull11p1}), and understanding such economic disasters in
general (see for example, Refs.~\cite{Bordo2009JMonetaryEcon57p1,
Boyd2009IMFWP09141, Claessens2009EconPolicy24p653, Cecchetti2009NBER15379,
BoysenHogrefe2010KielWP1586, Claessens2010JAsianEcon21p247}), less has been done
to find robust indicators of economic recovery (see for example,
Refs.~\cite{Chauvet2000JEmpiricalFin7p87, Biggs2009DNBWP218,
Mirestean2010IMFResBull11p1}).  In this work, we hope to address this important
gap.

Recently, we adapted the recursive entropic segmentation method
\cite{BernaolaGalvan1996PhysicalReviewE53p5181,
RomanRoldan1998PhysicalReviewLetters80p1344} developed by
Bernaola-Galv\'an and coworkers for biological sequence segmentation, and
applied it to financial time series segmentation \cite{Wong2009}.  Based on our
segmentation of the Dow Jones Industrial Average (DJIA) time series between 1997 and
2008, we saw that the US economy, as measured by the DJIA, switched between a
high-volatility crisis phase and a low-volatility growth phase.  The first
crisis phase lasted from mid-1998 to mid-2003, coinciding with the US technology
bubble and the ensuing economic recession.  The second crisis phase started
in mid-2007, coinciding with the US Subprime Crisis and the ensuing global
financial crisis.  More interestingly, we could also identify a year-long series
of precursor shocks prior to the mid-1998 and mid-2007 onsets of two crisis
phases, as well as a year-long series of inverted shocks prior to the mid-2003
economic recovery.  The series of inverted shocks started with the mid-2002 Dow
Jones low, so if we believe the internal dynamics of the US economy
had not changed from the previous financial crisis to the present financial
crisis, we would naively expect the US economy to recover one year after the
March 2009 lows, i.e. the second quarter of 2010, give and take.

Clearly, a single study of a single time series spanning only two financial
crises and one growth period is hardly enough statistical evidence in Bernanke's
favor.  To enhance the statistical significance of features seen in the
segmented DJIA time series, we carried out a cross-section study, comparing the
segmented time series of the ten Dow Jones US (DJUS) economic sector indices
\cite{Lee2009arXiv09114763}.  By identifying the sequences of onsets in the ten
DJUS indices, we find sectors in the US economy going first to last into the
present financial crisis in merely two months!  While we may or may not have an
extended sequence of precursor shocks to work with for predicting market crashes
and financial crisis (see the recent update \cite{Yan2010PhysProcedia00p1} on
the heroic efforts by Sornette and coworkers), when the dominoes are set in
motion policy makers will have a month or two to contain the crisis.  Since this
financial crisis eventually spread globally, we will have to wait for the next
potential crisis to find out if containment is at all possible.  We do know,
however, that the US Federal Reserve acted promptly, announcing the first of a
series of interest rate cuts in August 2007.  Unfortunately, as detailed in
Ref.~\cite{Lee2009arXiv09114763}, we saw these rate cuts rapidly losing
effectiveness.  A critical discussion on the actions taken by the US Federal
Reserve can be found in Ref.~\cite{Mamun2010ApplFinEcon20p15}.

In the same comparative study, we also identified the sequence of economic
recoveries in the different US economic sectors.  The excruciatingly slow
complete economic recovery from the previous financial crisis, defined as
consistent growth in the first sector to consistent growth in the last sector,
took one and a half years.  Given this long time scale, developing robust
indicators to detect economic recovery, and thereafter designing timely stimulus
packages, should be easier than finding sensitive indicators that would warn us
of an impending financial crisis.  We would imagine that tracking slow
month-to-month indicators should be enough to give us a confident forecast on
the start of growth, but all through the second half of 2009 and 2010 to date,
we hear commentators mostly urging caution \cite{InvestingDaily5May2009,
Telegraph8May2009, Bloomberg11May2009, Forbes14May2009,
ChannelNewsAsia2July2009, MarketOracle30Aug2009, Today25Mar2010, CNN6Apr2010}.
We believe this cautious outlook can be blamed partly on swings in the stock
markets, which always become strong when things are taking a turn for the better
or for the worst.  Perhaps the way to allay such market-driven fears is to
extract convincing signs from the high-frequency stock-market data itself.
Based on these signs, policy makers can then tell more confidently that the
economy will recover in a matter of months, and start planning measures to
further stimulate the recovery.

In this paper, which is organized into six sections, we report a minimal
spanning tree (MST) study of the segmented time series of the ten DJUS economic
sector indices.  In Section \ref{sect:datamethods}, we describe the data sets
studied, and the statistical methods used to analyze them.  In Section
\ref{sect:macroMST}, we examine the gross structure of the 10-sector MST over
the 2000 to 2009 period, as well as those over the 2002--2003 crisis period, the
2004--2005 growth period, and the 2008--2009 crisis period.  We explain the
macroeconomic significance of the core-fringe structure of the MSTs, and also
suggest why the MSTs organize themselves into a star topology during growth, and
into a chain topology during crisis.  Then, in Section \ref{sect:segmentMST}, we
construct MSTs within segments associated with distinct macroeconomic phases to
study the correlational dynamics within the US economy.  We again find that the
MST is star-like in low-volatility segments, and chain-like in high-volatility
segments.  This tells us that the star-like MST is a robust and reliable
character of economic well being.  By combining temporal information obtained
through statistical segmentation and clustering, we show that volatility shocks
always start at the fringe and propogate inwards.  Some of the links to leader
sectors have anomalously high cross-correlations.  We also check whether such
volatility shocks have a more domestic or more global origin.  Finally in
Section \ref{sect:MSTrearrangements}, by examining a nearly contiguous sequence
of corresponding segments, we look at how the MST rearranges in the pre-recovery
periods for both the previous and the current financial crises.  We found very
violent rearrangements prior to the previous economic recovery.  For the present
financial crisis, we can see clear signatures of star-to-chain and chain-to-star
rearrangements, accompanied by the expected changes in market volatilities and
cross-correlations.  This suggests that the US market has become more efficient,
as far as processing information is concerned, over the past 5--10 years.  After
predicting that the US economy will recover in early 2010, we summarize our
findings in Section \ref{sect:conclusions}.

\section{Data and methods}
\label{sect:datamethods}

\subsection{Data}

Tic-by-tic data for the ten Dow Jones US (DJUS) economic sector indices (see
Table \ref{table:DJUS} for the indexing scheme $i = 1, \dots, 10$ used) over the
period 14 February 2000 to 25 November 2009 were downloaded from the
Thomson-Reuters Tickhistory (formerly known as Taqtic) database \cite{Taqtic}.
These were then processed into time series $\bX_i = (X_{i,1}, \dots, X_{i,t},
\dots, X_{i,N})$ at fixed time intervals indexed by $1 \leq t \leq N$.  Since
financial markets are known to exhibit complex dynamics on multiple time scales,
the data frequency has to be carefully selected.  In the financial economics
literature, intervals ranging from 5 to 60 minutes have been used for estimating
realized or benchmark daily volatilities for foreign exchange or stock market
time series \cite{Taylor1997JEmpiricalFin4p317,
Dacorogna1998NonlinearModellingofHighFrequencyFinancialTimeSeries,
Schwert1998BrookingsWhartonPapersonFinancialServices, Andersen2001JASA96p42,
Andersen2001JFinEcon61p43}. In general, higher data frequencies are not
employed due to worries about the effects of market microstructures.

\begin{table}[htbp]
\centering\footnotesize
\caption{The ten Dow Jones US economic sector indices as defined by
the Industry Classification Benchmark (ICB).  These are float-adjusted
market capitalization weighted sums of variable numbers of component
stocks, introduced on February 14, 2000 to measure the performance of
US stocks in the ten ICB industries.  The makeup of these indices are
reviewed quarterly, and the number of components and float-adjusted
market capitalizations taken from
\protect\url{http://www.djindexes.com/mdsidx/downloads/fact_info/Dow_Jones_US_Indexes_Industry_Indexes_Fact_Sheet.pdf},
and are accurate as of November 30, 2010.  The top components of each
index are shown in Appendix \ref{app:components}.  Although they do not make up
any of the top spots, homebuilders and developers are components of the consumer
goods (NC) sector.}
\label{table:DJUS}
\vskip .5\baselineskip
\begin{tabular}{cclcc}
\hline
$i$ & symbol & sector & \parbox[c]{3.0cm}{\centering number of 
component stocks} & \parbox[c]{4.0cm}{\centering float-adjusted market
capitalization (billion USD)} \\
\hline
1 & BM & Basic Materials & 155 & 506.7 \\
2 & CY & Consumer Services & 484 & 1,649.1 \\
3 & EN & Oil \& Gas & 214 & 1,405.7 \\
4 & FN & Financials & 876 & 2,192.5 \\
5 & HC & Healthcare & 512 & 1,423.8 \\
6 & IN & Industrials & 692 & 1,725.7 \\
7 & NC & Consumer Goods & 326 & 1,351.1 \\
8 & TC & Technology & 509 & 2,158.1 \\
9 & TL & Telecommunications & 44 & 379.5 \\
10 & UT & Utilities & 96 & 470.9 \\
\hline
\end{tabular}
\end{table}

We chose to sample the time series at 30-minute intervals.  As explained in
Ref.~\cite{Wong2009}, the half-hourly data frequency allows us to confidently
identify statistically stationary segments as short as a day.  Higher data
frequency was not used, because in a macroeconomic study such as this, we are
not interested in segments shorter than a day, i.e. the intraday market
microstructure.  From the index time series $\bX_i$, we then prepare the
log-index movement time series $\bx_i = (x_{i,1}, \dots, x_{i,t}, \dots,
x_{i,N-1})$, where $x_{i, t} = \log X_{i,t+1} - \log X_{i, t}$.  We work with
log-index movements, because different indices have different magnitudes, and it
is more meaningful to compare their fractional changes.

\subsection{Segmentation and clustering as discovery tools in statistical
physics}

\begin{figure}[htbp]
\centering\footnotesize
\includegraphics[scale=0.35]{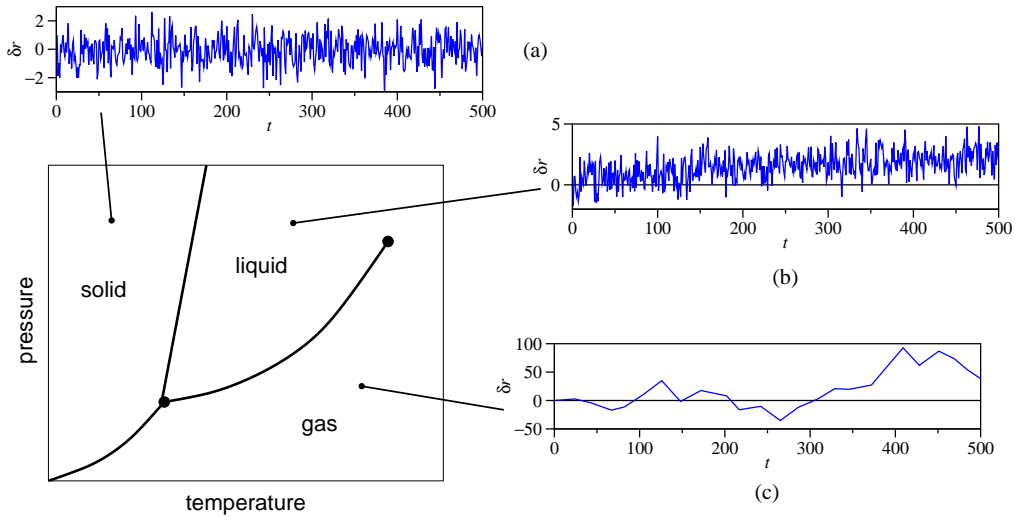}
\caption{A typical phase diagram showing where the solid, liquid, and gas phases
of a substance occurs in the pressure-temperature ($p$-$T$) plane.  Also shown
in the figure are the equilibrium fluctuations $\delta r$ in the displacement of
a given atom in the (a) solid phase, with time-independent variance $\langle
(\delta r)^2 \rangle \propto T$; (b) liquid phase, with a diffusive variance
$\langle (\delta r)^2 \rangle \propto t$; and (c) gas phase, with long ballistic
lifetimes.}
\label{fig:macromicro}
\end{figure}

Before we go on to describe in greater details the segmentation and clustering
methods we used in this study, let us disgress to discuss how segmentation and
clustering can be useful tools for making discoveries in statistical physics.
To begin, let us consider how the thermodynamic phase diagram of a given system,
like that shown in Fig.~\ref{fig:macromicro}, can be determined.  Typically, the
thermodynamic phases are characterized by order parameters, which are frequently
macroscopic quantities like density, magnetization, or electric polarization.
Order parameters have the property that they have different values in different
phases.  Hence, so long as the length and time scales of our system of interest
are not too large, we can measure all macroscopic quantities experimentally, see
which changes sharply as we go from one phase to another to identify the
macroscopic order parameters.  We can then perform even more careful
measurements on these macroscopic order parameters to construct the phase
diagram.

When the length or time scales of our system of interest are too large, for
example, in protein folding (small length scale, but long time scale), or a
nuclear weapon detonating in a city (small time scale, but long length scale),
or climate change (long length scale and long time scale), it may no longer be
possible to identify the order parameters empirically.  However, it may be
comparatively easy to obtain high-frequency time series data of any number of
microscopic quantities.  To see how these microscopic time series can be useful
towards our elucidation of the phase diagram, we must first understand that each
thermodynamic phase is associated with a low-dimensional manifold in the phase
space of our system.  Such low-dimensional manifolds arise because of
conservation laws, or through interaction-driven self-organization, or both.  In
this statistical mechanical picture, order parameters are the thermodynamic
coordinates of individual low-dimensional manifolds.  When the system is in a
given phase, microscopic variables fluctuate about the associated
low-dimensional manifold, which typically has slow dynamics.  When the system
makes a transition into another phase, the slow dynamics and the low-dimensional
manifold changes, and so does the fast fluctuations of the microscopic
variables.

In Fig.~\ref{fig:macromicro}, we illustrate this connection between the
character of fast fluctuations and the underlying slow dynamics, by showing the
equilibrium fluctuations $\delta r$ in the displacement of a given atom.  In
Fig.~\ref{fig:macromicro}(a), we show the non-diffusive equilibrium fluctuations
$\delta r$ in the solid phase.  In this phase, the variance $\langle (\delta
r)^2 \rangle$ of these fluctuations is proportional to the temperature $T$, but
otherwise time independent.  In Fig.~\ref{fig:macromicro}(b), we show the
diffusive equilibrium fluctuations $\delta r$ in the liquid phase.  In this
phase, the variance $\langle (\delta r)^2 \rangle$ of the fluctuations grows
with time.  Finally, in Fig.~\ref{fig:macromicro}(c), we show the diffusive
equilibrium fluctuations $\delta r$ in the gas phase.  Diffusive fluctuations in
the gas phase can be distinguished from those in the liquid phase by the long
ballistic lifetimes in the former. 

\begin{figure}[ht]
\centering\footnotesize
\includegraphics[scale=0.45]{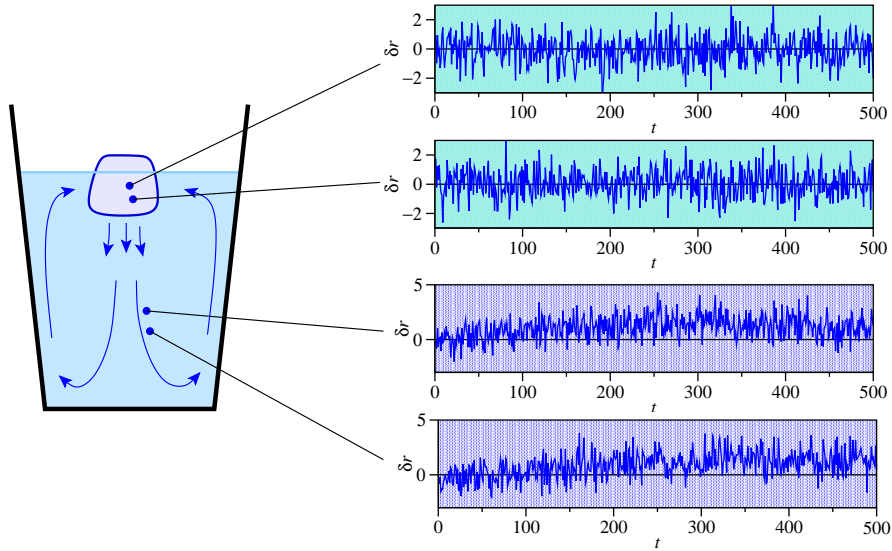}
\caption{Microscopic time series within a system with mixed phases, in this
example, ice and water.  Based on our understanding of the character of
microscopic fluctuations, the time series can be sorted into an ice cluster and
a water cluster.  Alternatively, these clusters can be discovered using
statistical clustering methods, without any prior knowledge of the phases and
their statistical properties.}
\label{fig:icewater}
\end{figure}

Now, suppose different phases are present in the system, like the ice-water
example shown in Fig.~\ref{fig:icewater}.  In this example, we can conclude that
two phases are present, by sorting the atomic displacement time series into two
groups, based on the fluctuation chacracteristics discussed above.
Alternatively, we can cluster the microscopic time series, to find the
statistically distinct ice and liquid phases.  We might even be able to detect
the convection currents present in the system.  Moreover, if the system
undergoes repeated phase transitions, the phases that appear in the history of
the system can be discovered by first segmenting the microscopic time series,
and then clustering the time series segments.  So long as there is adequate
data, and the microscopic fluctuations in different phases are sufficiently
distinct, we can always discover how many such phases are present through
statistical segmentation and clustering of the microscopic time series.  These
two generally robust procedures are very well suited to the study of complex
systems, for which we have high frequency time series data, but no idea how many
phases are present, and what their characteristics are.  In addition to the
macroeconomic study reported in this paper, and related studies on the Japanese
economy, the Asian and European economies, we are also applying the two methods
to understand protein folding dynamics and earthquake dynamics.

\subsection{Segmentation}

Financial time series are well known to be highly nonstationary.  In particular,
several recent studies revealed that the instantaneous volatility fluctuates
about a constant level, before switching over rapidly to fluctuations about a
different constant level \cite{Conlon2010arXiv10010497, Tseng2010arXiv10020284,
Spetiotopoulos2010arXiv10075274}.  Based on these, and similar earlier
observations, economics and finance practitioners explored various methods for
decomposing a nonstationary time series into stationary segments, which are
called \emph{regimes} or \emph{trends} in the economics and finance literatures.
In these literatures, segment boundaries are referred to as \emph{structural
breaks}, \emph{trend breaks}, or \emph{change points}.  The earliest works in
this field are by Goldfeld and Quandt \cite{Goldfeld1973JEconometrics1p3}, and
by Hamilton \cite{Hamilton1989Econometrica57p357}.  Since these pioneering
works, an enormous economics literature on structural breaks and change points
has been amassed, a few based on the original Markov switching models
\cite{Kim1999RevEconStat81p608}, and many others based on autoregressive models
and unit-root tests \cite{Bai1994JTimeSerAnal15p453,
Bai1995EconometricTheory11p403, Chong1995EconLett49p351,
Loader1996AnnStats24p1667, Bai1997RevEconStats79p551,
Lumsdaine1997RevEconStats79p212, Bai1998Econometrica66p47,
Lavielle2000JTimeSerAnal21p33, Chong2001EconometricTheory17p87,
Hansen2001JEconPerspec15p117, Zivot2002JBusEconStats20p25,
Bai2003JApplEconometrics18p1, Perron2005JEconometrics129p65,
Guo2006JFinRes29p79, CarrioniSilvestre2009EconometricTheory25p1754}.  In the
econophysics literature, apart from our own work, we are aware only of the work
by Vaglica \emph{et al.}, who broke the transaction histories of three highly
liquid stocks on the Spanish stock market into directional segments to study
trading strategies adopted in this market \cite{Vaglica2008PhysRevE77e036110},
and the recent preprint by T\'oth \emph{et al.}, who segmented the time series
of market orders on the London Stock Exchange, modeling each segment by a
stationary Poisson process \cite{Toth2010arXiv10012549}.

As with all model-driven segmentation of time series data, we assume that each
economic sector time series $\bx_i$ consist of $M_i$ segments, and that within
segment $m_i$, the log-index movements $x^{m_i}_{i, t}$ follows a stationary
statistical distribution.  From the seminal work by Mantegna and Stanley
\cite{Mantegna1995Nature376p46}, we know that high-frequency index movements can
be fitted very well to stable L\'evy distributions.  We also know from the study by
Kullmann \emph{et al.} \cite{Kullmann1999PhysicaA269p98} that the daily
log-index movements can be fitted well to a truncated L\'evy distribution, when
the sample size is small, but becomes normally distributed when the sample size
is large.  This suggests that the appropriate model for each stationary segment
ought to be a L\'evy stable process.  However, parameter estimation for L\'evy
stable distributions \cite{Fama1971JASA66p331, Hill1975AnnStats3p1163,
Koutrouvelis1980JASA75p918, Koutrouvelis1981CommStatSimulComp10p17,
McCulloch1986CommStatSimulComp15p1109, Zolotarev1986OneDimStableDistributions,
Nolan1998StatsProbLett38p187, Nolan2001LevyProcesses} is a computationally
expensive process, and computing the probability density
\cite{Worsdale1975JRoyalStatsSocC24p123, Panton1992CommStatSimulComp21p485,
McCulloch1997CompStatsDataAnal23p307, Nolan1997StochasticModels13p759,
Nolan1999MathCompModel29p229} is equally tedious.  From our experience
segmenting biological sequences, we know that segment boundaries that are
statistically very significant can be discovered by any segmentation procedure,
no matter what model we assumed for the underlying stationary segments.  We
believe that the most statistically significant segment boundaries in financial
time series would also be equally insensitive to choice of model, or model
mis-specification.  Indeed, when we compared segments of the 2002--2003 DJIA
half-hourly time series obtained assuming that the log-index movements are
normally distributed, against those obtained assuming the log-index movements
are L\'evy stable distributed, the strongest segment boundaries are in good
agreement (no more than two days apart) \cite{LisaFYP}.  With this reassurance,
we chose to intentionally mis-specify the model, and work instead with the
lognormal index movement model.  In this model, the log-index movements in
segment $m_i$ are assumed to follow a stationary Gaussian process with mean
$\mu_{i, m_i}$ and variance $\sigma_{i, m_i}^2$.  Unlike parameter estimation
for the L\'evy distribution, maximum-likelihood estimates of the Gaussian
parameters $\mu_{i, m_i}$ and $\sigma_{i, m_i}^2$ can be done very cheaply.

To find the unknown segment boundaries $t_{i, m_i}$, which separates segments
$m_i$ and $m_i + 1$, we use the recursive segmentation scheme introduced by
Bernaola-Galv\'an \emph{et al}.  \cite{BernaolaGalvan1996PhysicalReviewE53p5181,
RomanRoldan1998PhysicalReviewLetters80p1344}.  In this segmentation scheme, we
start with the time series $\bx = (x_{1}, \dots, x_{t},$ $x_{t+1}, \dots,
x_{n})$, and compute the Jensen-Shannon divergence
\cite{Lin1991IEEETransactionsonInformationTheory37p145}
\begin{equation}
\Delta(t) = \ln\frac{L_2(t)}{L_1},
\end{equation}
where within the log-normal index movement model,
\begin{equation}
L_1 = \prod_{s=1}^n \frac{1}{\sqrt{2\pi\sigma^2}}
\exp\left[-\frac{(x_{s} - \mu)^2}{2\sigma^2}\right]
\end{equation}
is the likelihood that $\bx$ is generated probabilistically by a
single Gaussian model with mean $\mu$ and variance $\sigma^2$, and 
\begin{equation}
L_2(t) = \prod_{s=1}^t \frac{1}{\sqrt{2\pi\sigma_L^2}}
\exp\left[-\frac{(x_s - \mu_L)^2}{2\sigma_L^2}\right]
\prod_{s=t+1}^n \frac{1}{\sqrt{2\pi\sigma_R^2}}
\exp\left[-\frac{(x_s - \mu_R)^2}{2\sigma_R^2}\right]
\end{equation}
is the likelihood that $\bx$ is generated by two statistically
distinct models: the left segment $\bx_L = (x_1, \dots, x_t)$ by a
Gaussian model with mean $\mu_L$ and variance $\sigma_L^2$, and the
right segment $\bx_R = (x_{t+1}, \dots, x_n)$ by a Gaussian model with
mean $\mu_R$ and variance $\sigma_R^2$.  In terms of the maximum
likelihood estimates $\hat{\mu}, \hat{\mu}_L, \hat{\mu}_R$ and
$\hat{\sigma}^2, \hat{\sigma}_L^2, \hat{\sigma}_R^2$, the
Jensen-Shannon divergence $\Delta(t)$, which measures how much better
a two-segment model fits the time series data compared to a one-segment
model, simplifies to
\begin{equation}
\Delta(t) = n\ln\hat{\sigma} - t \ln \hat{\sigma}_L - (n - t) \ln
\hat{\sigma}_R + \frac{1}{2} \geq 0.
\end{equation}

\begin{figure}[htbp]
\centering
\includegraphics[scale=0.4,clip=true]{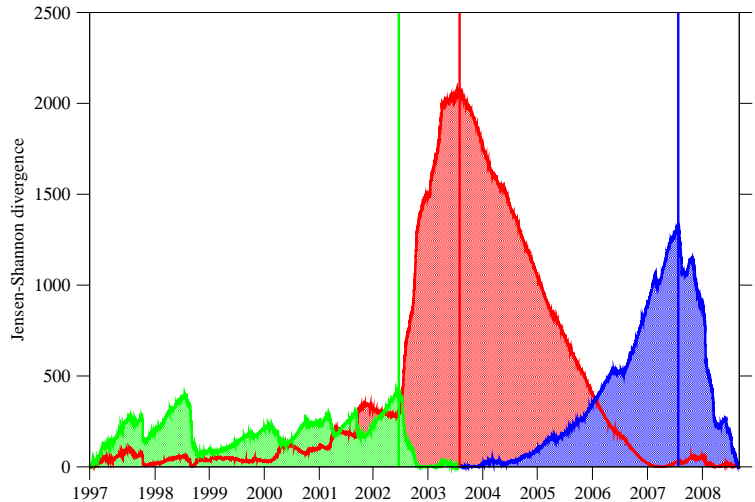}
\caption{The Jensen-Shannon divergence spectrum for the DJIA time series from
Jan 1997 to Aug 2008 (red).  This is a typical spectrum consisting of one very
strong peak, in this example, at mid-2003.  Also show are the Jensen-Shannon
divergence spectra for the left segment (green, 1997 to mid-2003) and the right
segment (blue, mid-2003 to Aug 2008) obtained at the second stage of the
recursive segmentation.  In this example, the two segments have divergence
maxima at mid-2002 and mid-2007 respectively.}
\label{fig:JSseg}
\end{figure}

Scanning through all possible times $t$, a cut is then placed at
$t^*$, for which the Jensen-Shannon divergence
\begin{equation}
\Delta^* = \Delta(t^*) = \max_t \Delta(t)
\end{equation}
is maximized, to break the time series $\bx = (x_1, \dots, x_n)$ into
two statistically most distinct segments $\bx^*_L = (x_1, \dots,
x_{t^*})$ and $\bx^*_R = (x_{t^*+1}, \dots, x_n)$ (see for example,
Fig.~\ref{fig:JSseg}).  This one-into-two
segmentation is then applied recursively onto $\bx^*_L$ and $\bx^*_R$
to obtain shorter and shorter segments (see also Fig.~\ref{fig:JSseg}, for
example).  At each stage of the
recursive segmentation, we also optimize the segment boundaries using
the first-order algorithm described in Ref.~\cite{CheongIRJSS}, where
we recompute the optimum position of segment boundary $m$, within the
time series subsequence bound by segment boundaries $m\pm 1$.  This is
done iteratively for all segment boundaries, until they have all
converged onto their optimum positions.

As the optimized recursive segmentation progresses, the Jensen-Shannon
divergence of newly discovered segment boundaries, as well as the previously
discovered segment boundaries, will in general become smaller and smaller.
Segment boundaries thus become less and less significant statistically, and at
some point, we must terminate the recursive segmentation.  There are three ways
to do so.  In the first approach, the Jensen-Shannon divergences of new segment
boundaries are tested for statistical significance against various $\chi^2$
distributions with the appropriate degrees of freedom
\cite{BernaolaGalvan1996PhysicalReviewE53p5181,
RomanRoldan1998PhysicalReviewLetters80p1344}.  The recursive segmentation
terminates when no new segment boundaries more significant than the chosen
confidence level $p$ can be found.  In the second approach, new segment
boundaries are accepted if the information criteria of the two-segment models
they imply are larger than the information criteria of the one-segment models we
are selecting against \cite{Li2001PhysicalReviewLetters86p5815,
Li2001ProcRECOMB01p204}.  Here, the recursive segmentation terminates when
further segmentation does not explain the data better.  In the third approach,
we define signal-to-noise ratios based on the Jensen-Shannon divergence
fluctuations within supersegments that new segment boundaries are supposed to
divide \cite{CheongIRJSS}.  The recursive segmentation terminates when the
signal-to-noise ratios of all new segment boundaries fall below a chosen
threshold value.  

Alternatively, we could also terminate the recursive
segmentation when no new optimized segment boundaries with Jensen-Shannon
divergence greater than a cutoff of $\Delta_0 = 10$ are found.  The short and
medium segments produced by this termination criterion are reasonable, but the
long segments obtained tend to have internal segment structures masked by their
context \cite{CheongCSP}.  We then recursively segment these long segments, by
progressively lowering the cutoff $\Delta_0$ until a segment boundary with
strength $\Delta > 10$ appears.  The final segmentation then consists of segment
boundaries discovered through the automated recursive segmentation, as well as
segment boundaries discovered through progressive refinement of overly long
segments.  Based on the experience in our previous works \cite{Wong2009,
Lee2009arXiv09114763}, this semi-automatic recursive segmentation appears to
produce acceptable results.

\subsection{Segment clustering}

After the segmentation is completed, we obtain a large number (typically $>
100$) of segments for each time series.  While successive segments are
statistically distinct from each other, segments that are far apart can actually
be statistically similar.  This observation suggests that the large number of
segments make up a small number of \emph{segment classes}.  By comparing
multiple indicators, economists classified different market periods into four
macroeconomic \emph{phases} or \emph{regimes}: (i) a growth phase; (ii) a
contraction phase; (iii) a correction phase; and (iv) a crash phase.  We
therefore expect the time series segments to also be organized into roughly four
classes.  A similar problem arise in biological sequences, where thousands of
segments can be organized into tens of segment types that differ in their
biological functions.  In the ground-breaking paper by Azad \emph{et al.}, the
248 segments of human chromosome 22 was classified into 53 domain types using
single-link hierarchical clustering \cite{Azad2002PhysRevE66e031913}.  Inspired
by this prospect of reducing the complexity of our segmentations, we performed
independent hierarchical agglomerative clusterings on the segments within each
US economic sector time series, using the complete link algorithm (see
Ref.~\cite{Jain1999} for details on the complete link algorithm, and also a
review on the broad area of statistical clustering).  We chose the complete link
algorithm, which is favored by social scientists for producing compact and
internally homogeneous clusters \cite{Baker1972RevEduRes42p345}, because our
goal is to discover macroeconomic phases with well-defined statistical
properties.  We do not use the far more popular single link algorithm
\cite{Sneath1957JGenMicrobiol17p201, Johnson1967Psychometrika32p241}, because it
tends to produce loose and elongated clusters
\cite{Baker1974JAmStatsAssoc69p440}.  Single link clustering is more meaningful
in the biological sciences, generating phylogenetic trees for example, because
the clustering procedure corresponds more closely with the nature of
evolutionary changes.  In general, if one expects to find highly homogeneous
collections of objects one would use complete link clustering, whereas if one
expects to find collections of objects that evolved from common ancestors one
would use single link clustering.

\begin{figure}[htbp]
\centering
\includegraphics[scale=0.35]{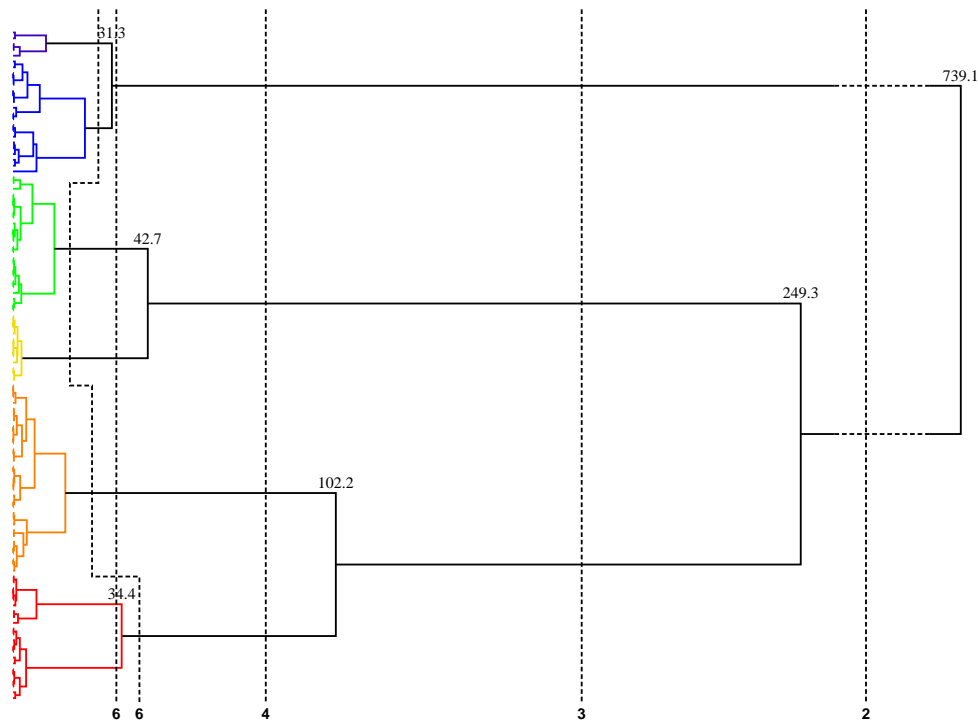}
\caption{The complete-link hierarchical clustering tree for the time series
segments of the Dow Jones Industrial Average between January 1997 and August
2008.  In this tree, we show the Jensen-Shannon divergence values at which the
top branches diverge.  We also show how uniform thresholds can be selected to
break the tree into two, three, four, or six clusters.  Finally, we show how
individual thresholds can be selected to obtain the six clusters reported in
Ref.~\cite{Wong2009}, which are colored in increasing order of market volatility
as deep blue, blue, green, yellow, orange, and red.}
\label{fig:cltree}
\end{figure}

In this segment clustering, we used the Jensen-Shannon divergences between
segments as their statistical distances.  Clustering of different periods within
a financial time series has been previously investigated
\cite{vanWijk1999ProcInfoVisualp4, Krawiecki2002PhysRevLett89e158701,
Fu2004ProcIntConfDataMiningp5}, in the absence of any segmentation analysis.
After complete-link hierarchical clustering of the segments within a given time
series, we typically end up with a dendrogram like that shown in
Fig.~\ref{fig:cltree} for the Dow Jones Industrial Average.  By varying a
uniform threshold, different number of clusters can be identified.  A small
number of clusters provide a coarser description, whereas a larger number of
clusters offer a finer description of the dynamics within the time series.  This
different numbers of clusters form a nested hierarchy of coarse-grained
descriptions of the US macroeconomic dynamics.  All these descriptions are
correct, but some are more useful than others, because they are statistically
more robust.  For example, in Fig.~\ref{fig:cltree}, we can identify four
clusters if the uniform threshold is $42.7 < \Delta < 102.2$, or five clusters
if the uniform threshold is $34.4 < \Delta < 42.7$, or six clusters if the
uniform threshold is $31.3 < \Delta < 34.4$.  Because of the broader range of
uniform thresholds, a four-cluster description is statistically more robust than
a five-cluster or six-cluster descriptions of the time series segments.

Once we understand statistical robustness as the primary criterion for cluster
selection, we can also employ local thresholds for each cluster.  In
Fig.~\ref{fig:cltree}, we show as an example the local thresholds used to pick
six clusters.  Other local thresholds can also be used, but so long as they are
statistically robust, one choice of clusters offer no advantage over another
choice of clusters.  These different choices of clusters tell the same story,
merely with different contrasts, very much like red-tinted and blue-tinted
versions of the same photograph.  With this in mind, we analyzed the
hierarchical complete-link clustering trees obtained for all ten DJUS economic
sectors, and selected between four to six clusters of segments for each US
economic sector.  These clusters represent different macroeconomic phases
(differentiated by their market volatilities) present in the time series data.
Once all segments have been assigned to their respective clusters, we use the
heat-map-like color scheme in Table~\ref{table:heatmap} to plot the
\emph{temporal distributions of clustered segments}.  All the analyses presented
in this paper are based on features identified from the temporal distributions
of clustered segments for the ten DJUS economic sector indices.

\begin{table}[ht]
\centering\footnotesize
\caption{Heat-map-like color scheme for the different volatility clusters, and
the macroeconomic phases they correspond to.  The crisis phase, which consists
of the high-volatility (yellow) and very-high-volatility (orange) clusters, is
significantly longer than the economic contraction phase accepted by economists.
In fact, economic contraction, as determined by successive quarters of
contraction in the GDP, typically occurs at the end of a crisis phase.  Also
shown are the average standard deviation in each phase for the various economic
sectors.  The extremely-low-volatility phase is seen only in CY, EN, and FN,
because the time series segments are organized into six clusters for these three
sectors only.  As we can see, within each macroeconomic phase, the average
volatilities discovered by the clustering procedure are fairly consistent
throughout most sectors.  The exceptions are HC and TL, which have consistently
lower volatilities.  It is possible to introduce a seventh cluster with
volatility $\sigma \approx 0.008$, and have all the clusters reclassified.  This
will produce a deterministic mapping between volatility and color, but we choose
not to, so as to achieve maximum visual contrast with the present color scheme.}
\label{table:heatmap}
\vskip .5\baselineskip
\begin{tabular}{|c|c|c|c|c|c|c|}
\hline
\emph{volatility} & 
extremely low & low & moderate & high & very high & extremely high \\
\hline
\emph{color} & black & blue & green & yellow & orange & red \\
\hline
\emph{phase} & \multicolumn{2}{|c|}{growth} & correction &
\multicolumn{2}{|c|}{crisis} & crash \\
\hline
BM & - & 0.0016 & 0.0037 & 0.0046 & 0.0069 & 0.0146 \\
CY & 0.0005 & 0.0015 & 0.0023 & 0.0031 & 0.0053 & 0.0121 \\
EN & 0.0010 & 0.0014 & 0.0027 & 0.0037 & 0.0058 & 0.0152 \\
FN & 0.0007 & 0.0016 & 0.0024 & 0.0039 & 0.0058 & 0.0134 \\
HC & - & 0.0006 & 0.0016 & 0.0023 & 0.0041 & 0.0076 \\
IN & - & 0.0013 & 0.0022 & 0.0035 & 0.0056 & 0.0140 \\
NC & - & 0.0009 & 0.0015 & 0.0022 & 0.0034 & 0.0085 \\
TC & - & 0.0019 & 0.0030 & 0.0042 & 0.0082 & 0.0121 \\
TL & - & 0.0008 & 0.0018 & 0.0024 & 0.0033 & 0.0078 \\
UT & - & 0.0014 & 0.0023 & 0.0030 & 0.0038 & 0.0088 \\
\hline
\end{tabular}
\end{table}

\subsection{Identifying corresponding segments}

Of the many features that we can identify from individual temporal
distributions, as well as across the panel of temporal distributions,
\emph{corresponding segments} that appear in all or most of the indices are the
most striking visually.  In the economics and finance literature, a mean or
volatility movement that occurs over multiple time series is called
\emph{comovement} \cite{Panton1976JFinQuantAnal11p415,
Stockman1995AmEconRev85p168, Karolyi1996JFin51p951, Croux2001RevEconStats83p232,
Forbes2002JFin57p2223, Barberis2005JFinEcon75p283, Baxter2005JMonEcon52p113},
\emph{common jumps} \cite{BarndorffNielsen2006JFinEconometrics4p1,
Bollerslev2008JEconometrics144p234, Jacod2009AnnStats37p1792}, \emph{common
shocks} \cite{Canova1998JIntEcon46p133, Rigobon2003JIntEcon61p261,
Andrews2005Econometrica73p1551}, or \emph{common breaks}
\cite{Bai1998RevEconStud65p395, CarrioniSilvestre2005EconometricsJ8p159,
Im2005OxfordBullEconStats67p393, Bai2009Econometrica77p1229,
Bai2009RevEconStud76p471, Kim2009WP}.  The consensus that arise from this body
of work is that the statistical significance of a change point is amplified by
the cross section it occurs concurrently over.  

In our study, the corresponding segments do not necessarily start at the same
time, because our use of high-frequency data allows us to identify the change
points that are individually optimum for the ten DJUS economic sector indices.
More importantly, our corresponding segments in the various indices do not end
at the same time.  As discussed in Ref.~\cite{Lee2009arXiv09114763}, the
durations of each corresponding segment, and the Jensen-Shannon divergence
values at the start of these segments, tell us how strongly the shock impacted
different sectors in the US economy.  Moreover, the different start times of the
corresponding segments allow us to roughly map out the progress of the shock.

Because of the different start times and different durations, we mark 
segments in the ten DJUS economic sector indices as corresponding segments if
they (i) have similar volatilities (high and high, or low and low); or (ii) are
flanked by volatility movements in the same directions(low-to-high and
moderate-to-high, or high-to-low and moderate-to-low).  For this, we took
advantage of the heat-map-like color scheme in the temporal distributions.

\subsection{Cross-correlations}
\label{sect:mst}

In performing segmentation and thereafter segment clustering, we have
selectively discarded information contained in the ten high-frequency time
series to obtain a coarse-grained picture of the US macroeconomic dynamics.
While this picture provides a useful bird's eye view of the dynamical processes
within the US economy, a significant amount of useful information has also been
thrown out.  To recover more of the information contained in the high-frequency
time series, and shed more light on the exciting stories unfolding before our
eyes, we compute the normalized cross-correlation matrix $C$, whereby the matrix
element
\begin{equation}
C_{ij} = \frac{\sum_{t=1}^T (x_{it} - \bar{x}_i)(x_{jt} -
\bar{x}_j)}{\sqrt{\sum_{t=1}^T (x_{it} - \bar{x}_i)^2 \sum_{t'=1}^T (x_{jt} -
\bar{x}_j)^2}}
\end{equation}
is the zero-lag cross-correlation between US economic sectors $i$ and $j$.

Cross-correlations between different stocks, and between different benchmark
indices have been widely studied in the finance literature.  Such studies have
been particularly popular in the bid to understand the meltdown of global
financial markets during the present financial crisis \cite{Frank2008IMFWP08200,
Lo2008Testimony, Tudor2009RomEconJ31p115, Cheung2010ApplFinEcon20p85,
Munnix2010arXiv10065847, Wong2010ApplFinEcon20p137}.  In the econophysics
literature, there have been attempts to understand the nontrivial
cross-correlations between different financial time series using random matrix
theory \cite{Laloux1999PhysRevLett83p1467, Plerou2002PhysRevE65e066126,
Utsugi2004PhysRevE70e026110, Wilcox2004PhysicaA344p294,
Wilcox2007PhysicaA375p584, Cukur2007PhysicaA376p555,
Kulkarni2007EurPhysJB60p101, Shen2009EuroPhysLett86e48005}.  In all these
studies, the cross-correlations were computed either over the entire data
period, or over sliding windows.  In our own study, we not only calculate the
cross-correlation matrix over the entire duration of the time series, but also
over two-year intervals strictly within the growth and crisis macroeconomic
phases, and over individual corresponding segments.

\begin{figure}[ht]
\centering
\includegraphics[scale=0.5]{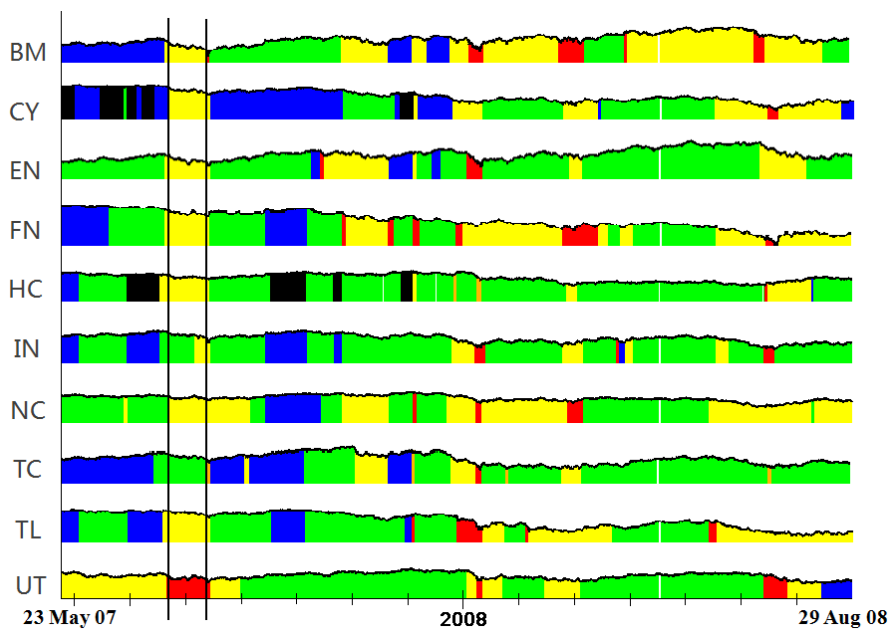}
\caption{Interval selected for the computation of the cross-correlation matrix
between the ten DJUS economic sector indices.  In this figure, the lower and
upper limits of the interval are chosen such that the interval covers a single
macroeconomic phase for nearly all indices.  Exception is made for IN, because
its high-volatility segment is too short, and thus the selected interval also
covers the preceding moderate-volatility segment.} 
\label{fig:selectlength}
\end{figure}

To compute the cross-correlation matrix over a given corresponding segment, we
select the largest interval within which most sectors can be found in a single
macroeconomic phase, as shown in Fig.~\ref{fig:selectlength}.  For the primarily
high-volatility corresponding segment identified in Fig.~\ref{fig:selectlength}, the
start dates and end dates in the different sectors are shown in Table
\ref{table:selectlength}.  Based on these dates, we chose our interval to start
on July 25, 2007, and end on August 14, 2007, so that all DJUS economic sectors,
with the exception of IN, TC, and UT, are strictly in the high-volatility 
phase.  Within this interval, UT is strictly in the extremely-high-volatility
phase.  The selected high-volatility interval overlaps with the
moderate-volatility phases in IN and TC.  This cannot be helped, because the
high-volatility phases in IN and TC started so late.  Furthermore, we believe it
is meaningful to allow this overlap, because the start dates of the
moderate-volatility segments in IN and TC are very close to the start dates of
the high-volatility segments in the other sectors.

\begin{table}[htbp]
\centering\footnotesize
\caption{Start dates and end dates in different DJUS economic sectors for the
primarily high-volatility corresponding segment identified in
Fig.~\ref{fig:selectlength}.  The high-volatility segments for IN and TC started
significantly later than those in other sectors, so we also show the start and
end dates for the preceding moderate-volatility segments for these two sectors.
For UT, it is a extremely-high-volatility segment that coincides with the
high-volatility segment in other sectors.}
\label{table:selectlength}
\vskip .5\baselineskip
\begin{tabular}{ccc}
\hline
Economic Sector & Start Date & End Date \\
\hline
BM & \cellcolor{yellow} July 23, 2007 & \cellcolor{yellow} August 14, 2007 \\
CY & \cellcolor{yellow} July 25, 2007 & \cellcolor{yellow} August 16, 2007 \\
EN & \cellcolor{yellow} July 23, 2007 & \cellcolor{yellow} August 15, 2007 \\
FN & \cellcolor{yellow} July 23, 2007 & \cellcolor{yellow} August 15, 2007 \\
HC & \cellcolor{yellow} July 20, 2007 & \cellcolor{yellow} August 15, 2007 \\
IN & \cellcolor{green} July 19, 2007 & \cellcolor{green} August 7, 2007 \\
   & \cellcolor{yellow} August 8, 2007 & \cellcolor{yellow} August 16, 2007 \\
NC & \cellcolor{yellow} July 25, 2007 & \cellcolor{yellow} September 9, 2007 \\
TC & \cellcolor{green} July 17, 2007 & \cellcolor{green} August 14, 2007 \\
   & \cellcolor{yellow} August 15, 2007 & \cellcolor{yellow} August 16, 2007 \\
TL & \cellcolor{yellow} July 20, 2007 & \cellcolor{yellow} August 16, 2007 \\
UT & \cellcolor{red} July 24, 2007 & \cellcolor{red} August 16, 2007 \\
\hline
\end{tabular}
\end{table}

In Table \ref{table:robustxcorr}, we show the cross-correlation matrix computed
using (a) this optimum interval, such that apart from IN and TC, all sectors are
in the high-volatility or extremely-high-volatility phases.  To show that the
criterion we used to select interval (a) produce statistically robust cross
correlations, we compare them against cross correlations computed using the
intervals (b) and (c).  Interval (b) is four trading days shorter than interval
(a), with two trading days taken off the latter's start and end.  As with
interval (a), all sectors apart from IN and TC are in the high-volatility or
extremely-high-volatility phases within this interval.  In contrast, interval
(c) is four trading days longer than interval (a), with two trading days added
to the latter's start and end.  Unlike for intervals (a) and (b), cross
correlations computed within this longer interval will contain contributions
from the lower-volatility phases adjacent to the high-volatility corresponding
segment for nearly all sectors.  As we can see from Table
\ref{table:robustxcorr}, the maximum positive difference between cross
correlations in (b) and cross correlations in (a) is $+0.063$, occurring for CY-TL,
and the maximum negative difference between cross correlations in (b) and cross
correlations in (a) is $-0.029$, occurring for HC-EN.  Similarly, the maximum
positive difference and maximum negative difference between cross correlations
in (c) and cross correlations in (a) is $+0.041$ and $-0.072$, occurring for
HC-EN and BM-FN, respectively.  The differences in cross correlations is
generally larger for the longer interval (c) than for the shorter interval (b),
because in (b), the interval contains a single statistically stationary segment
for most sectors, whereas in (c), the interval incorporated time series data
from more than one segments.  Nonetheless, these cross correlational differences
are all small.

\begin{table}[htbp]
\centering\footnotesize
\caption{Cross-correlation matrix computed from the half-hourly times series of
the ten DJUS economic sector indices over three intervals: (a) from July 25,
2007 to August 14, 2007; (b) from July 27, 2007 to August 10, 2007; and (c) from
July 23, 2007 to August 16, 2007.}
\label{table:robustxcorr}
\vskip .5\baselineskip
(a) 
\vskip .5\baselineskip
\begin{tabular}{|c|c|c|c|c|c|c|c|c|c|c|}
\hline
& BM & CY & EN & FN & HC & IN & NC & TC & TL & UT \\
\hline
BM & & 0.872 & 0.825 & 0.898 & 0.726 & 0.900 & 0.789 & 0.818 & 0.709 & 0.832 \\
\hline
CY & 0.872 & & 0.753 & 0.898 & 0.826 & 0.915 & 0.876 & 0.856 & 0.768 & 0.835 \\
\hline
EN & 0.825 & 0.753 & & 0.750 & 0.663 & 0.776 & 0.720 & 0.745 & 0.607 & 0.759 \\
\hline
FN & 0.898 & 0.898 & 0.750 & & 0.771 & 0.889 & 0.845 & 0.827 & 0.741 & 0.814 \\
\hline
HC & 0.726 & 0.826 & 0.663 & 0.771 & & 0.827 & 0.913 & 0.804 & 0.772 & 0.770 \\
\hline
IN & 0.900 & 0.915 & 0.776 & 0.889 & 0.827 & & 0.861 & 0.877 & 0.808 & 0.842 \\
\hline
NC & 0.788 & 0.876 & 0.720 & 0.845 & 0.913 & 0.861 & & 0.852 & 0.783 & 0.819 \\
\hline
TC & 0.818 & 0.856 & 0.745 & 0.827 & 0.804 & 0.877 & 0.852 & & 0.769 & 0.783 \\
\hline
TL & 0.709 & 0.768 & 0.607 & 0.741 & 0.772 & 0.808 & 0.783 & 0.769 & & 0.729 \\
\hline
UT & 0.832 & 0.835 & 0.759 & 0.814 & 0.770 & 0.842 & 0.819 & 0.783 & 0.729 & \\
\hline
\end{tabular}
\vskip .5\baselineskip
(b) 
\vskip .5\baselineskip
\begin{tabular}{|c|c|c|c|c|c|c|c|c|c|c|}
\hline
& BM & CY & EN & FN & HC & IN & NC & TC & TL & UT \\
\hline
BM & & 0.888 & 0.807 & 0.891 & 0.713 & 0.909 & 0.783 & 0.843 & 0.746 & 0.823 \\
\hline
CY & 0.888 & & 0.752 & 0.928 & 0.848 & 0.932 & 0.910 & 0.904 & 0.831 & 0.859 \\
\hline
EN & 0.807 & 0.752 & & 0.728 & 0.634 & 0.771 & 0.703 & 0.747 & 0.619 & 0.736 \\
\hline
FN & 0.891 & 0.928 & 0.728 & & 0.770 & 0.894 & 0.849 & 0.849 & 0.770 & 0.808 \\
\hline
HC & 0.713 & 0.848 & 0.634 & 0.770 & & 0.826 & 0.921 & 0.793 & 0.784 & 0.772 \\
\hline
IN & 0.909 & 0.932 & 0.771 & 0.894 & 0.826 & & 0.880 & 0.880 & 0.830 & 0.847 \\
\hline
NC & 0.783 & 0.910 & 0.703 & 0.849 & 0.921 & 0.880 & & 0.865 & 0.816 & 0.838 \\
\hline
TC & 0.843 & 0.904 & 0.747 & 0.849 & 0.793 & 0.880 & 0.865 & & 0.764 & 0.802 \\
\hline
TL & 0.746 & 0.831 & 0.619 & 0.770 & 0.784 & 0.830 & 0.816 & 0.764 & & 0.768 \\
\hline
UT & 0.823 & 0.859 & 0.736 & 0.808 & 0.772 & 0.847 & 0.838 & 0.802 & 0.768 & \\
\hline
\end{tabular}
\vskip .5\baselineskip
(c) 
\vskip .5\baselineskip
\begin{tabular}{|c|c|c|c|c|c|c|c|c|c|c|}
\hline
& BM & CY & EN & FN & HC & IN & NC & TC & TL & UT \\
\hline
BM & & 0.837 & 0.835 & 0.826 & 0.738 & 0.912 & 0.770 & 0.840 & 0.705 & 0.821 \\
\hline
CY & 0.837 & & 0.771 & 0.891 & 0.847 & 0.906 & 0.891 & 0.868 & 0.787 & 0.844 \\
\hline
EN & 0.835 & 0.771 & & 0.743 & 0.704 & 0.802 & 0.736 & 0.775 & 0.639 & 0.783 \\
\hline
FN & 0.826 & 0.891 & 0.743 & & 0.785 & 0.860 & 0.853 & 0.824 & 0.737 & 0.804 \\
\hline
HC & 0.738 & 0.847 & 0.704 & 0.785 & & 0.842 & 0.914 & 0.829 & 0.784 & 0.793 \\
\hline
IN & 0.912 & 0.906 & 0.802 & 0.860 & 0.842 & & 0.860 & 0.902 & 0.808 & 0.838 \\
\hline
NC & 0.770 & 0.891 & 0.736 & 0.853 & 0.914 & 0.860 & & 0.868 & 0.792 & 0.826 \\
\hline
TC & 0.840 & 0.868 & 0.775 & 0.824 & 0.829 & 0.902 & 0.868 & & 0.785 & 0.796 \\
\hline
TL & 0.705 & 0.787 & 0.639 & 0.737 & 0.784 & 0.808 & 0.792 & 0.785 & & 0.746 \\
\hline
UT & 0.821 & 0.844 & 0.783 & 0.804 & 0.793 & 0.838 & 0.826 & 0.796 & 0.746 & \\
\hline
\end{tabular}
\end{table}

\begin{figure}[htbp]
\centering
\includegraphics[scale=0.4]{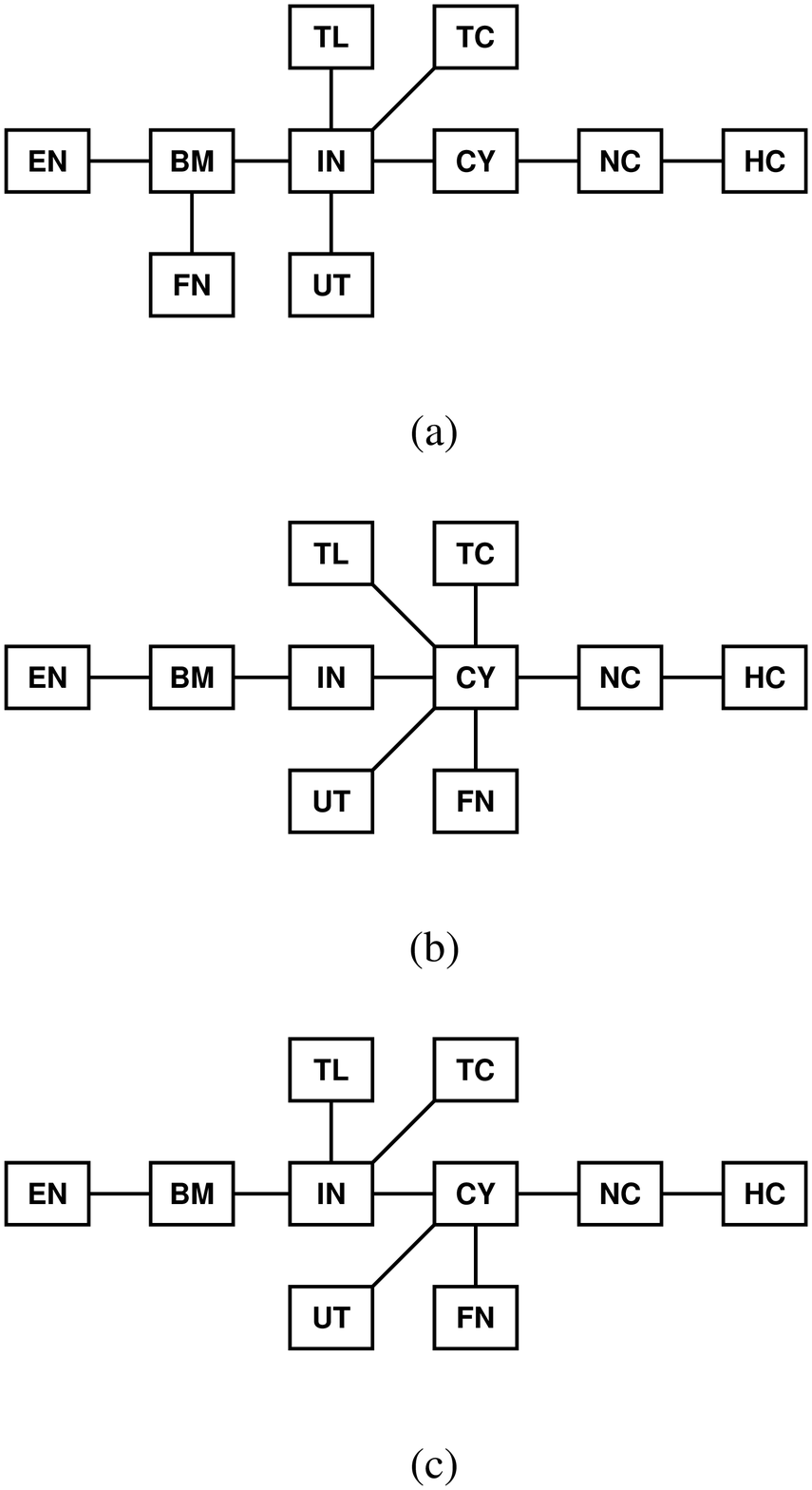}
\caption{The minimal spanning trees of the ten DJUS economic sectors,
constructed using half-hourly time series from (a) July 25, 2007 to August 14,
2007; (b) July 27, 2007 to August 10, 2007; and (c) July 23, 2007 to August 16,
2007.}
\label{fig:robustMST}
\end{figure}

To understand what impacts these small cross correlational differences have on
the minimal spanning trees (described in the next subsection) we generate for
this study, we show in Fig.~\ref{fig:robustMST} the minimal spanning trees
derived from the cross-correlation matrices for these three intervals.  In this
figure, we see the same EN-BM-IN-CY-NC-HC backbone, which is insensitive to how
we select the interval, and thus statistically robust.  We also see that the
sectors TC, TL, and UT, are linked sometimes to IN, and other times to CY.
Going back to the cross correlations shown in Table \ref{table:robustxcorr}
between TC, TL, UT and CY, IN, we find poor agreements over the three intervals
for the CY-TC, CY-TL and CY-UT cross correlations.  In contrast, the IN-TC,
IN-TL, and IN-UT cross correlations are in good agreement over the three
intervals.  This suggests that the minimal spanning links between TC, TL, UT and
CY, IN are not as statistically robust as the rest of the minimal spanning
links in this corresponding segment.  In any case, we shall see in Section
\ref{sect:macroMST} that CY and IN are core sectors, while TC, TL, and UT are
fringe domestic sectors of the US economy.  Whether they are to IN or CY, 
minimal spanning links directly connect TC, TL, UT from the fringe to CY, IN in
the core.  This direct core-fringe linkage in itself represents a statistically
robust characteristic of the US economy.

\subsection{Minimal spanning trees}

Even though our cross-correlation matrices are only $10 \times 10$ in size, the
information contained in the 36 independent matrix elements is still not easy
for a human to process.  To better understand the correlational dynamics between
the US economic sectors at different times, we look instead at simplified
graphical representations of the cross-correlation matrices.  For this study, we
work primarily with the \emph{minimal spanning tree} (MST) representation of the
cross-correlation matrix.  In Section \ref{sect:MSTvsPMFG}, We also also explore
the \emph{planar maximally filtered graph} (PMFG) representation, to understand
what kind of cross-correlational structures have been left out in the MST
representation, in which cycles are not admitted.  Ultimately, if we had
analyzed the cross correlations network of all stocks on the American stock
markets, community detection methods \cite{Girvan2002PNAS99p7821,
Newman2004EurPhysJB38p321, Duch2005PhysRevE72e027104, Newman2006PNAS103p8577,
Reichardt2006PhysRevE74e016110, Lancichinetti2009PhysRevE80e056117,
Fortunato2010PhysReps486p75} will allow us to develop a coarse grain description
of the US economy, and thereby shed more light on its dynamics.  In this study,
the use of such techniques are not necessary, since we are only looking at ten
DJUS economic sector indices, which are themselves already coarse grain
descriptions of the US markets.

The minimal spanning tree (also called \emph{minimum spanning tree}) approach to
understanding weighted graphs is frequently credited to Kruskal
\cite{Kruskal1956ProcAmMathSoc7p48} or Prim \cite{Prim1957BellSysTechJ36p1389},
although there were studies dating all the way back to 1926.  For a good reading
on the history of the minimal spanning tree method, see the article by Graham
and Hell \cite{Graham1985AnnHistoryComp7p43}.  In economics, the method is not
widely used \cite{Hill1999RevEconStats81p135, Hill2001IntEconRev42p167}.
However, since its first application in econophysics by Mantegna
\cite{Mantegna1999EuroPhysJB11p193}, and shown to be a robust caricature of the
underlying correlations \cite{Bonanno2000PhysRevE62pR7615,
Micciche2003PhysicaA324p66}, the MST has been incorporated into the basic tool
suite for statistical analysis of financial market data
\cite{Bonanno2003PhysRevE68e046130, Jung2006PhysicaA361p263,
Brida2007IntJModPhysC18p1783, Borghesi2007PhysRevE76e026104,
Brida2008PhysicaA387p5205, Eom2009PhysicaA388p900, Brida2010ComputEcon35p85}.
In particular, Onnela \emph{et al.} made extensive use of MSTs to study the
dynamics of cross correlations during market crashes
\cite{Onnela2003PhysicaA324p247, Onnela2003PhysScriptT105p48,
Onnela2003PhysRevE68e056110}.  Clustering techniques based on the MST have also
been used to discover different sectors in a stock market
\cite{Onnela2004EurPhysJB38p353, Bonanno2004EurPhysJB38p363,
Boginski2005CompStatDataAnal48p431, Tumminello2007EurPhysJB55p209,
Coelho2007PhysicaA373p615, Jung2008PhysicaA387p537}, how the interdependences of
the European economies are evolving \cite{Gligor2007EurPhysJB57p139,
Gilmore2008PhysicaA387p6319}, and how global markets are linked to each other
\cite{Miskiewicz2006IntJModPhysC17p317, Coelho2007PhysicaA376p455,
Eryigit2009PhysicaA388p3551}.  More recently, Eom \emph{et al.} used the MST as
a means to reduce the $N(N-1)/2$ linkages between $N$ stocks to $N-1$ links, for
studying the effects of market factors on the information flow between stocks
\cite{Eom2010PhysicaA389p1643}.

To construct an MST representation of the cross-correlation matrix, Mantegna
defined the distance metric \cite{Mantegna1999EuroPhysJB11p193}
\begin{equation}
d_{ij} = \sqrt{2(1 - C_{ij})},
\end{equation}
which measures the statistical distance between two financial time series $i$
and $i$, whose cross-correlation is $-1 \leq C_{ij} \leq 1$.  Applying Kruskal's
algorithm \cite{Kruskal1956ProcAmMathSoc7p48}, a link is first drawn connecting
the pair $(i_1, j_1)$ of time series with the smallest distance $d_{i_1 j_1} =
\min_{(i,j)} d_{ij}$.  Following this, a link is drawn connecting the pair
$(i_2, j_2)$ with the next smallest distance $d_{i_2 j_2} = \min_{(i, j) \neq
(i_1, j_1)} d_{ij}$.  This process is repeated with pairs $(i_k, j_k)$ with
increasingly larger distances $d_{i_k j_k}$, until all time series are
incorporated into the spanning graph.  There is one additional constraint: if
$(i_l, j_l)$ is the next pair of time series to be linked based on their
distance $d_{i_l j_l}$, but will create a cycle in the growing graph in so
doing, no link will be drawn between $i_l$ and $j_l$.  Instead, we will skip
$(i_l, j_l)$ and move on to the pair $(i_m, j_m)$ with the next smallest
distance $d_{i_m j_m}$.  The spanning graph obtained at the end contains no
cycles, hence the name minimal spanning tree.  Alternatively, since $d_{ij}$ is
a monotonically decreasing function of $C_{ij}$, we can get the same MST, if we
start by linking the pair of time series with the largest cross-correlation, and
then progressively linking pairs with smaller and smaller cross-correlations, so
long as we ensure the no-cycle constraint is satisfied at all times.

\section{Macroeconomic MSTs}
\label{sect:macroMST}

Before we move on to our MST analysis, let us first develop an intuitive picture
for the sectorial structure of the US economy.  As a significant fraction of
what the US produces is consumed domestically, the US market is a gigantic
consumption market.  We therefore expect the noncyclical consumer goods (NC) and
consumer services (CY) to be central players in the US economy.  Furthermore, CY
and NC consume products predominantly generated by the industrials (IN), thus we
expect IN, CY and NC (and perhaps also FN, since financing is an important
ingredient in US consumerism) to be the core sectors of the US economy.  In
contrast, emerging economic sectors such as telecommunications (TL) and
technologies (TC), along with less attractive economic sectors like healthcare
(HC) and utilities (UT), contribute less significantly to the GDP, and hence sit
at the fringe of the US economy.  Finally, the oil \& gas (EN) and basic
materials (BM) sectors are strongly driven by changes in the global supply and
demand cycle, and thus represent the US economy's connection to the global
market.

Indeed, this intuitive picture is supported by quantitative GDP data from the US
Bureau of Economic Analysis (BEA) \cite{BEA}.  The BEA uses an industry
classification scheme different from the ICB, so we map the BEA data onto the
ICB economic sectors using the subindustry descriptions available, as shown in
Table \ref{table:USGDP}, to figure out what their GDP contributions are.  Though
somewhat overestimated because of this mapping problem, we see that CY, IN, and
NC combined contributed on average about 7,000 billion USD to the US gross
domestic product (GDP), which was worth 14,200 billion USD on average between
2007 and 2009.  These three sectors therefore contribute about 50\% of the US
GDP.  FN, by itself, contributed about 3,000 billion USD to the US GDP, or about
21\%.  These contributions support our intuitive picture of these four sectors
playing a central role in the US economy.  Based on the subindustry breakdown in
Table \ref{table:USGDP}, HC probably contributes on average slightly more than
1,000 billion USD to the US economy annually, or about 7\% of the GDP.  The
contributions from TC and TL are hard to nail down, because their subindustries
are frequently lump with others that are assigned to CY, but we estimate that
their contributions to the GDP are around the 300 billion USD mark, or about 3\%
of the GDP.  From Table \ref{table:USGDP}, we see that UT contributes 260
billion USD (about 2\%) on average annually, while BM and EN both contribute
around the 400 billion USD mark, also around 3\%, to the US GDP.  Again, based
on these numbers, we are probably not too far off guessing that these sectors
are less critical to the US economy.

\begin{table}[htbp]
\centering\footnotesize
\caption{Contributions in billions of USD to the US gross domestic product (GDP)
by the various Bureau of Economic Analysis (BEA) industries for 2007, 2008, and
2009.  The industry classification used by the BEA differs from the ICB.  While
some BEA industries can be easily assigned to a unique DJUS economic sector,
other BEA industries contain subindustries from more than one DJUS economic
sector.  A BEA industry is marked with a dagger, if we assign it to one economic
sector, but some of its listed subindustries to other economic sectors.  In these
cases, the GDP contributions will be split between the economic sectors.  On the
other hand, if its subindustries belong to different economic sectors, but are
not finely divided enough for us to split its GDP contributions, the BEA
industry will be marked with an asterix.  For BEA industries with mixed
components in CY, IN, and NC, we do not break the GDP contributions
down by subindustries.  Data taken from Ref.~\cite{BEA}.}
\label{table:USGDP}
\vskip .5\baselineskip
\begin{tabular}{lllrrrl}
\hline
\multicolumn{3}{l}{Industry} & 2007 & 2008 & 2009 &
\parbox[c]{1.5cm}{\raggedright Economic Sector} \\
\hline
\multicolumn{3}{l}{\parbox[t]{5.7cm}{\raggedright Gross domestic product}} & 14,061.8 & 14,369.1 & 14,119.0 & \\
\hline
\multicolumn{3}{l}{Private industries} & 12,301.9 & 12,514.0 & 12,196.5 & \\
& \multicolumn{2}{l}{\parbox[t]{5.5cm}{\raggedright Agriculture, forestry, fishing, and
hunting*}} & 144.7 & 160.1 & 133.1 & NC \\
& \multicolumn{2}{l}{Mining\textdagger} & 91.3 & 106.3 & 99.1 & BM \\
& & \parbox[t]{5.1cm}{\raggedright Oil and gas extraction} & 162.9 & 210.8 & 141.7 & EN \\
& \multicolumn{2}{l}{Utilities} & 248.8 & 262.6 & 268.1 & UT \\
& \multicolumn{2}{l}{Construction} & 657.2 & 623.4 & 537.5 & IN/NC \\ 
& \multicolumn{2}{l}{Manufacturing\textdagger} & 1,138.9 & 1,119.9 & 1,092.1 & IN/NC \\
& & Primary metals & 59.0 & 61.5 & 43.4 & BM \\
& & Paper products & 58.6 & 53.8 & 56.1 & BM \\
& & \parbox[t]{5.1cm}{\raggedright Petroleum and coal products} & 149.7 & 151.9 & 120.0 & EN \\
& & \parbox[t]{5.1cm}{\raggedright Chemical products} & 223.2 & 201.1 & 216.5 & BM/HC \\
& & \parbox[t]{5.1cm}{\raggedright Plastics and rubber products} & 69.5 & 59.4 & 56.7 & BM \\
& \multicolumn{2}{l}{Wholesale trade} & 813.3 & 822.9 & 780.8 & CY \\
& \multicolumn{2}{l}{Retail trade} & 886.1 & 840.2 & 819.6 & CY \\
& \multicolumn{2}{l}{\parbox[t]{5.5cm}{\raggedright Transportation and warehousing}} & 405.4 & 418.7 & 389.5 & IN/CY \\
& \multicolumn{2}{l}{Information*\textdagger} & 285.6 & 293.4 & 283.5 & CY \\
& & \parbox[t]{5.1cm}{\raggedright Broadcasting and telecommunications} & 347.7 & 359.1 & 355.8 & TL \\
& \multicolumn{2}{l}{\parbox[t]{5.5cm}{\raggedright Finance, insurance, real estate, rental, and leasing}} & 2,891.3 & 2,974.9 & 3,040.3 & FN \\
& \multicolumn{2}{l}{\parbox[t]{5.5cm}{\raggedright Professional and business services}} & 1,700.5 & 1,768.8 & 1,701.3 & CY/IN \\
& \multicolumn{2}{l}{\parbox[t]{5.5cm}{\raggedright Educational services, health care, and social assistance}} & 1,078.3 & 1,148.9 & 1,212.9 & \\
& & \parbox[t]{5.1cm}{\raggedright Health care and social assistance} & 941.0 & 1,001.9 & 1,057.9 & HC \\
& \multicolumn{2}{l}{\parbox[t]{5.5cm}{\raggedright Arts, entertainment, recreation, accommodation, and food services}} & 545.2 & 535.4 & 513.1 & CY \\
& \multicolumn{2}{l}{\parbox[t]{5.5cm}{\raggedright Other services, except government}} &	344.6 & 340.9 & 335.4 & CY \\
\hline
\multicolumn{3}{l}{Government} & 1,759.9 & 1,855.1 & 1,922.5 & \\
\hline
\end{tabular}
\end{table}

\begin{table}[htbp]
\centering\footnotesize
\caption{Cross-correlation matrix computed from the half-hourly time series of
the ten DJUS economic sector indices over the period February 2000 to August
2008.  Also shown are the cross correlations $\langle C \rangle$ of each
economic sector averaged across the rest of the US economy.}
\label{table:xcorrexample}
\vskip .5\baselineskip
\begin{tabular}{|c|c|c|c|c|c|c|c|c|c|c|}
\hline
& BM & CY & EN & FN & HC & IN & NC & TC & TL & UT \\
\hline
BM & & 0.611 & 0.522 & 0.568 & 0.347 & 0.656 & 0.556 & 0.438 & 0.435 & 0.458 \\
\hline
CY & 0.611 & & 0.320 & 0.767 & 0.435 & 0.815 & 0.660 & 0.679 & 0.600 & 0.433 \\
\hline
EN & 0.522 & 0.320 & & 0.316 & 0.261 & 0.393 & 0.350 & 0.254 & 0.276 & 0.451 \\
\hline
FN & 0.568 & 0.767 & 0.316 & & 0.403 & 0.751 & 0.616 & 0.577 & 0.559 & 0.440 \\
\hline
HC & 0.347 & 0.435 & 0.261 & 0.403 & & 0.436 & 0.469 & 0.325 & 0.342 & 0.323 \\
\hline
IN & 0.656 & 0.815 & 0.393 & 0.751 & 0.436 & & 0.660 & 0.775 & 0.618 & 0.445 \\
\hline
NC & 0.556 & 0.660 & 0.350 & 0.616 & 0.469 & 0.660 & & 0.472 & 0.497 & 0.485 \\
\hline
TC & 0.438 & 0.679 & 0.254 & 0.577 & 0.325 & 0.775 & 0.472 & & 0.566 & 0.270 \\
\hline
TL & 0.435 & 0.600 & 0.276 & 0.559 & 0.342 & 0.618 & 0.497 & 0.566 & & 0.382 \\
\hline
UT & 0.458 & 0.433 & 0.451 & 0.440 & 0.323 & 0.445 & 0.485 & 0.270 & 0.382 & \\
\hline
$\langle C\rangle$ & 0.510 & 0.591 & 0.349 & 0.555 & 0.371 & 0.617 & 0.529 & 0.484 & 0.475 & 0.410 \\
\hline
\end{tabular}
\end{table}

\begin{figure}[htbp]
\centering\footnotesize
\includegraphics[scale=0.4]{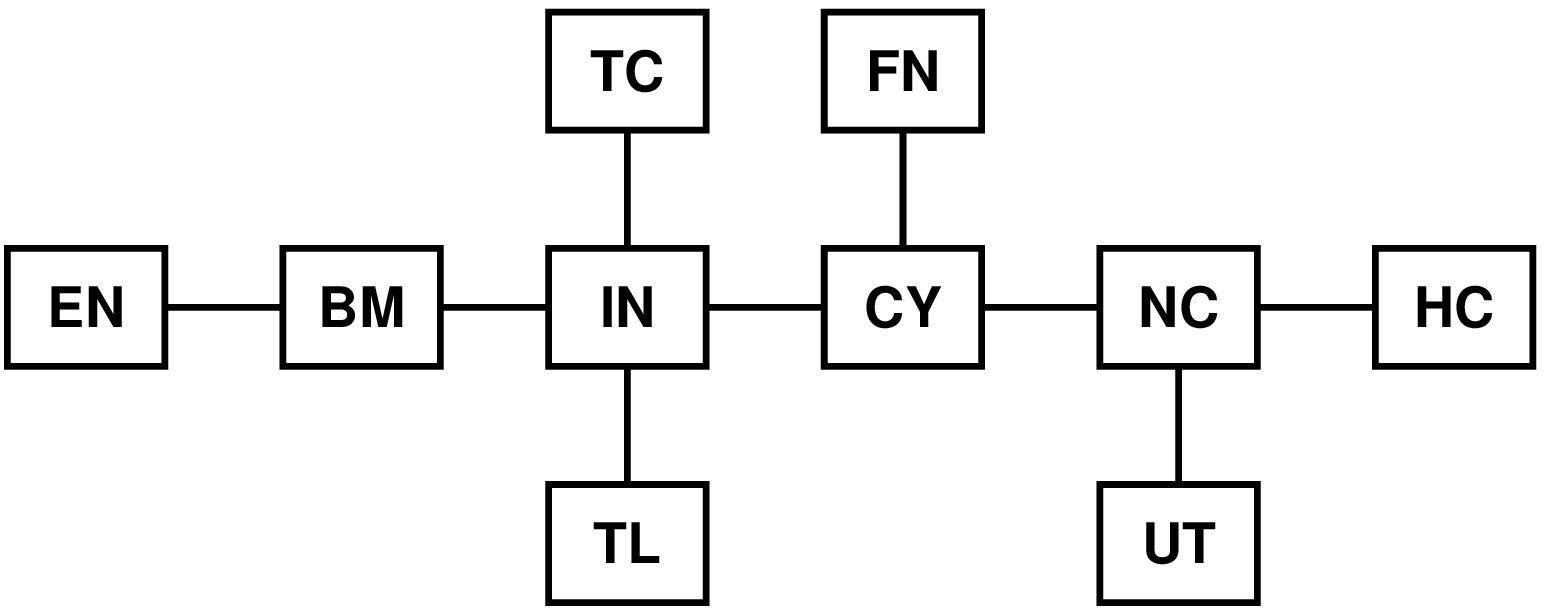}
\vskip .5\baselineskip
(a)
\vskip \baselineskip
\includegraphics[scale=0.4]{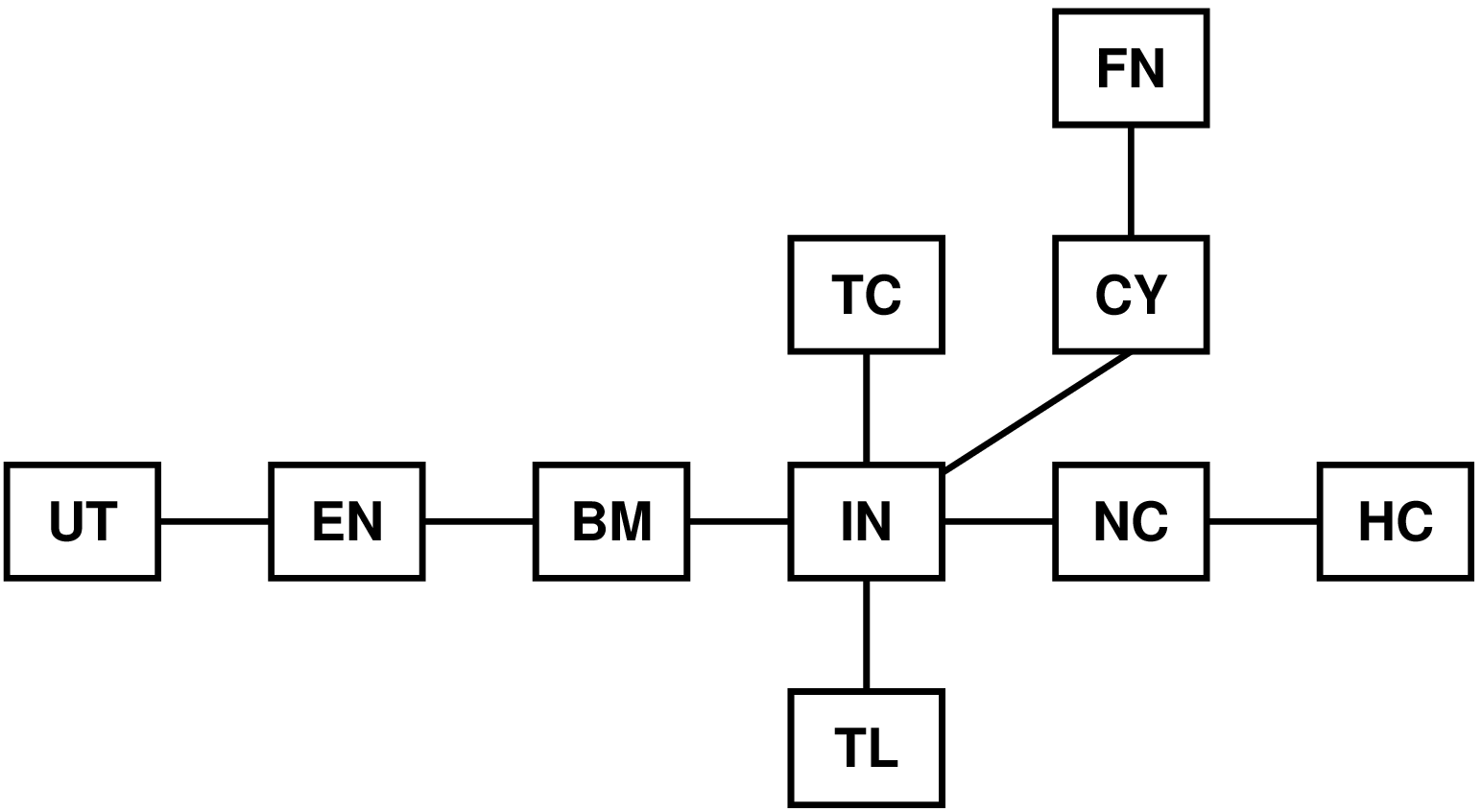}
\vskip .5\baselineskip
(b)
\vskip \baselineskip
\includegraphics[scale=0.4]{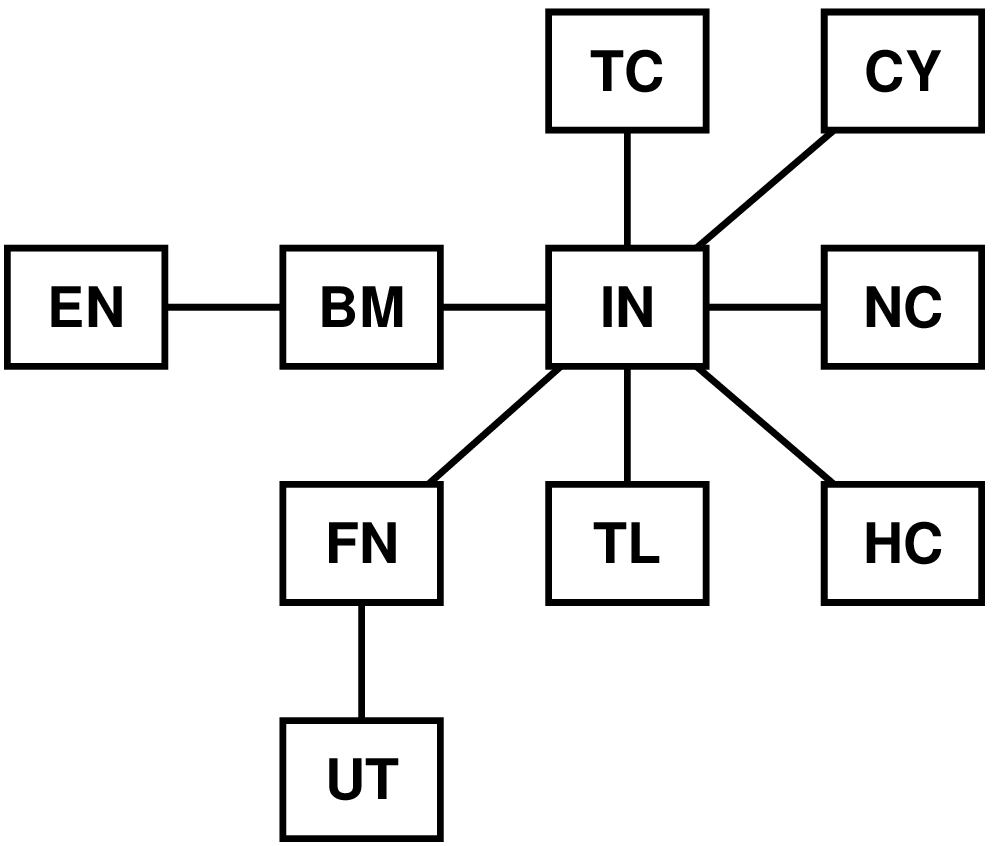}
\vskip .5\baselineskip
(c)
\vskip \baselineskip
\includegraphics[scale=0.4]{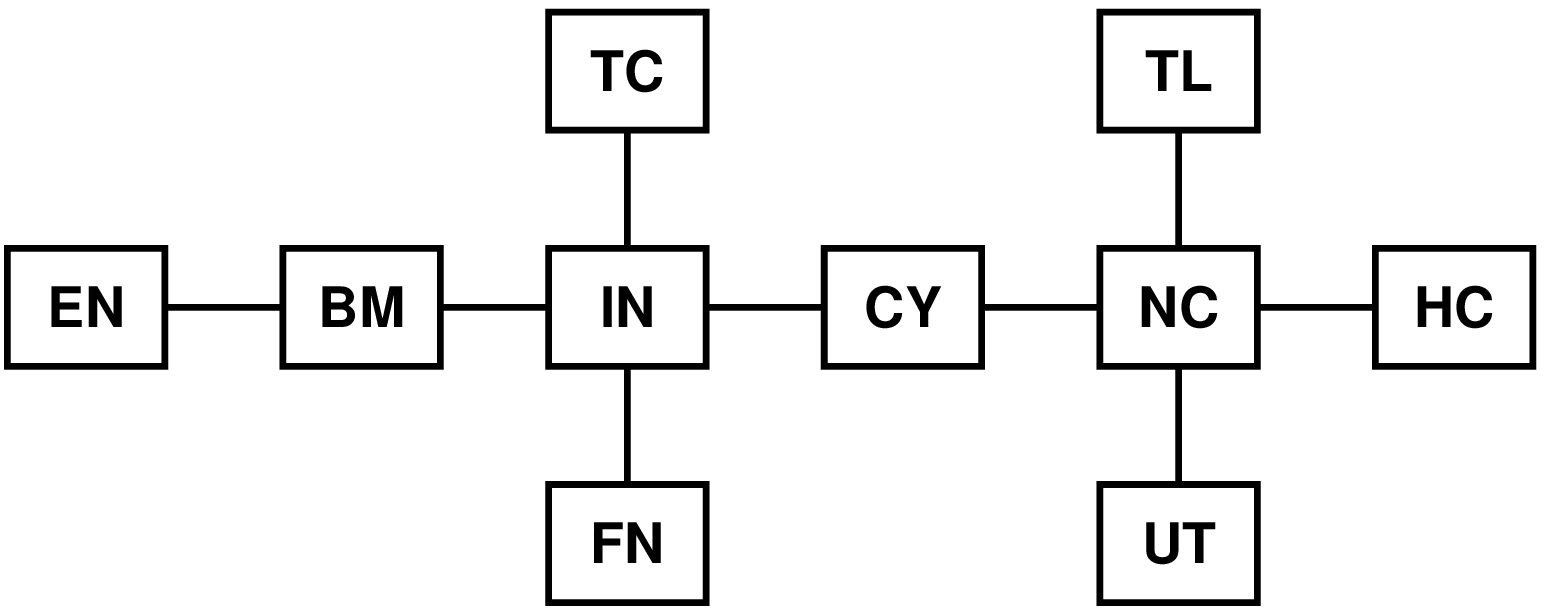}
\vskip \baselineskip
(d)
\caption{The MSTs of the ten DJUS economic sectors, constructed using
half-hourly time series from (a) February 2000 to August 2008, (b) 2001--2002,
(c) 2004--2005, and (d) 2008-2009.  The first and the third two-year windows,
(b) and (d), are entirely within an economic crisis, whereas the second two-year
window, (c), is entirely within an economic growth period.}
\label{fig:MSTgross}
\end{figure}

To further verify our intuitive picture of the US economy, we computed the
cross-correlation matrix of the time series from February 2000 to August 2008.
As shown in Table \ref{table:xcorrexample}, IN and CY are the most strongly
correlated, with $C(\rm IN, CY) = 0.815$, while EN and TC are the least strongly
correlated, with $C(\rm EN, TC) = 0.254$.  Based on the average cross
correlations $\langle C \rangle$, it also appears that IN is most strongly tied
in with the rest of the sectors ($\langle C \rangle(\rm IN) = 0.617$), while EN
is least strongly tied in with the rest of the US economy ($\langle C\rangle(\rm
EN) = 0.349$).  Based on this cross-correlation matrix, we constructed the MST
shown in Fig.~\ref{fig:MSTgross}(a).  As expected, the core sectors of the US
economy, IN, CY and NC, are at the centre of the MST.  The sectors EN and BM,
which represent the US economy's connection to the world market, sit on one end
of the MST, while the sectors HC, TC, TL, and UT, lies on the fringe of the MST,
consistent with their lesser importance to the US economy.  Heimo \emph{et
al.}~arrived at a similar conclusion, in their MST study of 116 NYSE stocks from
1982 to 2000 \cite{Heimo2009PhysicaA388p145}.

Over the period 2000 to 2009, the US National Bureau of Economic Research (NBER)
recorded two economic contractions \cite{NBER}.  The first was from March 2001
to November 2001.  The second was from December 2007 to June 2009.  Our own
studies showed that the US economy went from a crisis phase (mid-1998 to
mid-2003, which contained the March 2001 to November 2001 contraction) into a
growth phase (mid-2003 to mid-2007), and back into a crisis phase (mid-2007 to
present, which contained the December 2007 to June 2009 contraction)
\cite{Wong2009}.  We expect interesting structural differences between the MSTs
constructed entirely within the previous crisis (2001--2002,
Fig.~\ref{fig:MSTgross}(b)), the previous growth (2004--2005,
Fig.~\ref{fig:MSTgross}(c)), and the present crisis (2008--2009,
Fig.~\ref{fig:MSTgross}(d)).  Indeed, we see two topologically distinct MST
structures: a chain-like MST structure which occurs for both crises, and a
star-like MST structure which occurs for the growth phase.  Even though we only
have three data points (two crises and a growth), we believe the generic
association of chain-like MST and star-like MST to the crisis and growth phases
respectively is statistically robust.  Our reasons are two-fold.  First, the MST
is a representation based on order statistics (ranks of cross correlations).
Results derived based on order statistics, which are insensitive to noise, tend
to be highly robust statistically, as we have illustrated in Section
\ref{sect:mst}.  Second, the star-to-chain transition in the MST structure as
the US economy goes from growth into crisis cannot be brought about by a fixed
quantum increase, nor can it be caused by a proportional increase, in
correlations.  These two types of correlational changes do not change the
ordering of cross correlations among the ten economic sectors, and hence cannot
modify the MST.  

Since noise and global shifts in correlations cannot be responsible for the
star-to-chain or the chain-to-star transitions, correlational changes that
accompany these transitions must be highly significant.  Our assessment that the
topology change in the MST is statistically significant is further supported by
the observations by Onnela \emph{et al.}, who looked at the mean occupation
level around the most connected node in their MST, and found the mean occupation
level becoming low during market crashes \cite{Onnela2003PhysicaA324p247,
Onnela2003PhysScriptT105p48, Onnela2003PhysRevE68e056110}.  This is the same
phenomenon we see for the star-to-chain evolution, at the microscopic scale of
individual stocks.  In the next two sections, we will investigate the characters
of these correlational changes, and discuss the implications for early detection
of true economic recovery based on the chain-to-star transition.

From Fig.~\ref{fig:MSTgross}, we also see that in both the crisis and growth
phases, IN is found be the central industry of the US economy.  This is
understandable, since the United States is a highly industrialised country with
IN driving the rest of the sectors.  However, when the US economy went from the
mid-2003 to mid-2007 economic growth into the current crisis, the IN star center
shed the sectors NC, HC and TL, which shifted to other parts of the MST.  In the
restructured MST, NC formed the center of another cluster.  We believe this is a
signature of the trigger role played by NC in the Subprime Crisis, since
homebuilders and developers (who are most directly affected by the waves of
mortgage defaults) are classified under this economic sector.  More
interestingly, the cluster centered around NC consists of HC, TL, and UT, which
were part of the five sectors that went first into the crisis phase (see Fig.~5
in Ref.~\cite{Lee2009arXiv09114763}).  The last of these five sectors is IN, so
it appears that correlational changes within these five sectors in July 2007 is
responsible for the main difference between the growth MST
(Fig.~\ref{fig:MSTgross}(c)) and the crisis MST (Fig.~\ref{fig:MSTgross}(d)).
This gross restructuring of the MST thus provides an interesting way to
visualize how the current financial crisis propagated throughout the entire US
economy.

\section{Segment-by-segment analysis}
\label{sect:segmentMST}

\begin{figure}[ht]
\centering
\includegraphics[scale=0.5]{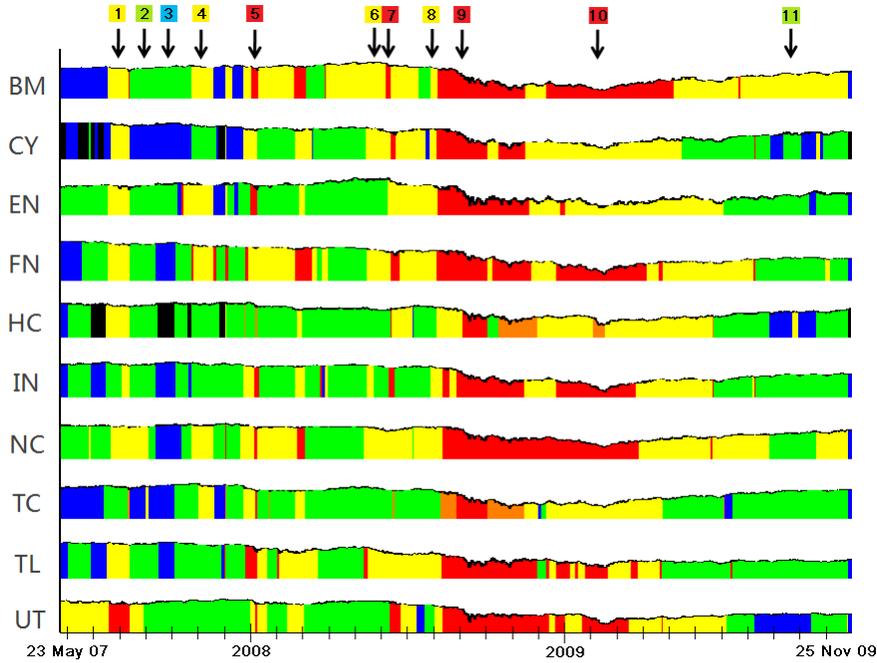}
\caption{Eleven corresponding segments identified in the time series of the ten
DJUS economic sector indices after the mid-2007 onset of the present global
financial crisis.  In this figure, the corresponding segments are numbered from
`1' to `11'.  In the rest of the paper, we label these corresponding segments
by their dominant volatility, and then by their order of appearance, so that `1'
= Y1, '2' = G1, '3' = B, '4' = Y2, '5' = R1, '6' = Y3, '7' = R2, '8' = Y4, '9' =
R3, `10' = R4, `11' = G2.  Note that Y1 is the same corresponding segment
identified in Fig.~\ref{fig:selectlength}, where we show the temporal
distributions of clustered segments of the ten DJUS economic sectors from May
23, 2007 to August 29, 2008, instead of from May 23, 2007 to November 25, 2009
in this figure.}
\label{fig:segments}
\end{figure}

Even within the macroeconomic growth and crisis phases, the DJUS economic sector
time series are highly nonstationary.  Both the cross correlations between the
ten sectors, and the MSTs they imply, are expected to be highly dynamic.  To
understand how cross correlations change with time, we extracted the average
cross correlations of the ten DJUS economic sectors in 11 corresponding segments
within the present financial crisis (see Fig.~\ref{fig:segments}).  All four
macroeconomic phases are represented in these 11 corresponding segments.
Ranking the average cross correlations from highest to lowest in Table
\ref{table:sectorxcorrranks}, we see that IN is always most strongly correlated
to the rest of the US economy, whatever the prevailing economic climate,
followed by CY and NC.  Meanwhile, EN is most weakly coupled to the rest of the
US economy, in most of the corresponding segments.  This is consistent with our
expectation that the oil \& gas industry's strong dependence on global supply
and demand makes it less susceptible to movements within the US economy.

\begin{table}[htbp]
\centering\footnotesize
\caption{Ranks of the ten DJUS economic sectors based on their average
half-hourly cross-correlations, over February 2000 to November 2009, as well as
over the 11 corresponding segments identified in Fig.~\ref{fig:segments}.  The
average cross correlations for EN in Y3, and those for BM, EN, and UT in Y4, are
anomalously low, even negative.  Also shown are the cross correlations
$\langle\!\langle C \rangle\!\rangle$ averaged over all ten sectors, for the
entire period from February 2000 to November 2009, as well as the 11
corresponding segments.}
\label{table:sectorxcorrranks}
\vskip .5\baselineskip
\begin{tabular}{|c|c|c|c|c|c|c|c|c|c|c|c|}
\hline
& BM & CY & EN & FN & HC & IN & NC & TC & TL & UT & $\langle\!\langle C
\rangle\!\rangle$ \\
\hline
Entire & 5 & 2 & 10 & 3  & 9 & 1 & 4 & 6 & 7 & 8 & 0.489 \\ \hline
Y1     & 6 & 2 & 10 & 4  & 8 & 1 & 3 & 5 & 9 & 7 & 0.811 \\ \hline
G1     & 6 & 2 & 10 & 5  & 7 & 1 & 3 & 4 & 9 & 8 & 0.738 \\ \hline
B      & 2 & 8 & 10 & 7  & 4 & 1 & 3 & 5 & 9 & 6 & 0.511 \\ \hline
Y2     & 4 & 3 & 10 & 8  & 6 & 1 & 2 & 5 & 7 & 9 & 0.700 \\ \hline
R1     & 7 & 9 & 3  & 10 & 8 & 1 & 2 & 4 & 5 & 6 & 0.797 \\ \hline
Y3     & 9 & 2 & 10 & 7  & 4 & 1 & 5 & 3 & 6 & 8 & 0.562 \\ \hline
R2     & 7 & 3 & 10 & 2  & 8 & 1 & 6 & 9 & 4 & 5 & 0.703 \\ \hline
Y4     & 9 & 4 & 10 & 2  & 5 & 1 & 7 & 3 & 6 & 8 & 0.559 \\ \hline
R3     & 5 & 2 & 10 & 10 & 4 & 1 & 3 & 6 & 9 & 8 & 0.863 \\ \hline
R4     & 4 & 3 & 5  & 7  & 6 & 2 & 1 & 8 & 10 & 9 & 0.796 \\ \hline
G2     & 4 & 2 & 6  & 7  & 9 & 1 & 5 & 3 & 10 & 8 & 0.709 \\ \hline
\end{tabular}
\end{table}

In general, we observe a positive correlation between the average market cross
correlation $\langle\!\langle C \rangle\!\rangle$ and the market volatility.  As
can be seen from Table \ref{table:sectorxcorrranks}, higher average market cross
correlations are generally associated with higher volatility phases.
Specifically, in the low-volatility economic growth phase (B), the average
market cross correlation is low, in the range $0.5 < \langle \!  \langle C
\rangle \! \rangle < 0.6$, whereas in the moderate-volatility market correction
phase (G1, G2), the average market cross correlation is also moderate, in the
range $0.6 < \langle \! \langle C \rangle \! \rangle < 0.7$.  In the
higher-volatility phases (Y1, Y2, Y3, Y4; R1, R2, R3, R4), the average market
cross correlation is high, in the range $0.7 < \langle \! \langle C \rangle \!
\rangle < 0.9$.  The higher correlations observed during the higher-volatility
phases is consistent with the tendency for traders to panic and to buy and sell
stocks from different sectors at the same time.  Conversely, when the market is
calm, stocks from different sectors tend to be bought and sold at different
times, explaining the lower correlations observed for the lower-volatility
phases.  These average market cross correlations are all higher than the average
market cross correlations computed over the entire time series, because cross
correlations within the US economy has been increasing over the years (see for
example, Ref.~\cite{GladysFYP2009}).

\subsection{MST structures}

As expected, changes in the MST can be seen going from one corresponding segment
to the next (see Fig.~\ref{fig:segmentMSTs}).  However, the sectors IN, CY and
NC remain at the cores of all 11 MSTs, whereas the sectors HC, TC, TL, and UT
are mostly found at the fringes of these MSTs.  Interestingly, the financials
(FN), which is frequently found close to the core, occasionally drifts out to
the fringe.  While the core-and-fringe structure of the MSTs remains well
defined as the market volatility changes, we observe shifting relative
importances between the different sectors.  We wll study these MST
rearrangements, which we believe are the US economy's response to shocks
originating within specific economic sectors, in Section
\ref{sect:MSTrearrangements}.  Here, let us make the remarkable observation
that, through the fluxes of correlational changes, the EN-BM-IN-CY-NC-TC-HC
backbone of the MSTs remained relatively unchanged throughout the entire crisis
period.  This robust correlational structure must therefore be a key to
understanding the performance of the present US economy.

In Fig.~\ref{fig:segmentMSTs}, we incorporate more visual information on the
cross-correlation matrix, by varying the widths of the bonds in the MSTs.  The
thicker the bond between two sectors, the stronger their correlations.  As we
can then see, sectors at the core are generally more strongly correlated than
those on the fringes of the MSTs.  This reinforces our intuitive picture that
sectors on the fringe are more detached from the overall economy, whereas those
at the core are most important to the US economy.  In this representation of the
MSTs, a correlational core consisting of thick bonds can also be seen.  Even as
the core and backbone of the MSTs remain more or less unchanged, the
correlational core of thick bonds expands and contracts with time.  We can think
of the correlational core defining the active participants in the US economy for
a given corresponding segment.  In the high-volatility phase, the correlational
core expands all the way out to the fringes, telling us that fringe sectors
become more involved in the US economy during a financial crisis.  A similar
phenomenon was observed by Onnela \emph{et al.}~in the MSTs of individual stocks
across market crashes \cite{Onnela2003PhysicaA324p247,
Onnela2003PhysScriptT105p48, Onnela2003PhysRevE68e056110}.

\begin{figure}[htbp]
\centering\footnotesize
\begin{minipage}[t]{.49\linewidth}
\centering
\includegraphics[scale=0.3]{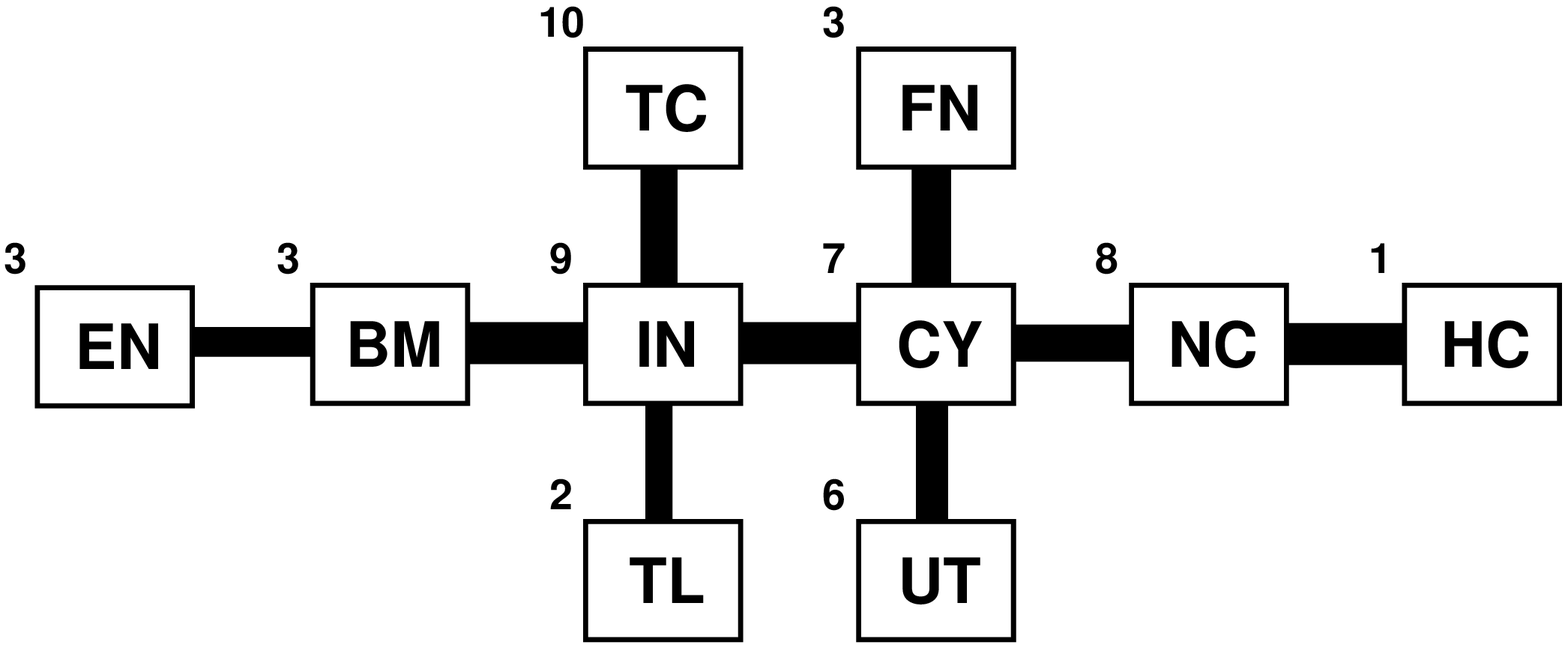}
\vskip .5\baselineskip
(a)
\end{minipage}
\hfill
\begin{minipage}[t]{.49\linewidth}
\centering
\includegraphics[scale=0.3]{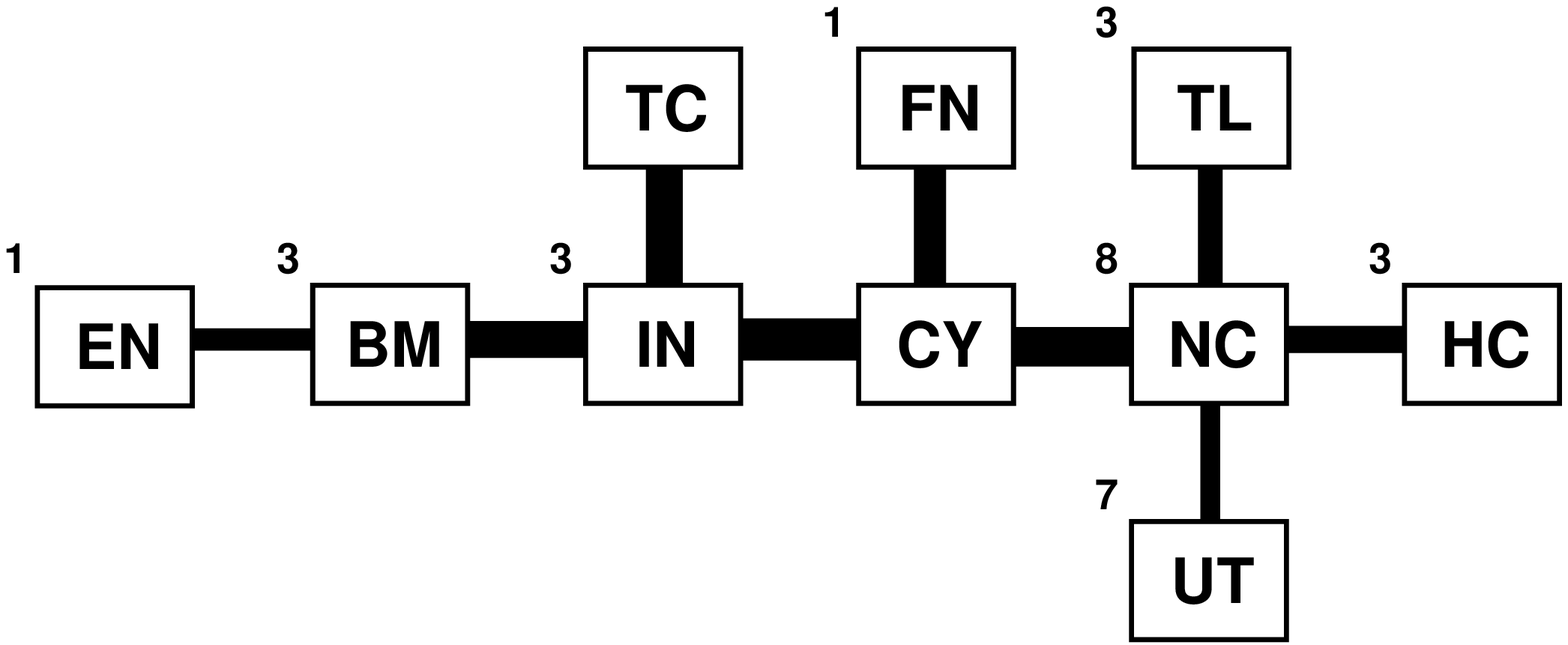}
\vskip .5\baselineskip
(b)
\end{minipage}
\vskip \baselineskip
\begin{minipage}[t]{.49\linewidth}
\centering
\includegraphics[scale=0.3]{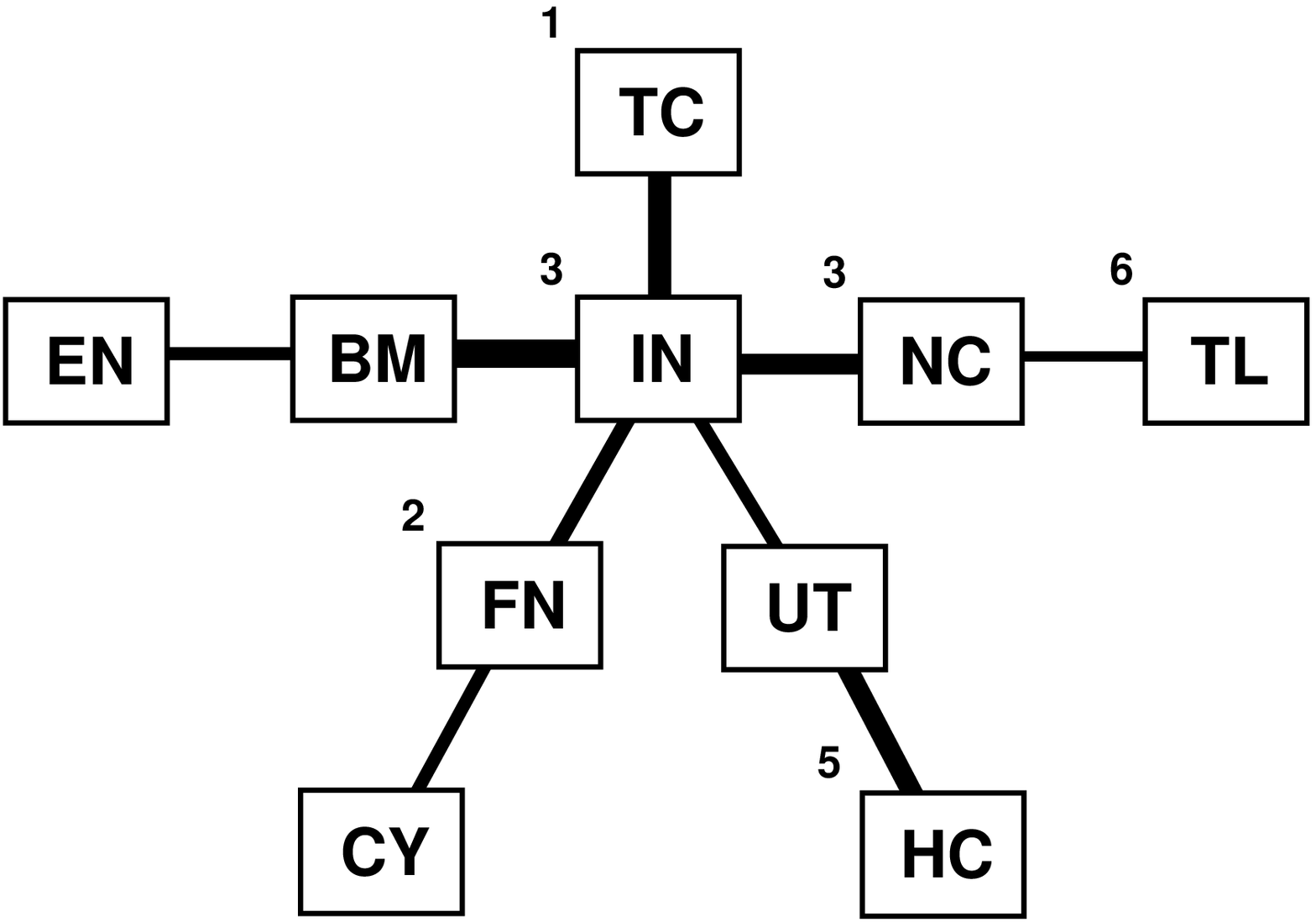}
\vskip .5\baselineskip
(c)
\end{minipage}
\hfill
\begin{minipage}[t]{.49\linewidth}
\centering
\includegraphics[scale=0.3]{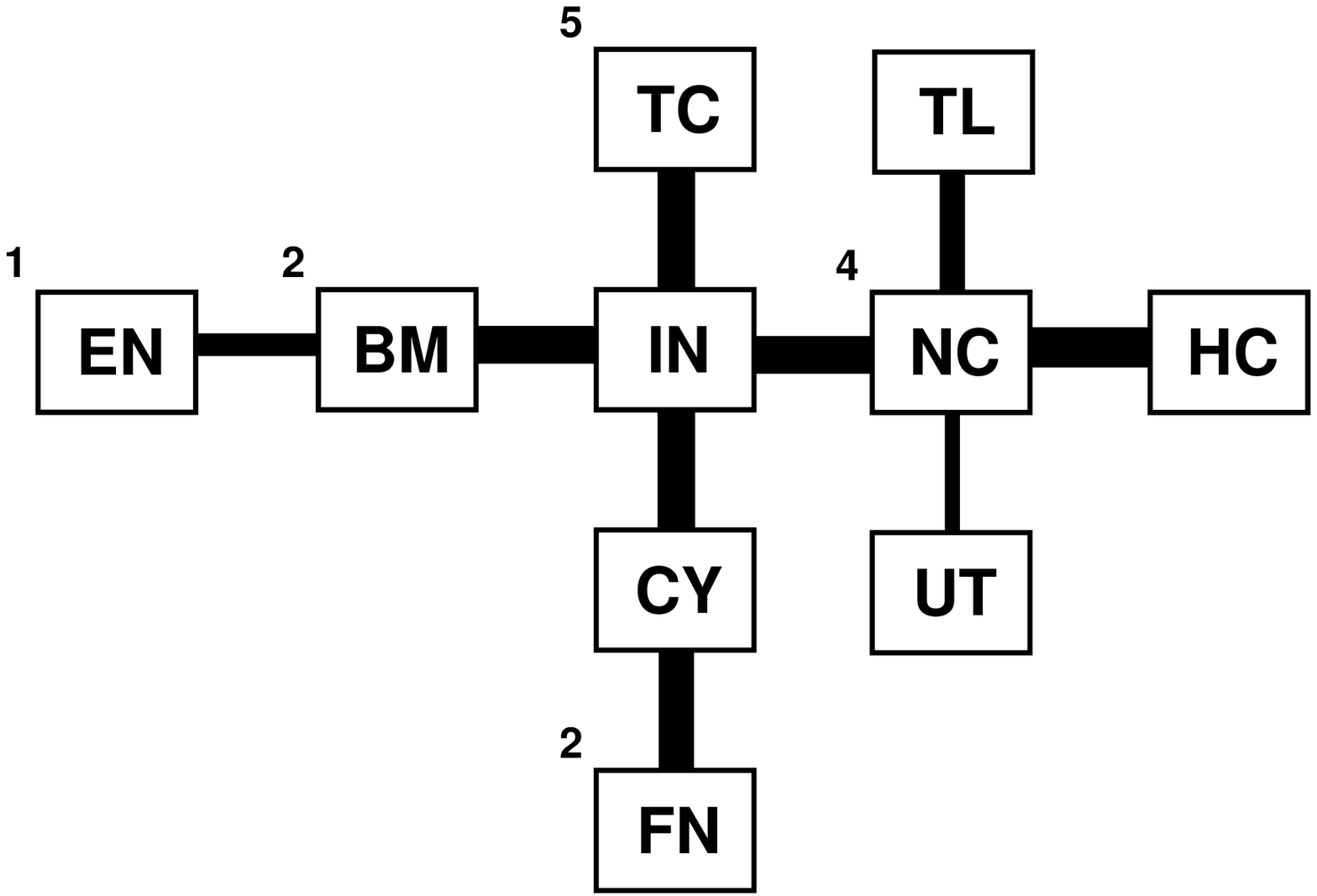}
\vskip .5\baselineskip
(d)
\end{minipage}
\vskip \baselineskip
\begin{minipage}[t]{.49\linewidth}
\centering
\includegraphics[scale=0.3]{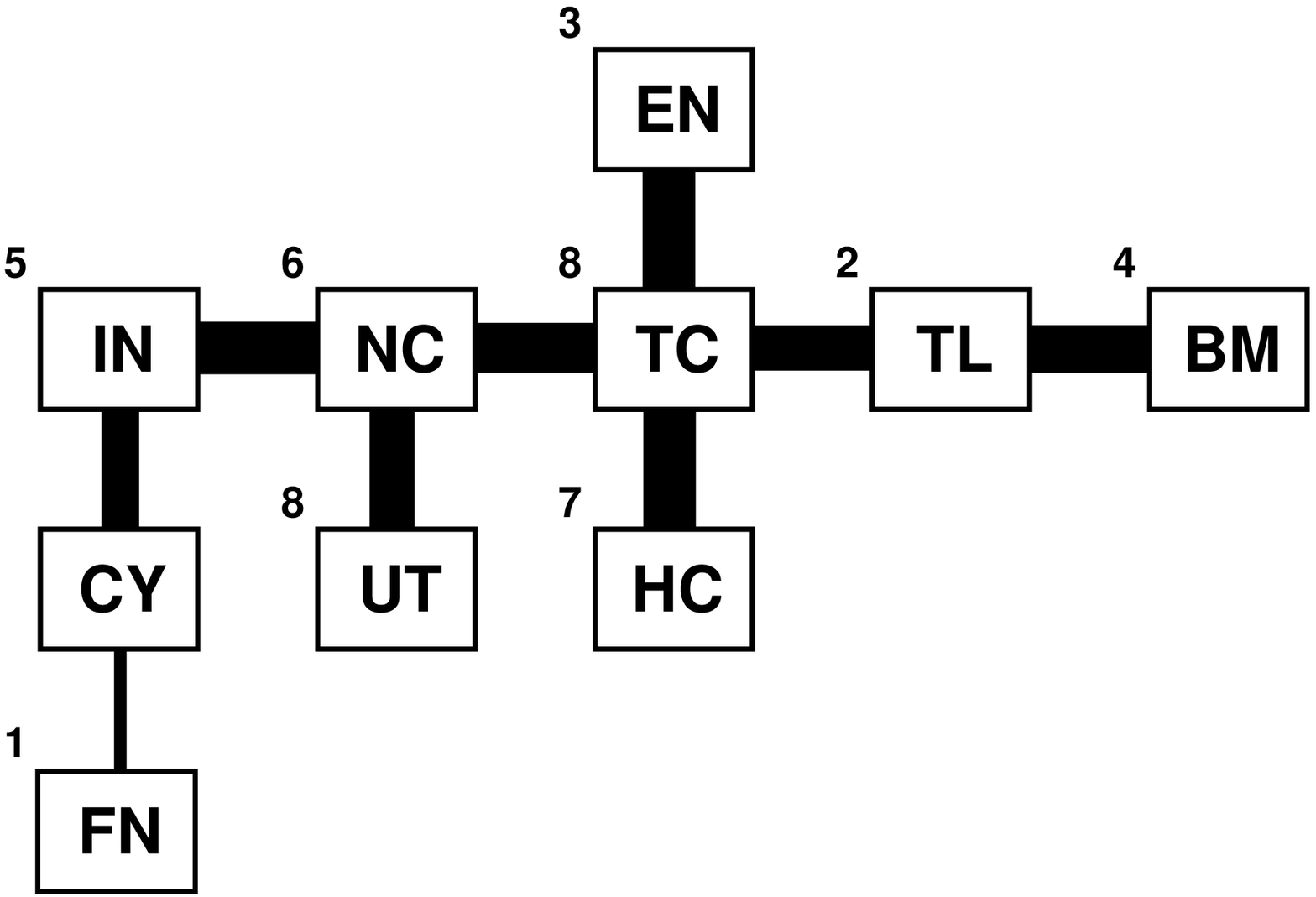}
\vskip .5\baselineskip
(e)
\end{minipage}
\hfill
\begin{minipage}[t]{.49\linewidth}
\centering
\includegraphics[scale=0.3]{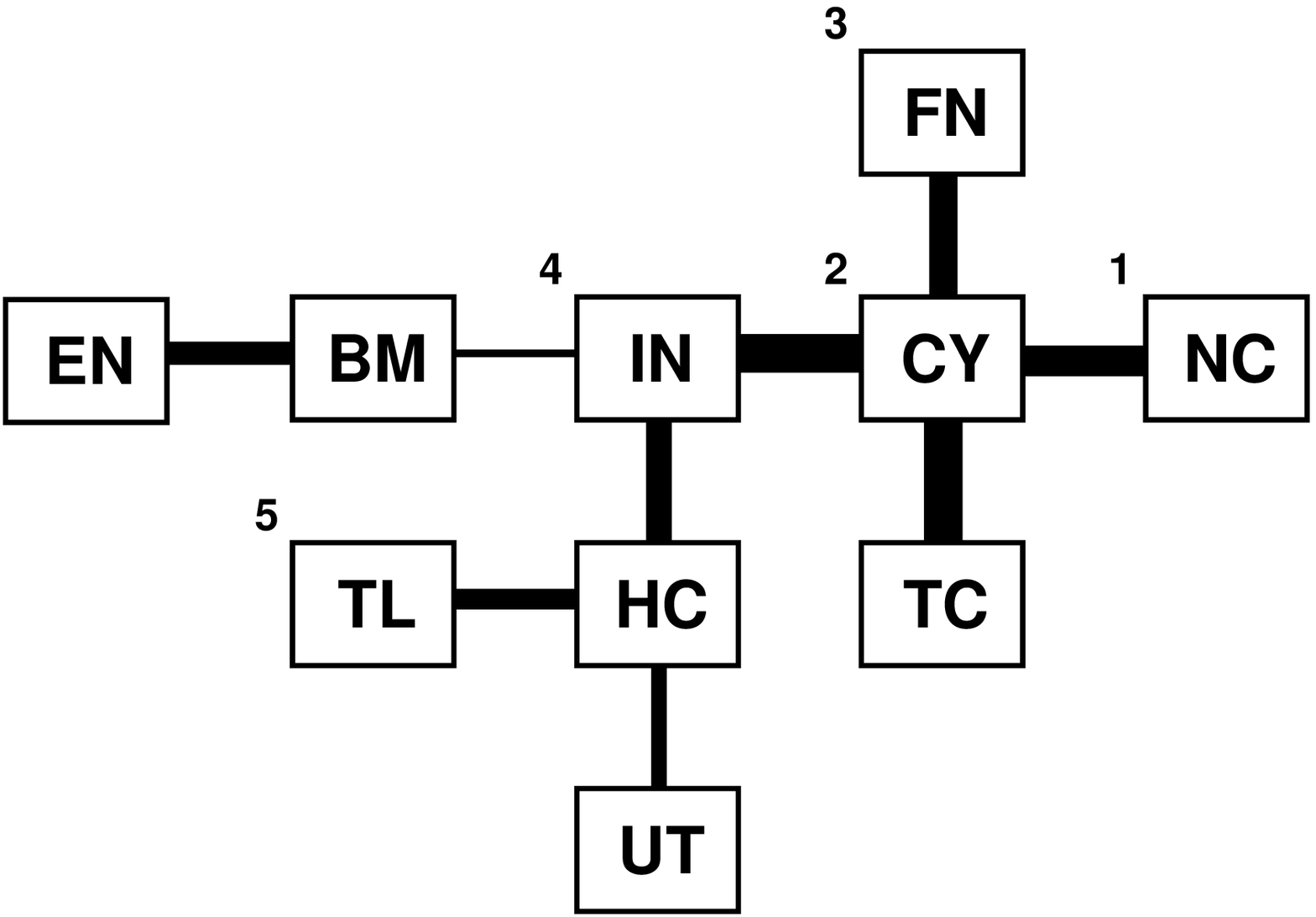}
\vskip .5\baselineskip
(f)
\end{minipage}
\caption{MSTs of the ten DJUS economic sectors for the corresponding segments
(a) Y1, (b) G1, (c) B, (d) Y2, (e) R1, (f) Y3, (g) R2, (h) Y4, (i) R3, (j) R4,
(k) G2, within the present financial crisis.  In this figure, thicker bonds
represent stronger cross correlations, whereas the number besides each sector
indicates the order with which the sector made the transition into the given
corresponding segment (whereever they can be identified).}
\label{fig:segmentMSTs}
\end{figure}

\begin{figure}[htbp]
\centering\footnotesize
\begin{minipage}[t]{.49\linewidth}
\centering
\includegraphics[scale=0.3]{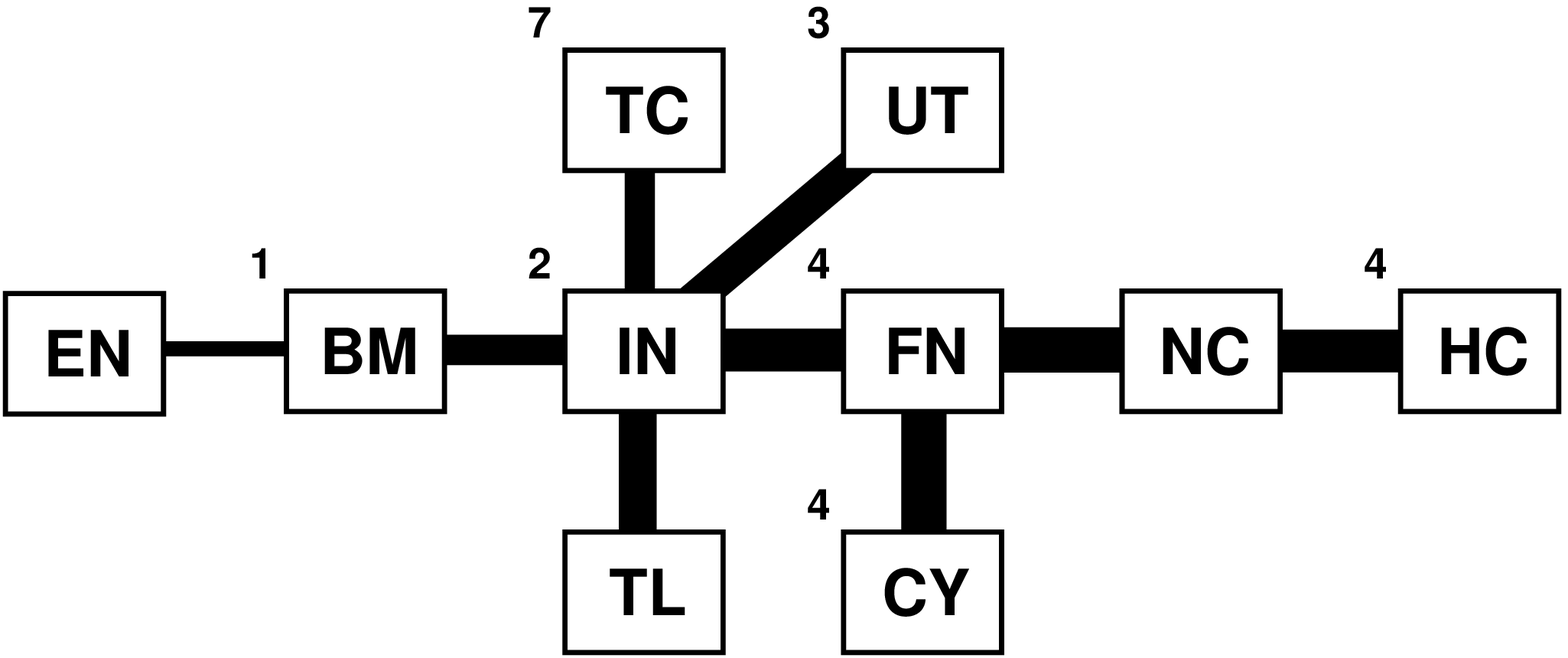}
\vskip .5\baselineskip
(g)
\end{minipage}
\hfill
\begin{minipage}[t]{.49\linewidth}
\centering
\includegraphics[scale=0.3]{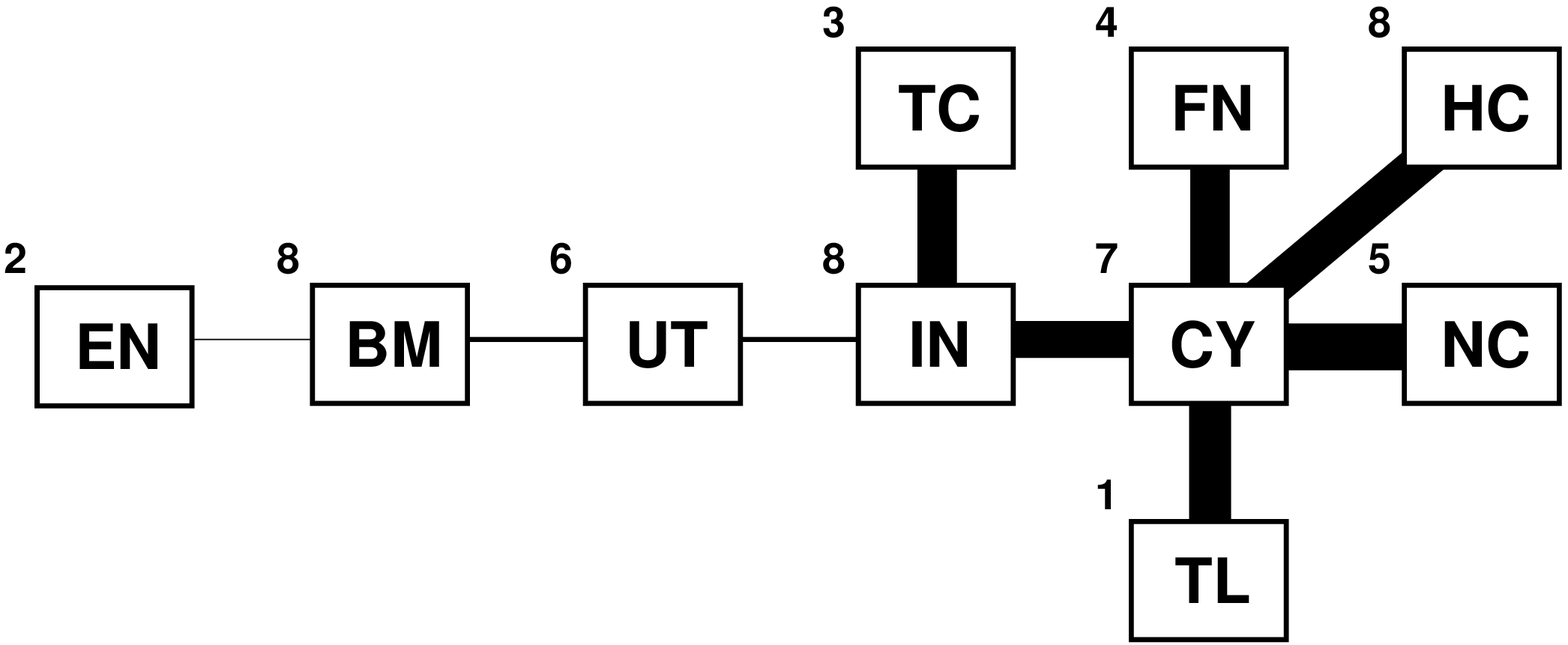}
\vskip .5\baselineskip
(h)
\end{minipage}
\vskip \baselineskip
\includegraphics[scale=0.3]{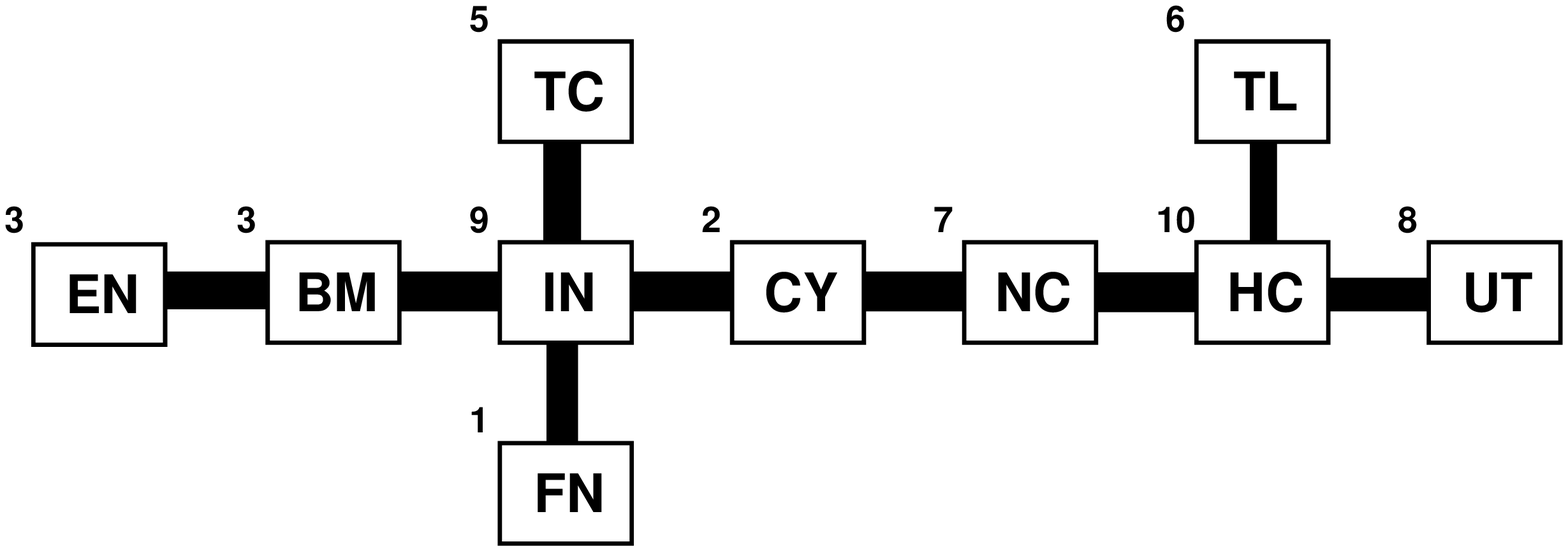}
\vskip .5\baselineskip
(i)
\vskip \baselineskip
\includegraphics[scale=0.3]{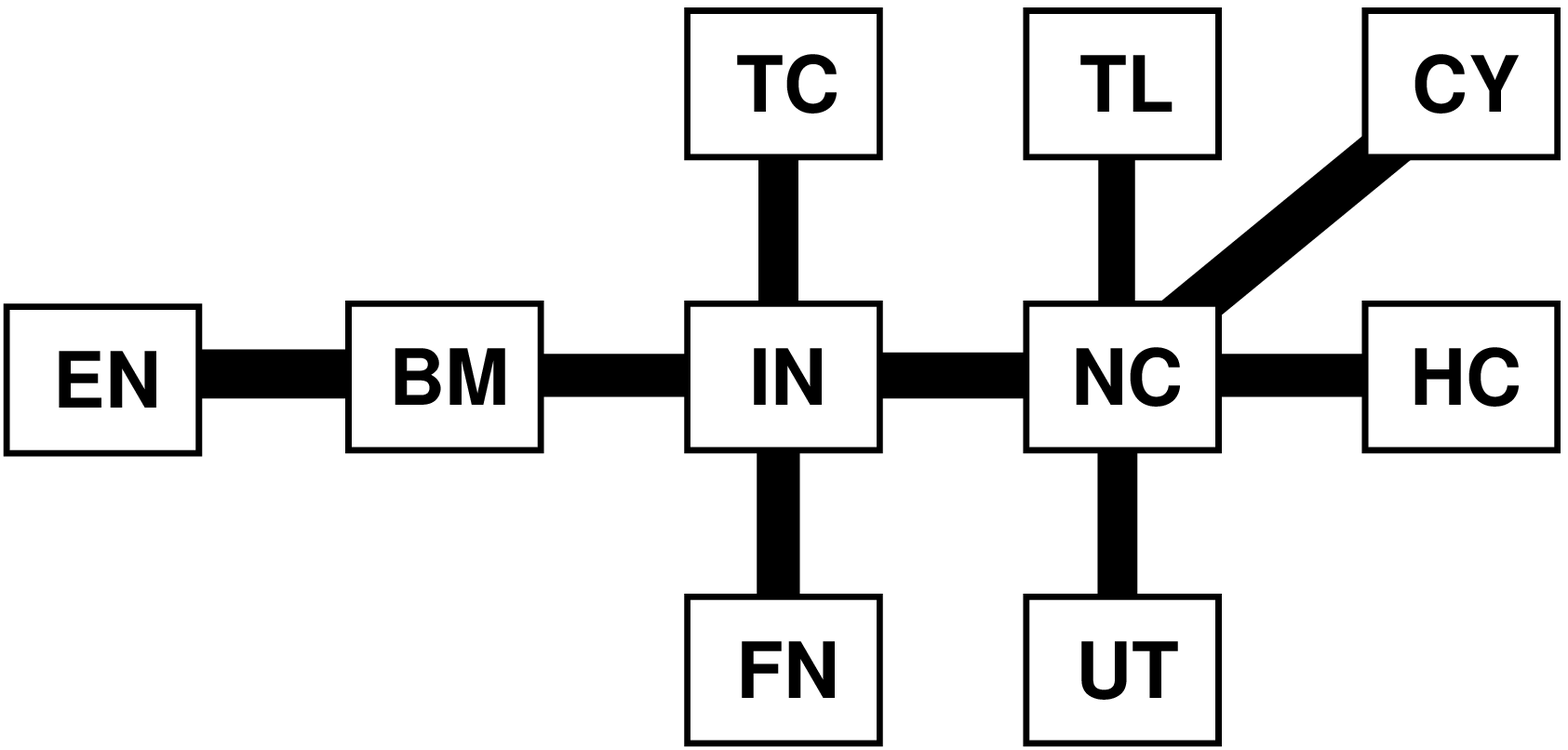}
\vskip .5\baselineskip
(j)
\vskip \baselineskip
\includegraphics[scale=0.3]{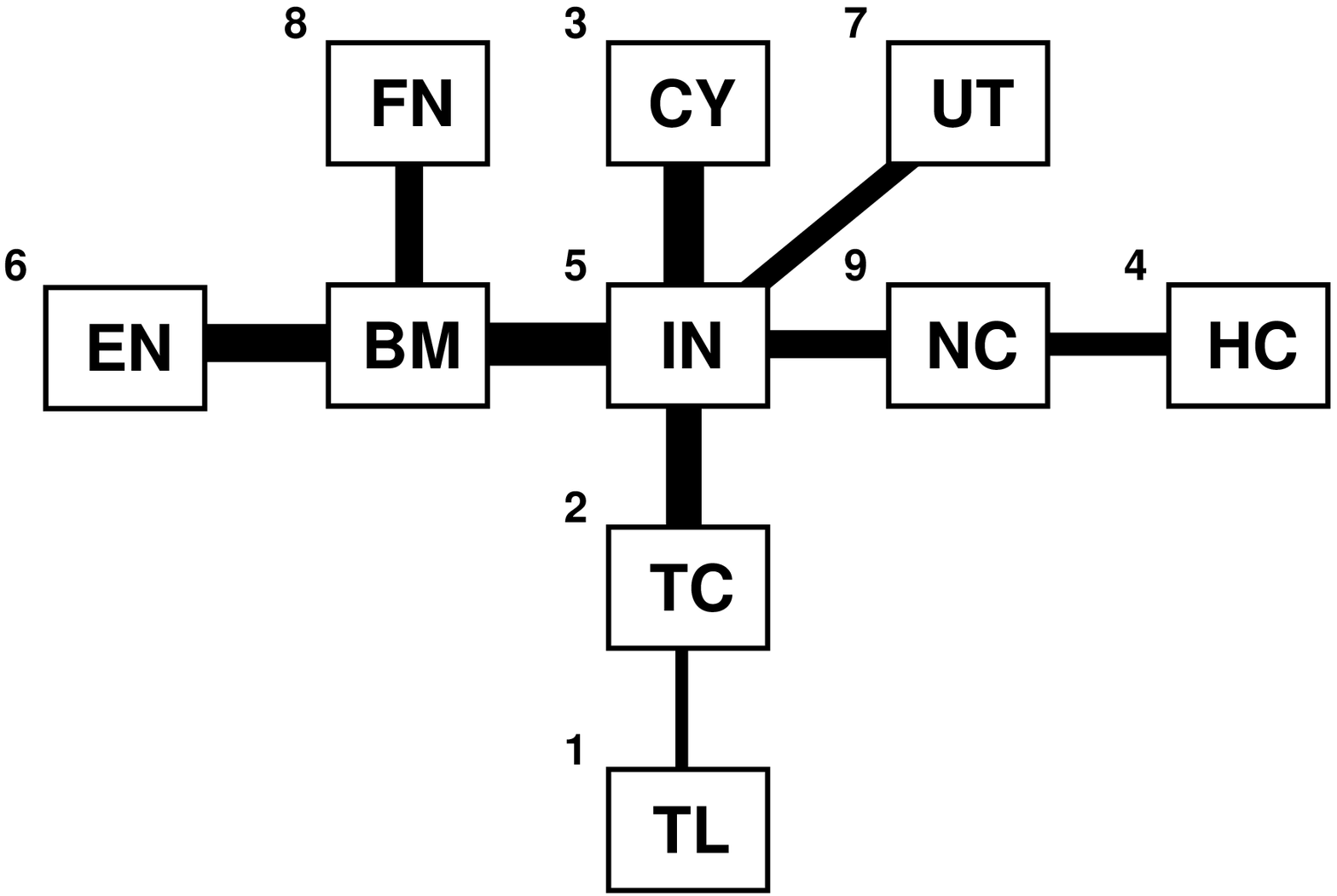}
\vskip .5\baselineskip
(k)
\setcounter{figure}{8}
\caption{(continued) MSTs of the ten DJUS economic sectors for the corresponding
segments (a) Y1, (b) G1, (c) B, (d) Y2, (e) R1, (f) Y3, (g) R2, (h) Y4, (i) R3,
(j) R4, (k) G2, within the present financial crisis.  In this figure, thicker
bonds represent stronger cross correlations, whereas the number besides each
sector indicates the order with which the sector made the transition into the
given corresponding segment (whereever they can be identified).}
\end{figure}

\subsection{MST dynamics}

In Ref.~\cite{Lee2009arXiv09114763}, we developed a causal tree analogy
speculating that exogenous shocks shaking the root of the tree will first be
felt strongly by branches closest to the root, and then weakly by branches
further from the root.  Naturally, now that we have a better picture of the
correlational structure of the US economy in the form of an MST, we expect
volatility shocks to propagate along the invariant backbone of the MST, since it
is along this backbone that we have the strongest cross correlations.  To
explore this idea, we make use of high-resolution temporal information available
from the segmentation/clustering analysis, to identify for each corresponding
segment the statistically significant start dates in the ten DJUS economic
sectors.  We then rank the start dates from earliest to latest, and in
Fig.~\ref{fig:segmentMSTs} label the sectors according to these ranks, omitting
those sectors for which the start date cannot be identified.  From the 11
corresponding segments identified within the present financial crisis, we find
that shocks always originate from the fringe of the MST, and propagate inwards.
However, contrary to our naive expectations, shocks do not necessarily propagate
along the MST.  For example, in Fig.~\ref{fig:segmentMSTs}(h), we see that the
corresponding segment Y4 started in TL, propagated to EN (which is not directly
connected to TL in the MST), and then onto TC and FN (both of which are not
directly connected to TL or EN), before propagating into the core of the MST.
This inward propagation of volatility shocks is seen even when the MST is
anomalous.  For example, in Fig.~\ref{fig:segmentMSTs}(e), where TC is at the
center of the MST, the corresponding segment R1 started first in FN, which has
moved to the fringe of the MST, then in TL, then in EN, and BM, before
propagating into the core of the MST.  In no case was a shock found to start at
the core of the MST.

\begin{table}[htbp]
\centering\footnotesize
\caption{Ranks of identifiable start dates in the ten DJUS economic sectors,
from earliest to latest, for each of the 11 corresponding segments between May
2007 and November 2009.}
\label{table:sectorstartdateranks}
\vskip .5\baselineskip
\begin{tabular}{|c|c|c|c|c|c|c|c|c|c|c|}
\hline
& BM & CY & EN & FN & HC & IN & NC & TC & TL & UT \\
\hline
Y1 & 3 & 7 & 3 & 3 & 1 & 9 & 8 & 10 & 2 & 6 \\ \hline
G1 & 3 & - & 1 & 1 & 3 & 3 & 8 & -  & 3 & 7 \\ \hline
B  & - & - & - & 2 & 5 & 3 & 3 & 1  & 6 & - \\ \hline
Y2 & 2 & - & 1 & 2 & - & - & 4 & 5  & - & - \\ \hline
R1 & 4 & - & 3 & 1 & 7 & 5 & 6 & 8  & 2 & 8 \\ \hline
Y3 & - & 2 & - & 3 & - & 4 & 1 & -  & 5 & - \\ \hline
R2 & 1 & 4 & - & 4 & 4 & 2 & - & 7  & - & 3 \\ \hline
Y4 & 8 & 7 & 2 & 4 & 8 & 8 & 5 & 3  & 1 & 6 \\ \hline
R3 & 3 & 2 & 3 & 1 & 10 & 9 & 7 & 5 & 6 & 8 \\ \hline
R4 & - & - & - & - & - & - & - & - & - & - \\ \hline
G2 & - & 3 & 6 & 8 & 4 & 5 & 9 & 2  & 1 & 7 \\ \hline
\end{tabular}
\end{table}

Looking at the leading sectors more closely, we find a mix between shocks
starting in EN and BM, and shocks starting in the fringe domestic sectors.  In
Table \ref{table:sectorstartdateranks}, we rank the start dates in the ten
sectors from earliest to latest, for each of the 11 corresponding segments.  In
cases where we have joint leaders, for example, EN and FN in G1, we split the
count between them.  In this way, we find that out of the 11 volatility shocks,
only two and a half originated from EN and BM.  The remaining eight and a half
shocks originated in fringe domestic sectors which are effectively not coupled
to the global market. This suggests that the US economy experiences internal
feedbacks that are stronger than its coupling to the global economy.  More
interestingly, we find in Fig.~\ref{fig:segmentMSTs} anomalously high cross
correlations at the fringe for some corresponding segments, for example, the
HC-NC link in Y1, the TC-IN link in B, and the TL-CY link in Y4.  As we can see
from Table ~\ref{table:sectorstartdateranks}, Y1 started in HC, B started in TC,
and Y4 started in TL.  This suggests that fringe cross correlations frequently
become anomalously high in the leading sector of a volatility shock.  This is
opposite to what we saw for the previous crisis, where there is a pronounced
`distancing-the-leader' effect, i.e.~the cross correlations between the leader
sector and all other sectors are smaller than the typical cross correlations
within the other sectors \cite{GladysFYP2009}.

Before we move on to compare the MST representation against the PMFG
representation of cross correlations between the ten DJUS economic sectors, let
us see what kind of macroeconomic significance we can attach to these
corresponding segments.  Clearly, until detailed studies of individual episodes
within the present financial crisis are completed and their full reports made
available, we have to rely on news reports for our macroeconomic interpretation.
Searching for and annotating news reports for this purpose is a very challenging
task, and thus we will only offer interpretations for the B, Y2, R1, R2, R3, and
R4 corresponding segments, where highly plausible news can be identified.  As it
turns out, both the B and Y2 corresponding segments were triggered by Federal
Reserve rate cuts \cite{FRB}, to 4.75\% on September 18, 2007, and to 4.50\% on
October 31, 2007 respectively.  In these two corresponding segments
(Fig.~\ref{fig:segmentMSTs}(c) and Fig.~\ref{fig:segmentMSTs}(d)), the
volatility shock is of a benevolent nature, and propagated very quickly into the
IN-NC core of the MSTs.  In contrast, the R1 corresponding segment was triggered
by a combination of the December 2007 employment situation report released on
January 4, 2008 by the US Bureau of Labor Statistics \cite{BLS4Jan2008}, and an
Institute for Supply Management report on the US service sector released in
December 2007 \cite{CNBC4Jan2008}.  As we can see in
Fig.~\ref{fig:segmentMSTs}(e), the volatility shock associated with the negative
news arrived at the IN-NC core later, suggesting that the more open chain-like
MST does indeed insulate its core from malign forces in the market.

Of the remaining three corresponding segments whose news trigger we managed to
identify, R2 and R3 are both related to the Lehman Brothers fiasco.  After the
US Bureau of Labor Statistics released their `worst employment report in five
years' on June 6, 2008 \cite{BLS6Jun2008}, Lehman Brothers announced on June 9,
2008 that it was expecting a 2.8 billion USD loss for 2008, and planned to raise
6 billion USD through sale of stock and convertible preferred stock
\cite{BBCNews9Jun2008}.  These events triggered a very short-lived R2
corresponding segment (Fig.~\ref{fig:segmentMSTs}(g)), which is unusual for how
rapidly the IN core responded to the volatility shock.  After R2, the market
calmed back down, but remained nervously in a high-volatility phase, until R3
started in mid-August 2008, as the demise of Lehman Brothers unfolded.  On
August 22, 2008, Lehman Brothers share prices soared when reports emerged that
the Korea Development Bank was considering buying the bank
\cite{MSNBC22Aug2008}.  When this acquisition fell through, Lehman Brothers
announced on August 28, 2008 its plans to lay off 1,500 employees
\cite{NYTimes28Aug2008}.  With no hope of government assistance on the horizon,
US Treasury Secretary Timoth Geithner arranged for last-minute talks over the
weekend of September 13 and 14, 2008, to have either the Bank of America or
Barclays buy over the entire Lehman Brothers.  When this last ditch effort
failed, Lehman Brothers file for Chapter 11 bankruptcy protection on Monday,
September 15, 2008, citing debts of 613 billion USD \cite{BBCNews15Sep2008}.
Lehman Brothers was then broken up.  Its brokerage part was sold to Barclays on
September 20, 2008 \cite{BBCNews20Sep2008}.  The rest of the company, which
includes franchise in the Asia Pacific region, Europe, and the Middle East, were
acquired by Nomura Holdings over a period spanning September 22, 2008 to October
13, 2008 \cite{JCNNetwork6Oct2008}.  The R3 corresponding segment, which
represents the tensest period in the present global financial crisis, ended in
mid-November 2008, before the prolonged debacle ended with Lehman Brothers'
investment management business, including Neuberger Berman, being sold on
December 3, 2008 to its management \cite{DealBook3Dec2008}.  Out of the 11
corresponding segments studied, the R3 MST (Fig.~\ref{fig:segmentMSTs}(i)) was
the most chain-like.  As expected, FN was the first to be hit by the saga.  But
while CY and BM were hit right after FN, the IN core of the MST was the second
last to succumb to the Lehman Brothers volatility shock.  This is another strong
testimony to the correlational insulation effect afforded by the chain-like
topology in the MST.

Finally, we identify the R4 corresponding segment, which started in late
December 2008 and ended in early May 2009, with the combination of crisis faced
by the US automobile industry, worries about liquidity levels in US banks, and
how the Federal Reserve was handling the financial crisis .  In mid-November,
Chrysler, Ford, and General Motors testified before the US Senate that they were in
urgent need of government assistance \cite{BBC19Nov2008}.  This was followed by
partisan politics delaying the rescue efforts.  While this crisis was playing
out on the American consciousness, investors were probably also getting worried
about the imminent January 30, 2009 expiry of temporary exceptions to sections
23A and 23B limitations \cite{PWCSep2008}.  These temporary exceptions allowed
existing funds to be more easily shared within financial groups, and were part
of the many measures announced by the Federal Reserve to ensure more efficient
use of liquidity by the banks.  The Federal Reserve extended temporary
exceptions to section 23A, and eventually allowed these to expire on October 30,
2009 \cite{NYFRB12150}.  But the news that most likely triggered the R4 segment
amidst this backdrop of negative news, was the December 30, 2008 press release
that the Federal Reserve will purchase toxic assets and make emergency loans
\cite{FRBPress30Dec2008}, and the market's realization that this will be to the
tune of 1.2 trillion USD \cite{WashingtonPost19Mar2009}.  The R4 corresponding
segment ended in early May 2009, as Chrysler filed for its inevitable Chapter 11
bankruptcy protection on May 1, 2009 \cite{Xinhuanet1May2009}, follwed by
General Motors one month later \cite{Telegraph1Jun2009}.  A US banks stress test
was also carried out by the Federal Reserve for 19 of the largest US banks.
This started in April 2009 and ended in early May 2009 \cite{CNBC7May2009}, but is
probably not strongly related to the end of R4, because this corresponding
segment ended in FN before the stress test started.  Instead, the stress test
showed up as a little blip in the temporal distribution of FN (see
Fig.~\ref{fig:segments}).  Looking at Fig.~\ref{fig:segmentMSTs}, we see also
that the R4 MST was already rather close to being star-like, suggesting that in
spite of the turmoil within R4, which included the March 2009 stock markets low,
the US economy was silently recovering from the crisis.

\subsection{Comparison between MST and PMFG}
\label{sect:MSTvsPMFG}

The \emph{planar maximally filtered graph (PMFG)} was introduced by Tumminello
\emph{et al.}~to extract a representative subgraph of the cross-correlation
matrix containing more information than the MST
\cite{Tumminello2005PNAS102p10421}.  Since then, the method has been applied for
sector identification \cite{Coronnello2005ActaPhysPolB36p2653}, to develop
hierarchically nested factor models \cite{Tumminello2007EuroPhysLett78e30006},
to understand the time horizon dependence of equity returns
\cite{Tumminello2007EurPhysJB55p209}, in portfolio optimization
\cite{Tola2008JEconDynControl32p235}, and to understand the network structure of
cross correlations among the world market indices
\cite{Eryigit2009PhysicaA388p3551}.  More recently, Pozzi \emph{et al.} computed
the MSTs and PMFGs for the daily returns of 300 of the most-capitalized stocks
on the NYSE for different window sizes between 2001 and 2003, and found that the
center is always populated by stocks from the financial sector, whereas other
sectors share the peripheral \cite{Pozzi2008AdvComplexSys11p927,
DiMatteo2010EurPhysJB73p3}.  This conclusion is different from what we arrived
at based on the DJUS economic sector indices between 2002 and 2003 (near the end
of the previous financial crisis), where IN remains central, and FN sits on the
periphery of the chain-like MST (Fig.~\ref{fig:MSTgross}(b)). 

\begin{figure}[htbp]
\centering\footnotesize
\includegraphics[scale=0.2]{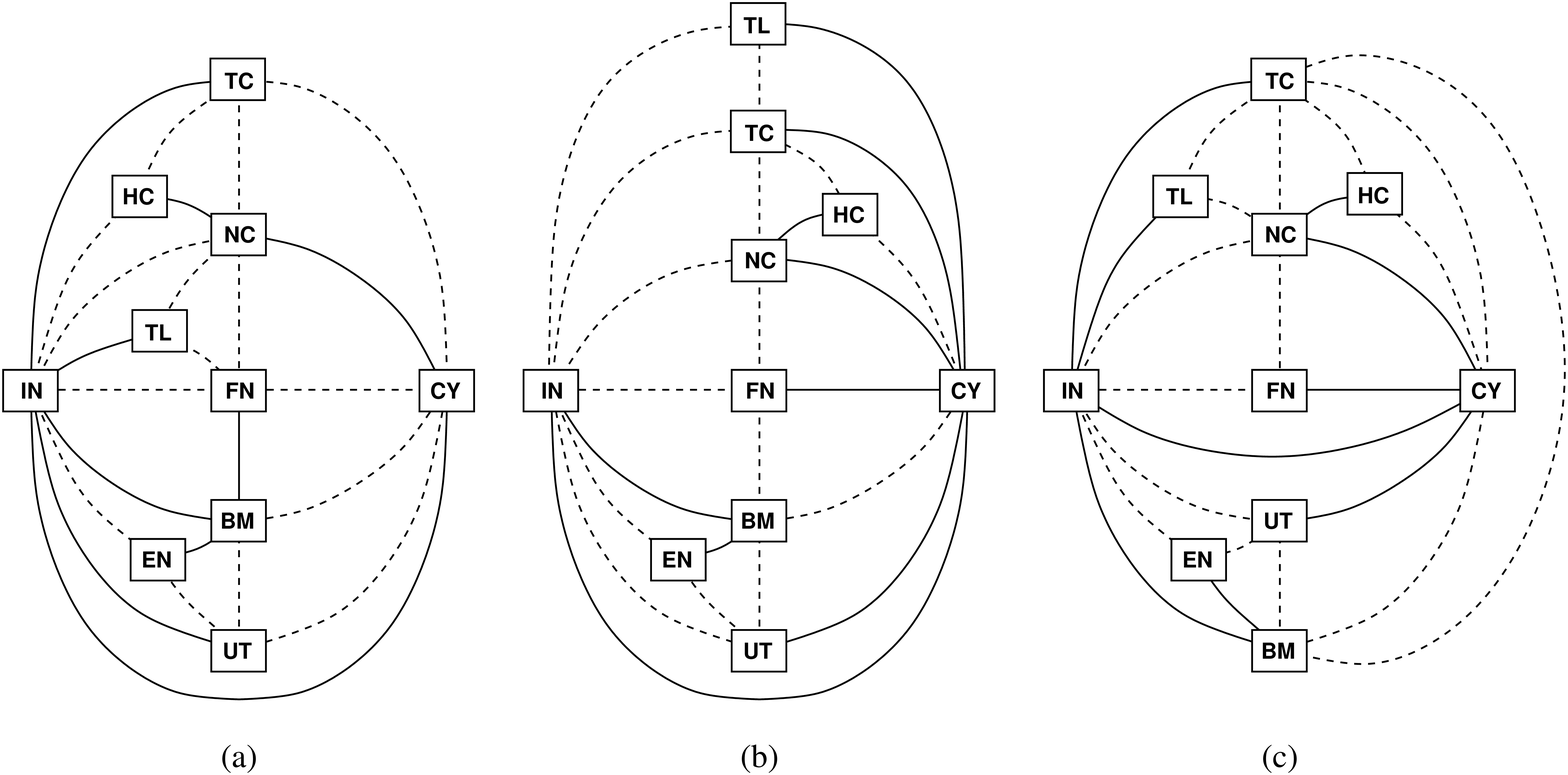}
\caption{Planar maximally filtered graphs (PMFGs) of the ten DJUS economic
sectors, constructed using half-hourly time series from (a) July 25, 2007 to
August 14, 2007; (b) July 27, 2007 to August 10, 2007; and (c) July 23, 2007 to
August 16, 2007.  In this figure, solid links are strong links making up the
MSTs, while dashed links are weaker links neglected in the MSTs.}
\label{fig:robustPMFG}
\end{figure}

Because our main interest in this study is the present financial crisis, we did
not construct the sectorial PMFG for the 2002--2003 period.  Instead, we
constructed the PMFGs for the three corresponding segments (Y1, G1, B) at the
start of the Subprime Crisis.  To check that the PMFG is also a robust
caricature of the cross correlations between the ten DJUS economic sectors, we
constructed the Y1 PMFGs for the intervals (a), (b), and (c) identified in
Section \ref{sect:mst}.  These are shown in Fig.~\ref{fig:robustPMFG}.  As we
can see, the only differences between the PMFGs for intervals (a) and (b) are
the positions and linkages of HC and TL.  On the other hand, the PMFG for
interval (c), which incorporates more than one segment for most economic
sectors, is quite different from the PMFG for interval (a).  In all three PMFGs,
we see IN and CY play the roles of primary and secondary centers.  This ability
to reveal secondary centers in the sectorial dynamics of the US economy is the
main advantage of the PMFG visualization has over the MST visualization.

\begin{figure}[ht]
\centering
\includegraphics[scale=0.27]{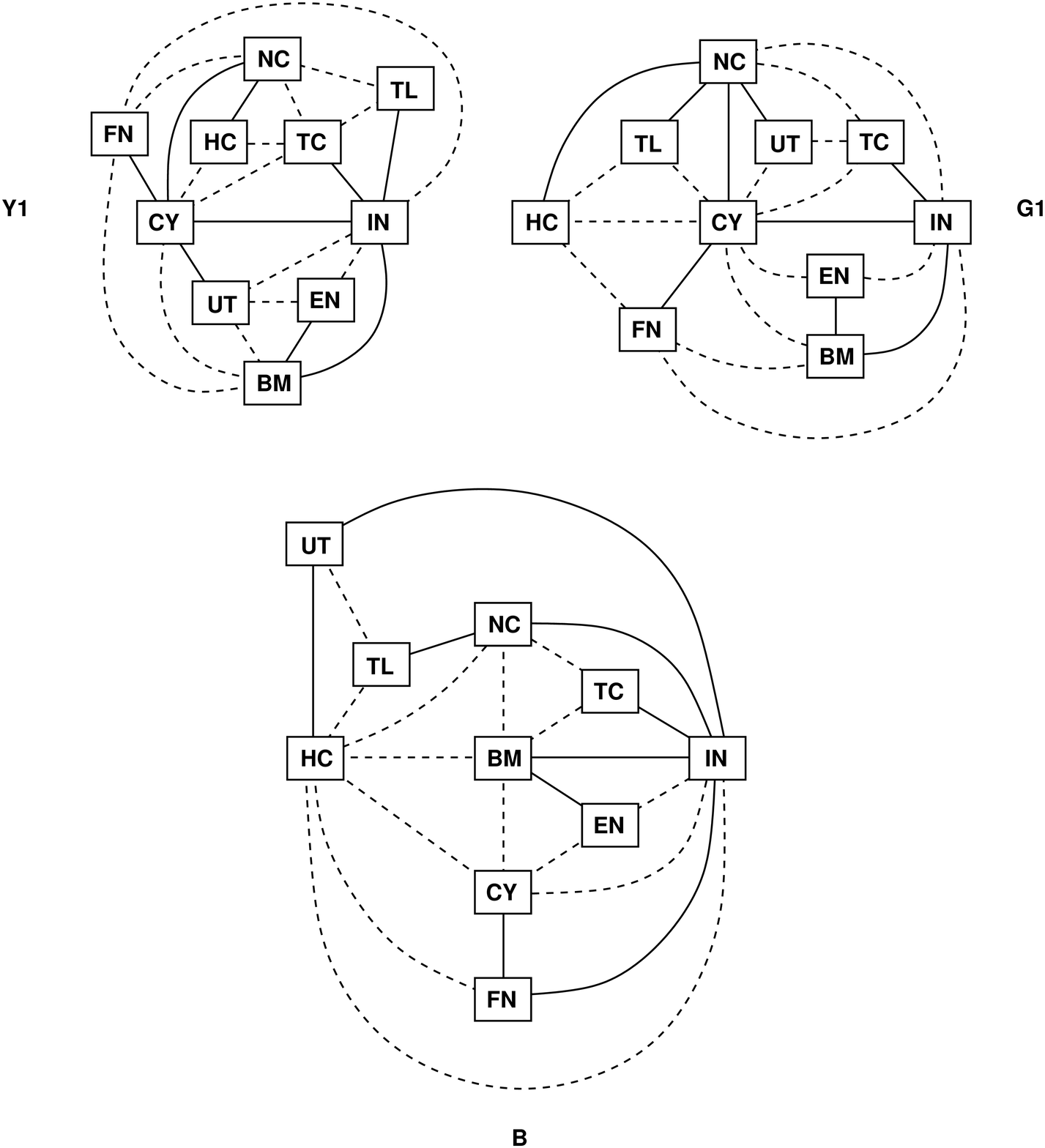}
\caption{Planar maximally filtered graphs (PMRGs) of the corresponding segments
Y1, G1, and B identified in Fig.~\ref{fig:segments}.  In this figure, solid
links are strong links making up the MSTs, while dashed links are weaker links
neglected in the MSTs.}
\label{fig:PMFG}
\end{figure}

For example, if we adopt the very simplistic criterion of having five or more
links to be a center in the PMFG, we see from Fig.~\ref{fig:PMFG} that BM (5),
CY (7), IN (7), NC (5), and TC (5) are all PMFG centers within the corresponding
segment Y1, CY (9), IN (6), NC (6) are the PMFG centers within the corresponding
segment G1, while BM (6), CY (5), IN (8), NC (5), HC (7) are the PMFG centers
within the corresponding segment B.  In particular, the PMFG structure of
corresponding segment B suggests that IN and HC are the two epicenters of
trading activities in October 2007.  Since IN is most strongly linked to growth
sectors (BM, CY, EN, FN, NC, TC) in the US economy, while HC is most strongly
linked to quality sectors (TL, UT), we believe we are seeing the signatures of a
`flight to quality' in the early stages of the Subprime Crisis.  Unlike the
`flights to quality' studied by economists (see for example, the recent works by
Phillips and Yu, who tracked the massive flow of funds from US technology stocks
to the US property market to commodities to the bond market, each time
generating a bubble that crashed when the funds leave
\cite{Phillips2010IntEconRev, Phillips2009SMUWP18}), the phenomenon we are
seeing is within the same asset class.

\section{MST rearrangements}
\label{sect:MSTrearrangements}

Up till this point, we understood from our combined segmentation/clustering and
cross-correlational analyses that the MST of the ten DJUS economic sectors
presents a star-like topology during economy growth, and a chain-like topology
during financial crisis (see Fig.~\ref{fig:MSTgross}).  These two limiting MSTs,
along with those of intermediate topologies, can also be seen at the mesoscopic
scale of corresponding segments within the present financial crisis (see
Fig.~\ref{fig:segmentMSTs}).  For each corresponding segment, we then looked at
the sectorial distribution of strong cross correlations, and the temporal order
in which sectors made the transition, to find that strong cross correlations are
frequently found at the fringe of the MST, where the volatility shocks always
originate.  In this section, we address the most natural question that follows:
what are the natures of the correlational changes, visualized as MST
rearrangements, that accompany these transitions?

\begin{figure}[htbp]
\centering
\includegraphics[scale=0.3]{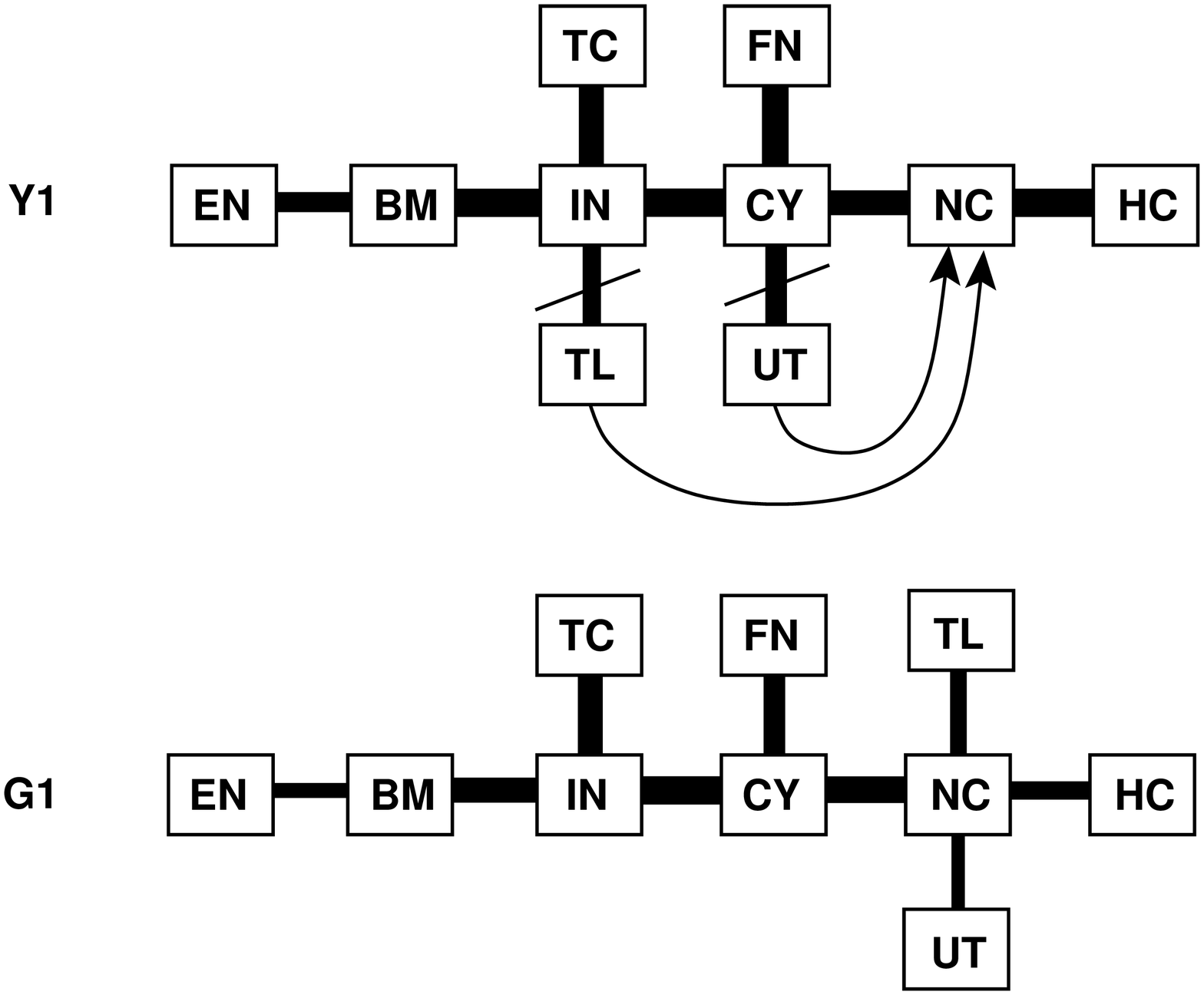}
\caption{The MST of the corresponding segment G1 can be obtained from the MST of
the corresponding segment Y1 preceding it, by breaking the TL-IN and UT-CY
bonds, and forming new bonds between TL-NC and UT-NC.}
\label{fig:MSTY1G1}
\end{figure}

\subsection{Minimal MST rearrangements}

If we treat the MST like a molecule, the MST rearrangements that occur from one
corresponding segment to the next can be described using the chemical language
of bond breaking and bond formation.  This analogy is useful, because it allows
us to focus on identifying the minimal set of primitive rearrangements that
occur in the MST, an example of which is shown in Fig.~\ref{fig:MSTY1G1}.
Between the successive corresponding segments Y1 and G1 identified in
Fig.~\ref{fig:segments}, we first note that the EN-BM-IN(-TC)-CY(-FN)-NC-HC
backbone remains unchanged.  We then note that TL, which is bonded to IN in Y1,
and UT, which is bonded to CY in Y1, are both bonded to NC in G1.  This tells us
that the minimal set of primitive rearrangements necessary to get from the Y1
MST to the G1 MST consists of the breaking of the TL-IN and UT-CY bonds, and the
formation of the TL-NC and UT-NC bonds.  We also see from Fig.~\ref{fig:MSTY1G1}
that, as expected, all MST cross correlations decreased going from Y1 to G1.  In
fact, all cross correlations decreased going from Y1 to G1.  Therefore, to have
the above rearrangements, we need $C(\rm TL, NC)$ to weaken slower than $C(\rm
TL, IN)$, or have $C(\rm TL, IN)$ weaken faster than $C(\rm TL, NC)$.
Similarly, we need $C(\rm UT, NC)$ to weaken slower than $C(\rm UT, IN)$, or
have $C(\rm UT, CY)$ weaken faster than $C(\rm UT, NC)$.  In any case, we need
at least one cross correlation within the (TL-IN, TL-NC) and (UT-CY, UT-NC)
pairs of cross correlations to be anomalous, for the rearrangement to occur.

\begin{figure}[ht]
\centering
\includegraphics[scale=0.3]{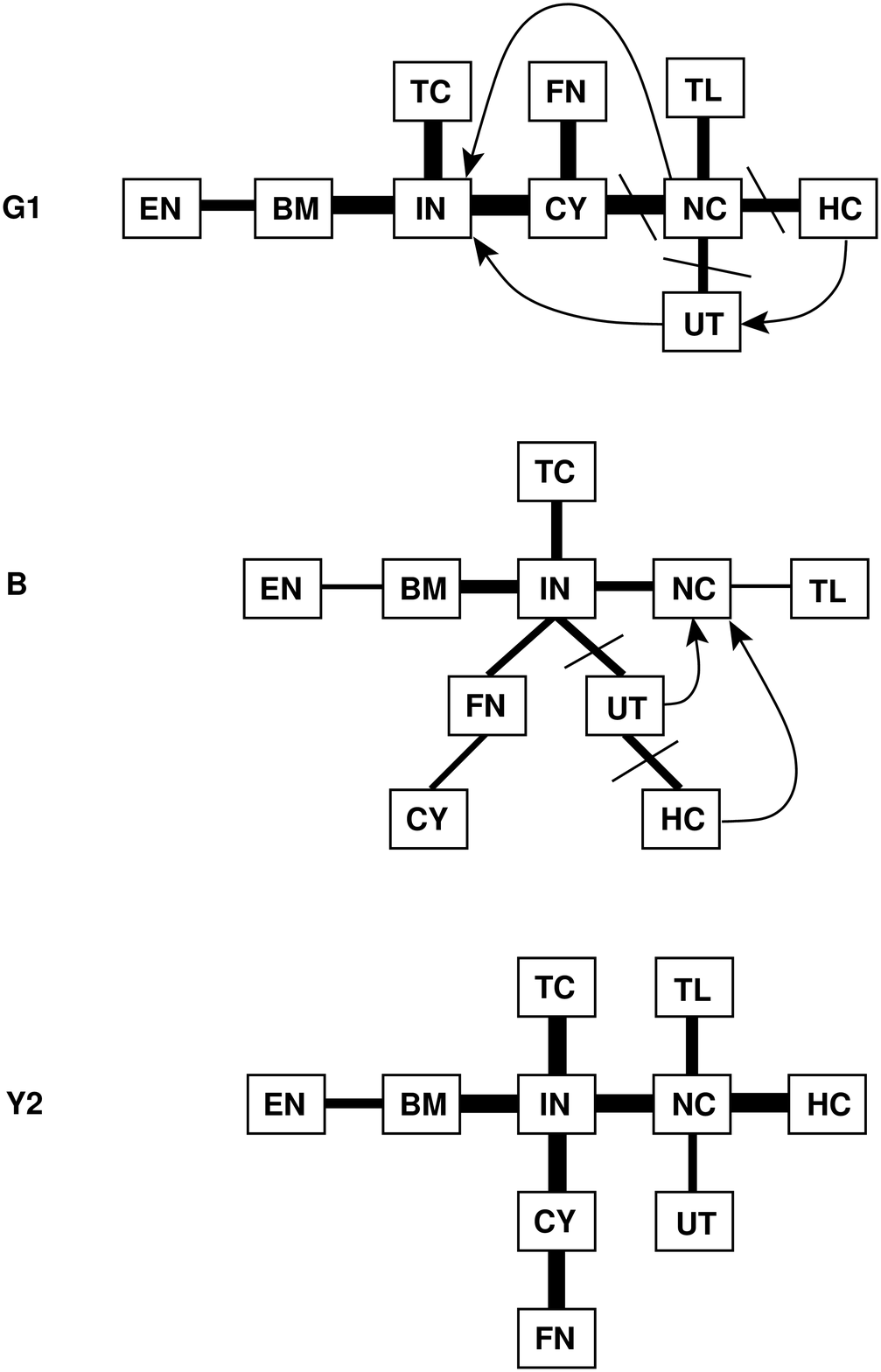}
\caption{The primitive MST rearrangements going from G1 to B to Y2.  The MST
went from chain-like in G1 to star-like in B, to an intermediate topology in Y2.}
\label{fig:MSTG1B}
\end{figure}

With this `chemical' understanding of minimal MST rearrangements, we now proceed
to investigate the cross-correlational changes going from corresponding segments
G1 to B to Y2, as shown in Fig.~\ref{fig:MSTG1B}.  Accompanying the Y1 to G1
transition, we saw a chain-like MST rearranging into another chain-like MST.
For the G1 to B to Y2 transitions, we see the more interesting MST
rearrangements from chain-like to star-like, and then to a topology intermediate
between a chain and a star.  As expected, more primitive rearrangements are
needed to bring about the chain-to-star transition going from the
moderate-volatility G1 to the low-volatility B.  Ignoring the change in sector
directly bonded to IN within the CY-FN pair, we see that three bonds have to be
broken and reformed.  These three bonds are significant, because NC is nearly a
star center in the G1 MST, but loses the status after the three bonds are
broken.  Of course, the bonds reformed around IN, making it the star center of
the B MST.  A similar interaction between NC and IN occurs again for the B to Y2
transition, where NC becomes central again, with the breaking of the UT-IN and
HC-UT bonds to reform around NC.  Since these corresponding segments are right
after the start of the Subprime Crisis, it is no wonder that NC features so
prominently in the MST rearrangements.

Quantitatively, we expect cross correlations to fall generically between
all sectors, when the US economy progressed from the moderate-volatility G1
segment to the low-volatility B segment.  This can be seen easily from the
thick bonds in the G1 MST compared to the thin bonds in the B MST in
Fig.~\ref{fig:MSTG1B}.  In fact, the drop in average cross correlations of CY is
anomalously large, from $\langle C\rangle(\rm CY) = 0.80$ in G1, to $\langle
C\rangle(\rm CY) = 0.45$ in B.  In addition, when all other cross correlations
were falling, that between UT and HC \emph{increased} slightly.  This bucking of the
trend makes the correlational changes between UT and HC highly significant
statistically.  Subsequently, when the US economy progressed from the
low-volatility B segment to the high-volatility Y2 segment, the cross
correlation between UT and HC \emph{decreased}, when the cross correlations
between all other sectors increased.

\subsection{Early detection of economic recovery}

Speaking of `green shoots' of economic revival that were evident at the time,
Federal Reserve chairman Ben Bernanke predicted that ``America's worst recession
in decades will likely end in 2009 before a recovery gathers steam in 2010''
\cite{Reuters16Mar2009}.  After learning that the MST of the ten DJUS economic
sectors is star-like and chain-like during the low-volatility economic growth
phase and high-volatility economic crisis phase respectively, we look out for a
star-like MST in the time series data of 2009 and 2010.  Star-like MSTs can also
be found deep inside an economic crisis phase.  However, within the crisis
phase, these star-like MSTs very quickly unravel to become chain-like MSTs.  On
the other hand, the star-shape topology is extremely robust and stable within
the growth phase.  Therefore, a persistent star-like MST, if it can be found,
may be interpreted as the statistical signature that the US economy is firmly on
track to full recovery (which may take up to two years across all sectors).

\begin{figure}[htbp]
\centering\footnotesize
\begin{minipage}[t]{.49\linewidth}
\centering
\includegraphics[scale=0.4]{MST2004growth}
\vskip .5\baselineskip
(a)
\end{minipage}
\hfill
\begin{minipage}[t]{.49\linewidth}
\centering
\includegraphics[scale=0.4]{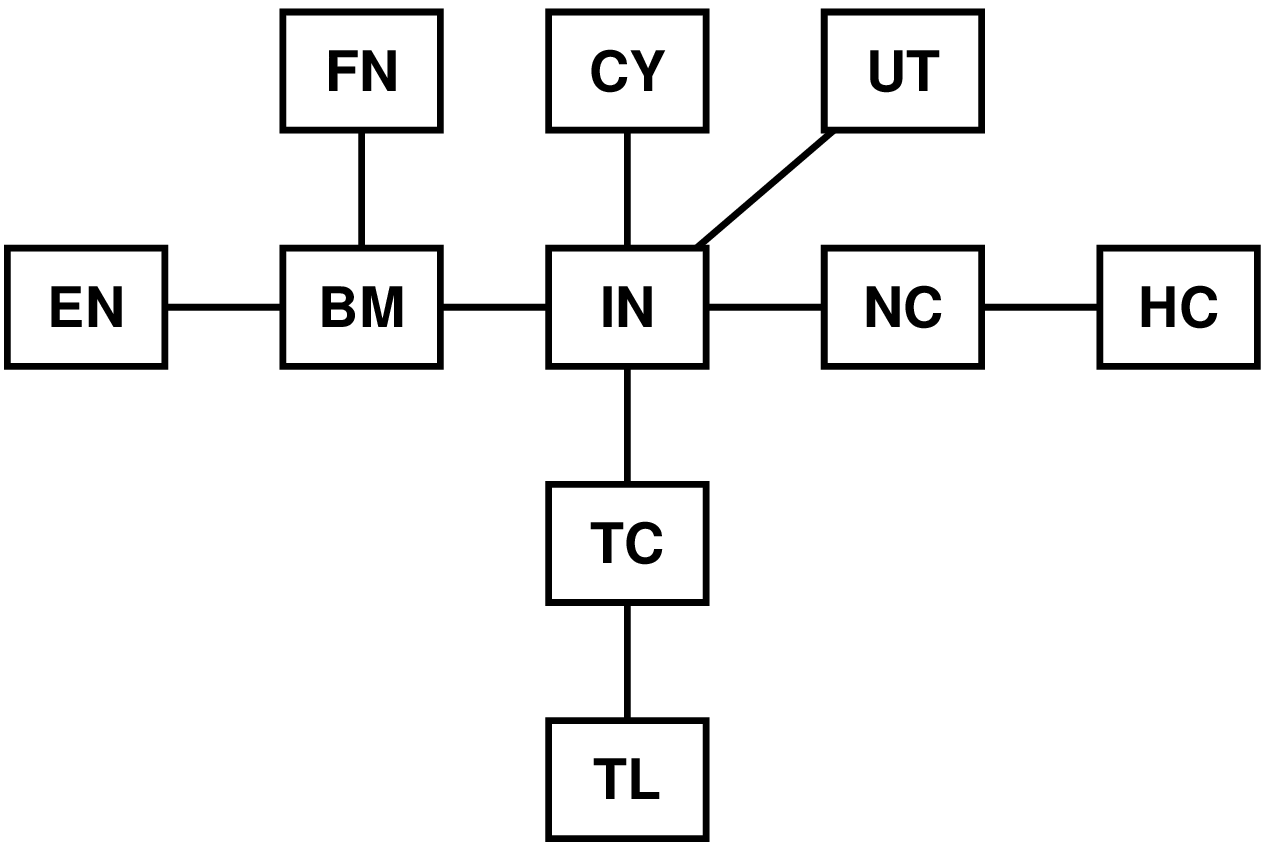}
\vskip .5\baselineskip
(b)
\end{minipage}
\caption{Comparison of the MSTs for (a) the 2004--2005 growth period, and (b) the
moderate-volatility segment around September 2009.}
\label{fig:MSTstarcompare}
\end{figure}

More importantly, the number of primitive rearrangements needed to transform the
MST of a given period into a star-like MST indicates how close we are to the
actual recovery.  We can use this feature of the prerecovery MST for the early
detection of economic recovery.  This should be possible whether the star-like
MST is a cause, in the sense that such a correlational structure within the US
economy promotes growth, or an effect, in the sense that economic growth
naturally results in this MST topology.  Indeed, when we inspect the MST
structure of the moderate-volatility G2 segment in Sep 2009, we find that it is
already star-like, with IN as the star center.  From
Fig.~\ref{fig:MSTstarcompare}, we see that it is two to three primitive
rearrangements away from the growth MST of 2004--2005.  Therefore, based on the
time series data up till 25 Nov 2009, the statistical evidence summarized in the
MST suggests that the US economy was already in the pre-recovery stage, and
Bernanke might be prophetic to call for an actual economic recovery in 2010.

\subsection{Comparison between previous and present recoveries}

\begin{figure}[htbp]
\centering\footnotesize
\begin{minipage}[t]{.49\linewidth}
\centering
\includegraphics[scale=0.25]{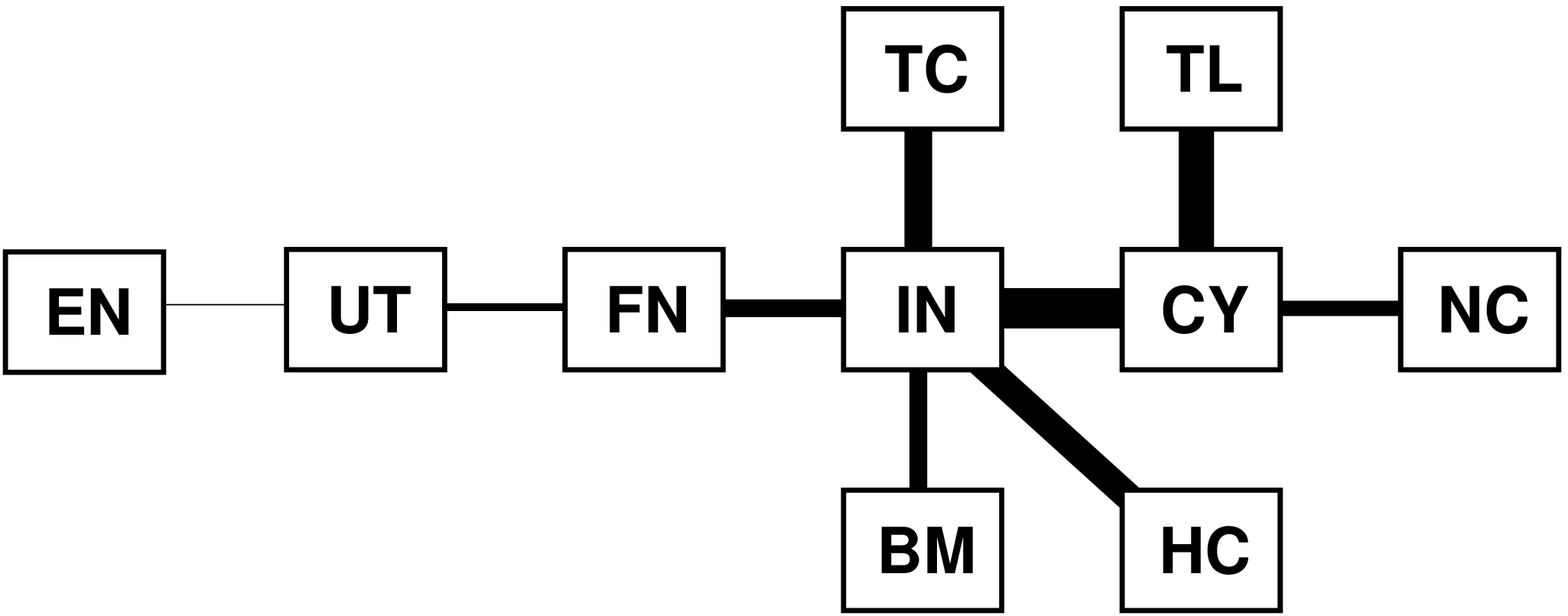}
\vskip .5\baselineskip
(a)
\end{minipage}
\hfill
\begin{minipage}[t]{.49\linewidth}
\centering
\includegraphics[scale=0.25]{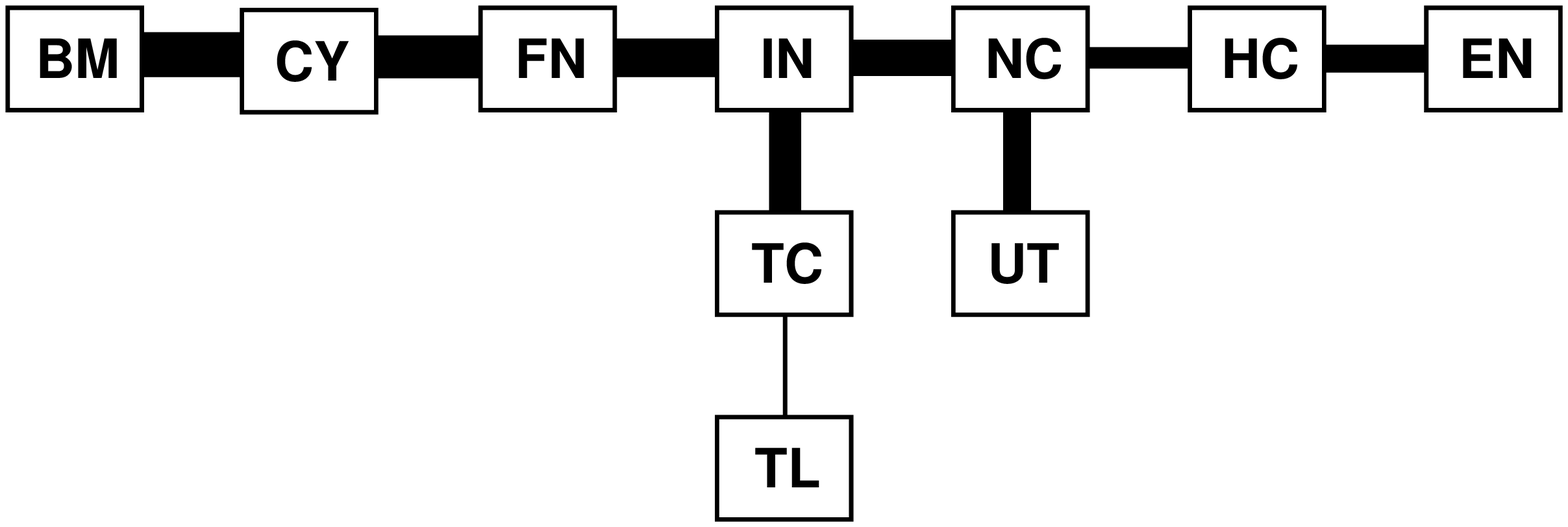}
\vskip .5\baselineskip
(b)
\end{minipage}
\vskip\baselineskip
\begin{minipage}[t]{.49\linewidth}
\centering
\includegraphics[scale=0.25]{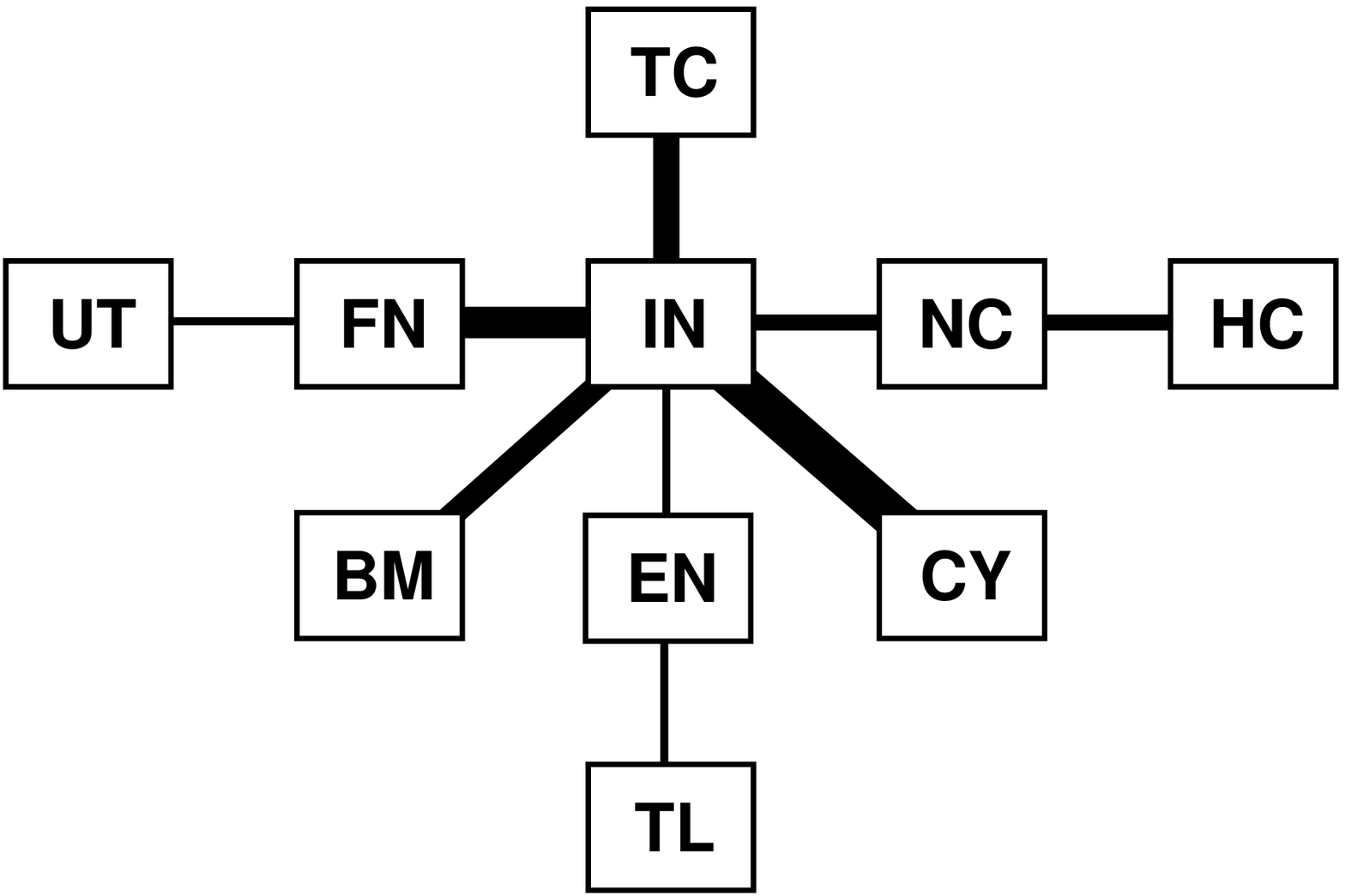}
\vskip .5\baselineskip
(c)
\end{minipage}
\hfill
\begin{minipage}[t]{.49\linewidth}
\centering
\includegraphics[scale=0.25]{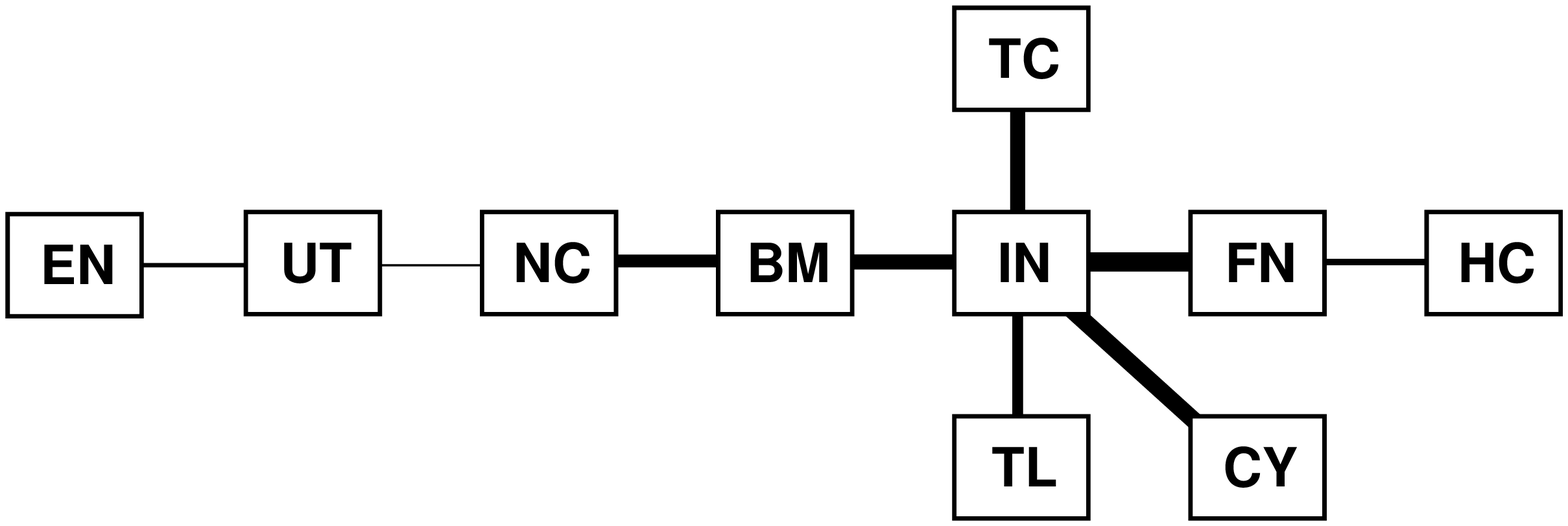}
\vskip .5\baselineskip
(d)
\end{minipage}
\caption{MSTs for successive corresponding segments: (a) moderate-volatility
segment before the 11 Sep 2001 attack on the World Trade Center; (b)
extremely-high-volatility segment right after the 11 Sep 2001 attack; (c)
high-volatility segment following; and (d) moderate-volatility segment
following.}
\label{fig:MST2001}
\end{figure}

Since the time series data we have covers the recovery periods for both
financial crises, we wanted also to compare the sequence of pre-recovery MSTs
for the previous crisis against the ones we see for the present crisis.  We
immediately encountered two problems.  First, volatility movements in the
various sectors between 2002 and 2004 are much less coordinated than those we
find in the present financial crisis, and thus it is difficult to find
corresponding segments.  Second, for successive corresponding segments that we
can find for the 2002--2004, successive MSTs are structurally very different
from each other, suggesting very violent rearrangements within the MSTs.  In
Fig.~\ref{fig:MST2001}, we show the MSTs of four successive corresponding
segments identified before and after the 11 Sep 2001 attack, from August 2001 to
December 2001.  These corresponding segments, an August 2001 moderate-volatility
segment before the 11 Sep 2001 attack, a two-week extremely-high-volatility
segment right after the attack, a October 2001 high-volatility segment following
this, and a November--December 2001 moderate-volatility segment afterwards, are
amongst the most well-defined ones that we can identified through the previous
financial crisis.  As we can see, it is impossible to assign a small number of
bonds that must be broken and reformed to go from one MST to the next.
Throughout the violent rearrangements, IN remained at the center of the MSTs.
We also see that the MST is chain-like before and after the 11 Sep 2001 attack,
became briefly star-like in the October 2001 high-volatility segment, before
unravelling again to a chain-like MST for the rest of the year.

\section{Conclusions}
\label{sect:conclusions}

To summarize, we performed a cross-section analysis on the high-frequency time
series of the ten DJUS economic sector indices between February 2000 and
November 2009, to discover statistical signatures that can be used to forecast
economic recovery.  The half-hourly time series of these indices are first
segmented individually using a recursive entropic segmentation scheme.  The
segments of each economic sector are then hierarchically clustered into between
four and seven clusters, representing the growth, crisis, correction, and crash
macroeconomic phases.  In our previous study \cite{Lee2009arXiv09114763}, we
compared the temporal distributions of clustered segments across all ten
economic sectors, to see that the US economy emerged from the previous
technology bubble financial crisis starting mid-2003, enjoyed a four-year period
of growth, and then succumbed to the present property bubble financial crisis
starting mid-2007.  From this cross-section of temporal distributions of
clustered segments, we also see the US economy taking one and a half years to
completely recover from the previous financial crisis, but only two months to
completely enter the present financial crisis.  More interestingly, for the
present financial crisis, we find the volatility dynamics within the US economic
sectors to be strongly driven by interest rate cuts by the Federal Reserve.  Of
the seven interest rate cuts made over 2007 and 2008, the first two lowered
market volatilities, the next two raised market volatilities, while the last
three had no permanent effect on market volatilities.

In this paper, we extended the cross-section analysis, by constructing the
cross-correlation matrices and therefrom the MSTs of the ten DJUS economic
sectors first over February 2000 to August 2008, and the two-year intervals
2002--2003, 2004--2005, 2008--2009, as well as the 11 corresponding segments
identified in Fig.~\ref{fig:segments}.  In general, we find stronger cross
correlatons when the market volatility is high, and weaker cross correlations
when the market volatility is low.  We also find evidence that cross
correlations within the US economy have been increasing over the years.  In all
MSTs, we find a core-fringe structure, with CY, IN, and NC forming the core, and
HC, TL, UT residing on the fringe.  In spite of the supposed market turmoil we
expect throughout the present financial crisis, a highly conserved
EN-BM-IN-CY-NC-TC-HC backbone can be identified in most of the MSTs.  Through an
enhanced visualization scheme for the MSTs, we see a dynamic core of
strongly-correlated sectors, which expands and contracts in tandem with changes
in the overall market volatility.  In addition, for all 11 corresponding
segments studied, we find the volatility shocks starting always at the fringe,
frequently accompanied by anomalously high cross correlations here, and
propagating inwards towards the core of the MST.  These volatility shocks
originate mostly from the US domestic fringe sectors, which are weakly coupled
to the world market, instead of coming from EN and BM, which are most strongly
coupled to the global supply and demand cycles.

More importantly, we see that the MSTs of the ten DJUS economic sectors can be
classified into two distinct topologies: star-like and chain-like.  The MST is
robustly star-like during economic growth, with IN at the center, and robustly
chain-like within an economic crisis.  For the present financial crisis, the MST
of a corresponding segment can be obtained from the MST of the preceding
corresponding segment through a small set of primitive rearrangements.  In
contrast, very violent rearrangements are implied going from the MST of one
corresponding segment to the MST of the next corresponding segment within the
previous, mid-1998 to mid-2003, financial crisis.  This suggests that the US
economy has become more efficient in processing information arising from
volatility shocks.  Combining these two observations, we postulated that the
star-like MST seen in the Sep 2009 G2 corresponding segment indicates that the
US economy was in the early stages of economic recovery.

After this study was completed, US market volatilities remained moderate to high
until the start of May 2010, when investor confidence was again tested, first by
the glitch in the NYSE electronic trading platform, and then by the unfolding
Greek Debt Crisis.  Market volatilities skyrocketed, and even after the European
Union announced their bailout plan for Greece, the atmosphere of economic
uncertainty lingered.  When interviewed in July 2010 on NBC's ``Meet the Press''
programme, US Treasury Secretary Timothy Geithner acknowledged the slow recovery
of the US economy, but added that it is gradually gaining strength
\cite{Yahoo25Jul2010}.  A commentary that appears the same day Geithner's
interview was aired complicates the mood, by citing economists who warn that
recent gains in the stock market need not be an indicator of economic recovery
\cite{YahooFinance25Jul2010}.  In fact, on 24 Aug 2010, world stock markets fell
over concerns that the yen is too strong for the good of the Japanese economy,
and also over more bad news anticipated from the US economic reports due to be
released the same week \cite{YahooFinance24Aug2010}.  

\begin{figure}[htbp]
\centering
\includegraphics[scale=0.25]{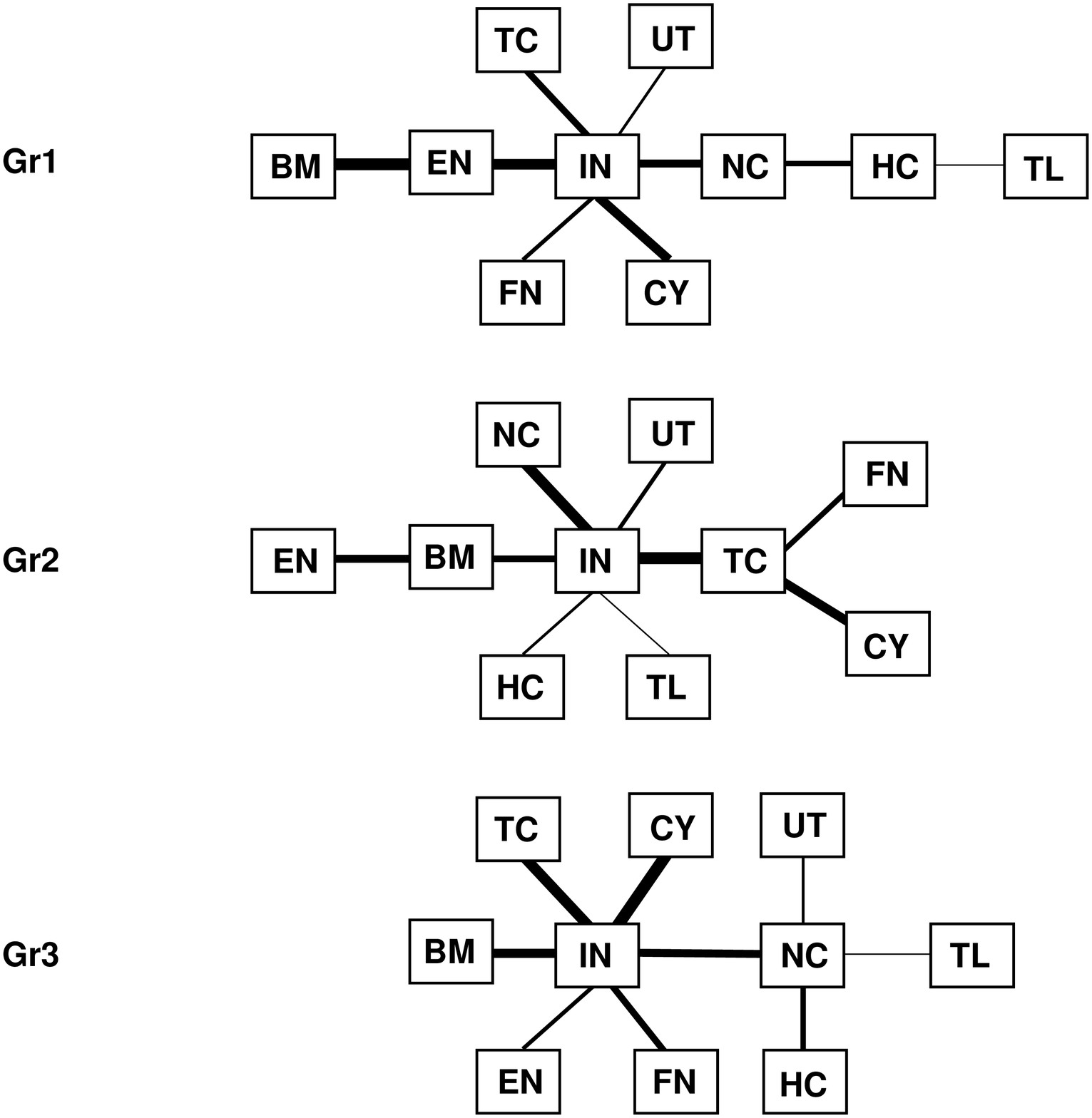}
\caption{MSTs for four successive corresponding segments straddling the Greek
debt crisis: Gr1 (extremely high volatility, 21 Jan to 15 Feb 2010), Gr2
(moderate to high volatility, 1--31 Mar 2010), and Gr3 (extremely high
volatility, 1 May to 15 Jul 2010).}
\label{fig:MSTGreek}
\end{figure}

To check if the US economic recovery might have been derailed by the Greek Debt
Crisis, we segmented the DJUS economic sector time series from January to July
2010, and constructed MSTs for three corresponding segments.  The MSTs for these
three corresponding segments, the extremely-high-volatility Gr1 segment (21
Jan--15 Feb 2010) and moderate-volatility Gr2 segment (1--31 Mar 2010) before
the Greek Debt Crisis, and the extremely-high-volatility Greek Debt Crisis Gr3
segment itself (1 May--15 Jul 2010), are shown in Fig.~\ref{fig:MSTGreek}.  As
we can see, even though market volatilities are high, the MST presented a very
robust star shape in all three corresponding segments.  While they are nervous,
it appears that investor sentiments during the Greek Debt Crisis are distinctly
different from those seen over 2008 and the first half of 2009.  Judging from
the increasingly star-like MST, it appears that the US economy is staying its
course to the long-awaited economic recovery.

\section*{Acknowledgements}

This research is supported by the Nanyang Technological University startup grant
SUG 19/07.  We have had helpful discussions with Chong Hui Tan.  We also thank
the referees for their comments, based on which we improved how the paper is
presented.

\appendix

\section{Top components of Dow Jones US economic sector indices}
\label{app:components}

\subsection{Basic Materials}

\begin{center}
\begin{tabular}{clc}
\hline
ISIN/Ticker & \parbox[c]{7cm}{\centering Company} & Adjusted Weight \\
\hline
FCX & Freeport-McMoRan Copper \& Gold Inc. & 10.29\% \\
DD & E.I. DuPont de Nemours \& Co. & 9.12\% \\
DOW & Dow Chemical Co. & 7.70\% \\
NEM & Newmont Mining Corp. & 6.14\% \\
PX & Praxair Inc. & 6.13\% \\
APD & Air Products \& Chemicals Inc. & 3.67\% \\
BTU & Peabody Energy Corp. & 3.38\% \\
AA & Alcoa Inc. & 2.89\% \\
PPG & PPG Industries Inc. & 2.78\% \\
ECL & Ecolab Inc. & 2.41\% \\
\hline
\end{tabular}
\end{center}

\subsection{Consumer Services}

\begin{center}
\begin{tabular}{clc}
\hline
ISIN/Ticker & \parbox[c]{7cm}{\centering Company} & Adjusted Weight \\
\hline
WMT & Wal-Mart Stores Inc. & 7.28\% \\
MCD & McDonald's Corp. & 5.33\% \\
DIS & Walt Disney Co. & 4.15\% \\
AMZN & Amazon.com Inc. & 3.90\% \\
HD & Home Depot Inc. & 3.28\% \\
CVS & CVS Caremark Corp. & 2.70\% \\
CMCSA & Comcast Corp. Cl A & 2.64\% \\
TGT & Target Corp. & 2.44\% \\
DTV & DIRECTV Group Inc. & 2.30\% \\
WAG & Walgreen Co. & 2.18\% \\
\hline
\end{tabular}
\end{center}

\subsection{Oil \& Gas}

\begin{center}
\begin{tabular}{clc}
\hline
ISIN/Ticker & \parbox[c]{7cm}{\centering Company} & Adjusted Weight \\
\hline
XOM & Exxon Mobil Corp. & 25.57\% \\
CVX & Chevron Corp. & 11.68\% \\
SLB & Schlumberger Ltd. & 7.59\% \\
COP & ConocoPhillips & 6.01\% \\
OXY & Occidental Petroleum Corp. & 5.14\% \\
APA & Apache Corp. & 2.96\% \\
HAL & Halliburton Co. & 2.47\% \\
APC & Anadarko Petroleum Corp. & 2.28\% \\
DVN & Devon Energy Corp. & 2.08\% \\
NOV & National Oilwell Varco Inc. & 1.84\% \\
\hline
\end{tabular}
\end{center}

\subsection{Financials}

\begin{center}
\begin{tabular}{clc}
\hline
ISIN/Ticker & \parbox[c]{7cm}{\centering Company} & Adjusted Weight \\
\hline
JPM & JPMorgan Chase \& Co. & 7.50\% \\
WFC & Wells Fargo \& Co. & 6.73\% \\
BAC & Bank of America Corp. & 5.50\% \\
C & Citigroup Inc. & 5.04\% \\
GS & Goldman Sachs Group Inc. & 3.40\% \\
BRK/B & Berkshire Hathaway Inc. Cl B & 3.38\% \\
AXP & American Express Co. & 2.32\% \\
USB & U.S. Bancorp & 2.30\% \\
V & VISA Inc. Cl A & 1.85\% \\
BK & Bank of New York Mellon Corp. & 1.65\% \\
\hline
\end{tabular}
\end{center}

\subsection{Healthcare}

\begin{center}
\begin{tabular}{clc}
\hline
ISIN/Ticker & \parbox[c]{7cm}{\centering Company} & Adjusted Weight \\
\hline
JNJ & Johnson \& Johnson & 12.56\% \\
PFE & Pfizer Inc. & 9.68\% \\
MRK & Merck \& Co. Inc. & 7.82\% \\
ABT & Abbott Laboratories & 5.28\% \\
AMGN & Amgen Inc. & 3.72\% \\
BMY & Bristol-Myers Squibb Co. & 3.20\% \\
UNH & UnitedHealth Group Inc. & 3.03\% \\
MDT & Medtronic Inc. & 2.68\% \\
LLY & Eli Lilly \& Co. & 2.44\% \\
GILD & Gilead Sciences Inc. & 2.26\% \\
\hline
\end{tabular}
\end{center}

\subsection{Industrials}

\begin{center}
\begin{tabular}{clc}
\hline
ISIN/Ticker & \parbox[c]{7cm}{\centering Company} & Adjusted Weight \\
\hline
GE & General Electric Co. & 10.45\% \\
UTX & United Technologies Corp. & 4.03\% \\
MMM & 3M Co. & 3.39\% \\
UPS & United Parcel Service Inc. Cl B & 3.10\% \\
CAT & Caterpillar Inc. & 2.98\% \\
UNP & Union Pacific Corp. & 2.77\% \\
BA & Boeing Co. & 2.58\% \\
EMR & Emerson Electric Co. & 2.57\% \\
HON & Honeywell International Inc. & 2.15\% \\
DE & Deere \& Co. & 1.95\% \\
\hline
\end{tabular}
\end{center}

\subsection{Consumer Goods}

\begin{center}
\begin{tabular}{clc}
\hline
ISIN/Ticker & \parbox[c]{7cm}{\centering Company} & Adjusted Weight \\
\hline
PG & Procter \& Gamble Co. & 13.34\% \\
KO & Coca-Cola Co. & 10.35\% \\
PM & Philip Morris International Inc. & 8.02\% \\
PEP & PepsiCo Inc. & 7.91\% \\
F & Ford Motor Co. & 4.09\% \\
MO & Altria Group Inc. & 3.85\% \\
KFT & Kraft Foods Inc. Cl A & 3.73\% \\
CL & Colgate-Palmolive Co. & 2.90\% \\
MON & Monsanto Co. & 2.50\% \\
NKE & Nike Inc. Cl B & 1.97\% \\
\hline
\end{tabular}
\end{center}

\subsection{Technology}

\begin{center}
\begin{tabular}{clc}
\hline
ISIN/Ticker & \parbox[c]{7cm}{\centering Company} & Adjusted Weight \\
\hline
AAPL & Apple Inc. & 13.57\% \\
MSFT & Microsoft Corp. & 9.33\% \\
IBM & International Business Machines Corp. & 8.58\% \\
GOOG & Google Inc. Cl A & 6.52\% \\
INTC & Intel Corp. & 5.65\% \\
CSCO & Cisco Systems Inc. & 5.31\% \\
ORCL & Oracle Corp. & 4.98\% \\
HPQ & Hewlett-Packard Co. & 4.73\% \\
QCOM & Qualcomm Inc. & 3.61\% \\
EMC & EMC Corp. & 2.11\% \\
\hline
\end{tabular}
\end{center}

\subsection{Telecommunications}

\begin{center}
\begin{tabular}{clc}
\hline
ISIN/Ticker & \parbox[c]{7cm}{\centering Company} & Adjusted Weight \\
\hline
T & AT\&T Inc. & 44.32\% \\
VZ & Verizon Communications Inc. & 24.29\% \\
AMT & American Tower Corp. Cl A & 5.45\% \\
CTL & CenturyLink Inc. & 3.47\% \\
S & Sprint Nextel Corp. & 2.99\% \\
CCI & Crown Castle International Corp. & 2.75\% \\
Q & \parbox[c]{7cm}{Qwest Communications International Inc.} & 2.67\% \\
FTR & Frontier Communications Corp. & 2.44\% \\
VMED & Virgin Media Inc. & 2.03\% \\
NIHD & NII Holdings Inc. & 1.72\% \\
\hline
\end{tabular}
\end{center}

\subsection{Utilities}

\begin{center}
\begin{tabular}{clc}
\hline
ISIN/Ticker & \parbox[c]{7cm}{\centering Company} & Adjusted Weight \\
\hline
SO & Southern Co. & 6.68\% \\
EXC & Exelon Corp. & 5.61\% \\
D & Dominion Resources Inc. (Virginia) & 5.28\% \\
DUK & Duke Energy Corp. & 4.94\% \\
NEE & NextEra Energy Inc. & 4.52\% \\
PCG & PG\&E Corp. & 3.96\% \\
AEP & American Electric Power Co. Inc. & 3.67\% \\
PEG & Public Service Enterprise Group Inc. & 3.38\% \\
SE & Spectra Energy Corp. & 3.32\% \\
ED & Consolidated Edison Inc. & 2.93\% \\
\hline
\end{tabular}
\end{center}

\end{document}